%% file: HIG-19-010_temp.tex
\pdfoutput=1
\documentclass[11pt,twoside,a4paper,cmspaper,final,collab]{cms-tdr}
\def\svnVersion{173e910}\def\svnDate{2023/07/06}\def\cmsCernNoTag{CERN-EP-2022-027}\def\cmsCernDate{\today}\def\cmsMessage{Published in the European Physical Journal C as \href{http://dx.doi.org/10.1140/epjc/s10052-023-11452-8}{\doi{10.1140/epjc/s10052-023-11452-8}.}}
\begin{document}\cmsNoteHeader{HIG-19-010}

\newlength\cmsTabSkip\setlength{\cmsTabSkip}{1ex}
\newlength\cmsFigWidthA
\ifthenelse{\boolean{cms@external}}{\setlength\cmsFigWidthA{0.49\textwidth}}{\setlength\cmsFigWidthA{0.70\textwidth}}
\newcommand{\HTT}{\ensuremath{\PH\to\PGt\PGt}\xspace}
\newcommand{\ggH}{\ensuremath{\Pg\Pg\PH}\xspace}
\newcommand{\VH}{\ensuremath{\PV\PH}\xspace}
\newcommand{\ttH}{\ensuremath{\ttbar\PH}\xspace}
\newcommand{\qqH}{\ensuremath{\PQq\PQq\PH}\xspace}
\newcommand{\Irelem}{\ensuremath{I_{\text{rel}}^{\Pe(\PGm)}}\xspace}
\newcommand{\ptem}{\ensuremath{\pt^{\Pe(\PGm)}}\xspace}
\newcommand{\Irelm}{\ensuremath{I_{\text{rel}}^{\PGm}}\xspace}
\newcommand{\Irele}{\ensuremath{I_{\text{rel}}^{\Pe}}\xspace}
\newcommand{\VHanapl}{\VH-analyses\xspace} 
\newcommand{\yDT}{\ensuremath{y^{\text{DT}}}\xspace}
\newcommand{\Dj}{\ensuremath{D_{\text{jet}}}\xspace}
\newcommand{\De}{\ensuremath{D_{\Pe}}\xspace}
\newcommand{\Dm}{\ensuremath{D_{\PGm}}\xspace}
\newcommand{\WH}{\ensuremath{\PW\PH}\xspace}
\newcommand{\ZH}{\ensuremath{\PZ\PH}\xspace}
\newcommand{\VHana}{\VH-analysis\xspace}
\newcommand{\NNana}{NN-analysis\xspace}
\newcommand{\Wjets}{\ensuremath{\PW}{+}\text{jets}\xspace}
\newcommand{\mtt}{\ensuremath{m_{\PGt\PGt}}\xspace}
\newcommand{\ZTT}{\ensuremath{\PZ\to\PGt\PGt}\xspace}
\newcommand{\ZLL}{\ensuremath{\PZ\to\Pell\Pell}\xspace}
\newcommand{\jettau}{\ensuremath{\text{jet}\to\tauh}\xspace}
\newcommand{\jetell}{\ensuremath{\text{jet}\to\Pell}\xspace}
\newcommand{\FF}{\ensuremath{F_{\mathrm{F}}}\xspace}
\newcommand{\FFi}{\ensuremath{\FF^{i}}\xspace}
\newcommand{\tautau}{\ensuremath{\tauh\tauh}\xspace}
\newcommand{\FFQCD}{\ensuremath{\FF^{\text{QCD}}}\xspace}
\newcommand{\FFWjets}{\ensuremath{\FF^{\Wjets}}\xspace}
\newcommand{\FFttbar}{\ensuremath{\FF^{\ttbar}}\xspace}
\newcommand{\Njet}{\ensuremath{N_{\text{jet}}}\xspace}
\newcommand{\mH}{\ensuremath{m_{\PH}}\xspace}
\newcommand{\mutau}{\ensuremath{\PGm\tauh}\xspace}
\newcommand{\mutautau}{\ensuremath{\PGm\tauh\tauh}\xspace}
\newcommand{\etau}{\ensuremath{\Pe\tauh}\xspace}
\newcommand{\etautau}{\ensuremath{\Pe\tauh\tauh}\xspace}
\newcommand{\mvis}{\ensuremath{m_\text{vis}}\xspace}
\newcommand{\emu}{\ensuremath{\Pe\PGm}\xspace}
\newcommand{\emutau}{\ensuremath{\Pe\PGm\tauh}\xspace}
\newcommand{\mumutau}{\ensuremath{\PGm\PGm\tauh}\xspace}
\newcommand{\TF}{\ensuremath{F_{\mathrm{T}}}\xspace}
\newcommand{\CBana}{CB-analysis\xspace}
\newcommand{\ptH}{\ensuremath{\pt^{\PH}}\xspace}
\newcommand{\NNanapl}{NN-analyses\xspace} 
\newcommand{\Dzeta}{\ensuremath{D_{\zeta}}\xspace}
\newcommand{\HWW}{\ensuremath{\PH\to\PW\PW}\xspace}
\newcommand{\ptV}{\ensuremath{\pt^{\PV}}}
\newcommand{\mjj}{\ensuremath{m_{\mathrm{jj}}}\xspace}
\newcommand{\detajj}{\ensuremath{\Delta\eta_\mathrm{{jj}}}\xspace}
\newcommand{\ptHhat}{\ensuremath{\hat{p}_{\mathrm{T}}^{\PH}}}
\newcommand{\ltau}{\ensuremath{\Pell\tauh}\xspace}
\newcommand{\yl}{\ensuremath{y_{l}}\xspace}
\newcommand{\yprimepk}{\ensuremath{y_{p}^{\prime(k)}}\xspace}
\newcommand{\Nbtag}{\ensuremath{N_{\text{Btag}}}\xspace}
\newcommand{\talpha}{\ensuremath{\langle t_{\alpha}\rangle}\xspace}
\newcommand{\yggH}{\ensuremath{y_{\ggH}}\xspace}
\newcommand{\yqqH}{\ensuremath{y_{\qqH}}\xspace}
\newcommand{\LT}{\ensuremath{L_{\mathrm{T}}}\xspace}
\newcommand{\ST}{\ensuremath{S_{\mathrm{T}}}\xspace}
\newcommand{\ZEE}{\ensuremath{\PZ\to\Pe\Pe}\xspace}
\newcommand{\ZMM}{\ensuremath{\PZ\to\PGm\PGm}\xspace}
\newcommand{\ptVhat}{\ensuremath{\hat{p}_{\mathrm{T}}^{\PV}}\xspace}
\newcommand{\FFjet}{\ensuremath{\FF^{\tauh}}\xspace}
\newcommand{\FFe}{\ensuremath{\FF^{\Pe}}\xspace}
\newcommand{\FFmu}{\ensuremath{\FF^{\PGm}}\xspace}
\newcommand{\ZZ}{\ensuremath{\PZ\PZ}\xspace}
\newcommand{\WW}{\ensuremath{\PW\PW}\xspace}
\newcommand{\HBB}{\ensuremath{\PH\to\PQb\PQb}\xspace}
\newcommand{\PGtPgt}{\ensuremath{\PGt\PGt}\xspace}
\newcommand{\PGmPGm}{\ensuremath{\PGm\PGm}\xspace}
\newcommand{\PePe}{\ensuremath{\Pe\Pe}\xspace}
\newcommand{\pp}{\ensuremath{\Pp\Pp}\xspace}
\newcommand{\sumCharged}{\ensuremath{\sum\pt^{\text{charged}}}\xspace}
\newcommand{\sumNeutral}{\ensuremath{\sum\et^{\text{neutral}}}\xspace}
\newcommand{\sumGamma}{\ensuremath{\sum\et^{\gamma}}\xspace}
\newcommand{\ptPU}{\ensuremath{\pt^{\text{PU}}}\xspace}
\newcommand{\ptvecE}{\ensuremath{\ptvec^{\kern1pt\Pe}}\xspace}
\newcommand{\ptvecM}{\ensuremath{\ptvec^{\kern1pt\PGm}}\xspace}
\newcommand{\pzetamiss}{\ensuremath{p_{\zeta}^\text{miss}}\xspace}
\newcommand{\pzetavis}{\ensuremath{p_{\zeta}^\text{vis}}\xspace}
\newcommand{\mTemu}{\ensuremath{\mT^{\emu}}\xspace}
\newcommand{\NSR}{\ensuremath{N_{\text{SR}}}\xspace}
\newcommand{\NAR}{\ensuremath{N_{\text{AR}}}\xspace}
\newcommand{\muinc}{\ensuremath{\mu_{\text{incl}}}\xspace}
\newcommand{\muggh}{\ensuremath{\mu_{\ggH}}\xspace}
\newcommand{\muqqh}{\ensuremath{\mu_{\qqH}}\xspace}
\newcommand{\muvh}{\ensuremath{\mu_{\VH}}\xspace}
\newcommand{\kappaV}{\ensuremath{\kappa_{\PV}}\xspace}
\newcommand{\kappaF}{\ensuremath{\kappa_{\mathrm{F}}}\xspace}
\ifthenelse{\boolean{cms@external}}{\providecommand{\cmsTable}[1]{#1}}{\providecommand{\cmsTable}[1]{\resizebox{\textwidth}{!}{#1}}}

\cmsNoteHeader{HIG-19-010}
\title{Measurements of Higgs boson production in the decay channel with a pair of \texorpdfstring{\PGt}{tau} leptons in proton-proton collisions at \texorpdfstring{$\sqrt{s}=13\TeV$}{sqrt(s)=13 TeV}}
\titlerunning{Higgs boson production in the \texorpdfstring{\PGt{}\PGt}{tau-tau} channel}
\date{\today}

\abstract{
Measurements of Higgs boson production, where the Higgs boson decays into a pair of \PGt leptons, are presented, using a sample of proton-proton collisions collected with the CMS experiment at a center-of-mass energy of 13\TeV, corresponding to an integrated luminosity of 138\fbinv. Three analyses are presented. Two are targeting Higgs boson production via gluon fusion and vector boson fusion: a neural network based analysis and an analysis based on an event categorization optimized on the ratio of signal over background events. These are complemented by an analysis targeting vector boson associated Higgs boson production. Results are presented in the form of signal strengths relative to the standard model predictions and products of cross sections and branching fraction to \PGt leptons, in up to 16 different kinematic regions. For the simultaneous measurements of the neural network based analysis and the analysis targeting vector boson associated Higgs boson production signal strengths are found to be $0.82\pm0.11$ for inclusive Higgs boson production, $0.67\pm0.19$ ($0.81\pm0.17$) for the production mainly via gluon fusion (vector boson fusion), and $1.79\pm0.45$ for vector boson associated Higgs boson production.
}

\hypersetup{
pdfauthor={CMS Collaboration},
pdftitle={Measurement of Higgs boson production in the decay channel with a pair of tau leptons in proton-proton collisions at sqrt(s)=13 TeV},
pdfsubject={CMS},
pdfkeywords={CMS, Higgs, taus, fermion couplings, STXS, simplified template cross section}}

\maketitle
\section{Introduction}

In the standard model (SM) of particle physics~\cite{Glashow:1961tr,Weinberg:1967tq, 
Salam:1968rm}, the masses of the \PW and \PZ bosons are obtained through their 
interaction with a fundamental field that enters the theory via the
Brout--Englert--Higgs mechanism~\cite{Englert:1964et,Higgs:1964ia,Higgs:1964pj,
Guralnik:1964eu,Higgs:1966ev,Kibble:1967sv}, in a process known as electroweak 
symmetry breaking. The Higgs boson (\PH) is the quantized manifestation of this 
field. A particle compatible with \PH was observed at the CERN LHC by the ATLAS 
and CMS experiments in the $\Pgg\Pgg$, \ZZ, and \WW final states using data 
collected in 2011--2012 at center-of-mass energies of $\sqrt{s}=7$ and 
8\TeV~\cite{Aad:2012tfa,Chatrchyan:2012ufa,Chatrchyan:2013lba}. The properties 
of the new particle, including its couplings, spin, and CP eigenstate are so far
consistent with those expected for a Higgs boson with a mass of $125.35\pm0.14
\GeV$~\cite{CMS:2020xrn} as predicted by the SM~\cite{CMS:2013fjq,
Khachatryan:2016vau,Sirunyan:2018koj,Aad:2019mbh,ATLAS:2020rej,CMS:2019ekd,       
CMS:2020cga,ATLAS:2020ior,CMS:2021nnc,ATLAS:2021pkb,ATLAS:2020evk,CMS:2021sdq}.      

In the SM, the mass generation of fermions is introduced in the form of Yukawa 
couplings to the Brout--Englert--Higgs field. Extensions of the SM, like 
supersymmetry~\cite{Golfand:1971iw,Wess:1974tw}, predict deviations of the \PH 
couplings, particularly to down-type fermions, such as the \PGt lepton or \PQb 
quark, which further increases interest in the \HTT decay~\cite{ATLAS:2017eiz,Sirunyan:2018zut}.

The \HTT decay, which in the SM and its extensions offers a much larger branching 
fraction than $\PH\to\PGm\PGm$ and reduced background compared to \HBB, is the 
most promising channel to study \PH decays to fermions. Accordingly, the first 
evidence for the \PH coupling to (down-type) fermions was found in the \HTT decay 
channel using data collected at $\sqrt{s}=7$ and 8\TeV~\cite{Chatrchyan:2014nva,
Chatrchyan:2014vua,Aad:2015vsa}. A combination of the measurements performed by 
the ATLAS and CMS experiments at the same $\sqrt{s}$ led to the first measurement 
of the \PH coupling to the \PGt lepton with a statistical significance of more 
than 5 standard deviations (s.d.)~\cite{Khachatryan:2016vau}. The first observation 
of \HTT decays with a single experiment was achieved by the CMS experiment adding 
data collected at 13\TeV in 2016 to the data collected at 7 and 
8\TeV~\cite{Sirunyan:2017khh}.

In recent years, \PH production rates have been investigated in the framework of 
the simplified template cross section (STXS) scheme, introduced by the LHC Higgs 
Working Group~\cite{deFlorian:2016spz}. This scheme defines a set of kinematic 
and topological phase space regions, referred to as STXS bins, for differential 
measurements. The STXS scheme supports the investigation of each production mode 
individually. It facilitates the combination of measurements across different \PH 
decay channels and across experiments. The STXS bins have been chosen to reduce 
the dependence on any underlying theoretical models embodied in the measurements. 
They have been defined in stages, corresponding to the anticipated statistical 
power of the data required to perform the measurements. In STXS stage-0, which 
was used in the analyses of the LHC Run-1 data, events have been assigned to 
basic categories according to their main \PH production mechanisms: 

\begin{enumerate}
\item gluon fusion;
\item vector boson fusion (VBF);  
\item quark-initiated \PH production in association with a vector boson \PV 
(\VH); 
\item \PH production in association with a top anti-top quark pair (\ttH).
\end{enumerate}

{\tolerance=800 Subsequent refinements to the STXS scheme resulted in division 
of the basic categories into finer bins, called stage-1.1~\cite{Berger:2019wnu}, 
and a redefinition of some of the stage-0 categories: Gluon fusion and gluon-initiated 
$\Pg\Pg\to\PZ(\PQq\PQq)\PH$ production with hadronic \PZ boson decays, which can 
be viewed as part of inclusive \PH production via gluon fusion at higher order 
in perturbation theory, are defined as a combined category, referred to as \ggH. 
The VBF and quark-initiated $\PQq\PQq\to\PV(\PQq\PQq)\PH$ production with hadronic 
\PV decays are defined as a combined category, referred to as \qqH. The \VH 
category in turn refers to leptonic \PV decays with at least one charged lepton. 
One further update resulted in the STXS stage-1.2 scheme, which forms the basis 
for this paper. With respect to the STXS stage-1.1 this update introduces a finer 
split of the \ggH bin with the \PH transverse momentum (\ptH) larger than 200\GeV, 
which is the only difference of relevance for this paper.\par}

The first STXS measurements combining the $\PH\to\Pgg\Pgg$, \ZZ, \WW, \PQb\PQb, 
\PGtPgt, and \PGmPGm decay modes have been performed by the ATLAS~\cite{Aad:2019mbh} 
and CMS~\cite{Sirunyan:2018koj} Collaborations. First STXS measurements in the 
\HTT decay channel alone have been performed by the ATLAS 
Collaboration~\cite{Aaboud:2018pen,ATLAS:2022yrq}. In this paper, we report three 
STXS analyses in the \HTT decay channel performed by the CMS Collaboration. All 
analyses are based on the LHC data sets of the years 2016--2018, corresponding 
to an integrated luminosity of 138\fbinv. The events are analyzed in the \emu, 
\etau, \mutau, and \tautau final states based on the number of electrons, muons, 
and \tauh candidates in the event, where \tauh refers to a hadronic \PGt lepton 
decay. Results are reported in the form of signal strengths relative to the SM 
predictions and products of cross sections and branching fraction for the decay 
into \PGt leptons. They are provided in the redefined STXS stage-0 and -1.2 
scheme. Some of the STXS stage-1.2 bins, to which the analyses are not yet 
sensitive, have been combined into supersets. 

Two analyses target the simultaneous measurement of the \ggH and \qqH processes. 
One, referred to as the ``cut-based'' (CB) analysis, exploits  an event 
categorization optimized on the ratio of signal over background events, together 
with one- (1D) and two-dimensional (2D) discriminants related to the properties 
of the \PGtPgt and jet final state to distinguish between signal and backgrounds. 
The other, referred to as the neural network (NN) analysis, exploits NN event 
multiclassification, with the output of the NNs as the only discriminating 
observables. The \NNana is found to be more sensitive with a 30\% stronger 
constraint on the signal strength for the STXS stage-0 \qqH bin and 30\% (40\%) 
stronger constraints, on average, on the STXS stage-1.2 \ggH (\qqH) bins, relative 
to the \CBana. In addition, a third analysis, referred to as the \VHana, targets 
\VH production. The inclusive, STXS stage-0, and -1.2 signal strengths and products 
of the cross sections and branching fraction for the decay into \PGt leptons 
obtained from the combination of the NN- and \VHanapl are considered the main 
results of the paper. A description of the \CBana is given as it facilitates 
comparisons of the selections documented in the paper with alternative simulations 
or theory models. It also serves as reference for a previous publication of 
differential cross section measurements in the \HTT decay channel~\cite{CMS:2021gxc}, 
with which it shares the same object selections, strategy for signal extraction, 
and methods for the estimation of data-driven backgrounds, and as a detailed 
verification for the statistical methods exploited by the \NNana. 

The remainder of this paper is organized as follows. In Sections~\ref{sec:detector}
and~\ref{sec:eventreco} the CMS detector and the reconstruction of the objects 
and event variables used in the analyses are introduced. The event selections 
commonly used in all three analyses are described in Section~\ref{sec:selection}. 
Modelings of signals and backgrounds are described in Section~\ref{sec:model}. 
The STXS scheme used to classify events is introduced in Section~\ref{sec:stxs-scheme}. 
In Sections~\ref{sec:NN-based},~\ref{sec:cut-based}, and~\ref{sec:VH}, the 
selection steps and  event classifications specific to the individual analyses 
are described in more detail. Systematic uncertainties are discussed in 
Section~\ref{sec:uncertainties}. The results of the analyses are discussed in 
Section~\ref{sec:results}. The paper is summarized in Section~\ref{sec:summary}. 

\section{The CMS detector}
\label{sec:detector}

The central feature of the CMS apparatus is a superconducting solenoid of 6\unit{m}
internal diameter, providing a magnetic field of 3.8\unit{T}. Within the solenoid 
volume are a silicon pixel and strip tracker, a lead tungstate crystal 
electromagnetic calorimeter (ECAL), and a brass and scintillator hadron calorimeter 
(HCAL), each composed of a barrel and two endcap sections. Forward calorimeters 
extend the pseudorapidity ($\eta$) coverage provided by the barrel and endcap 
detectors. Muons are detected in gas-ionization chambers embedded in the steel 
flux-return yoke outside the solenoid.

Events of interest are selected using a two-tiered trigger system. The first 
level (L1), composed of custom hardware processors, uses information from the 
calorimeters and muon detectors to select events at a rate of around 100\unit{kHz} 
within a fixed latency of about 4\mus~\cite{Sirunyan:2020zal}. The second level, 
known as the high-level trigger (HLT), consists of a farm of processors running 
a version of the full event reconstruction software optimized for fast processing, 
and reduces the event rate to around 1\unit{kHz} before data 
storage~\cite{Khachatryan:2016bia}. 

A more detailed description of the CMS detector, together with a definition of 
the coordinate system used and the relevant kinematic variables, can be found 
in Ref.~\cite{Chatrchyan:2008zzk}. 

\section{Event reconstruction}
\label{sec:eventreco}

{\tolerance=800 The reconstruction of the proton-proton (\pp) collision products 
is based on the particle-flow (PF) algorithm~\cite{Sirunyan:2017ulk}, which 
combines the information from all CMS subdetectors to reconstruct a set of particle 
candidates (PF candidates), identified as charged and neutral hadrons, electrons, 
photons, and muons. In the 2016 (2017--2018) data sets the average number of 
interactions per bunch crossing was 23 (32). The fully recorded detector data of 
a bunch crossing defines an event for further processing. The primary vertex (PV) 
is taken to be the vertex corresponding to the hardest scattering in the event, 
evaluated using tracking information alone, as described in Ref.~\cite{CMS-TDR-15-02}. 
Secondary vertices, which are detached from the PV, might be associated with 
decays of long-lived particles emerging from the PV. Any other collision vertices 
in the event are associated with additional mostly soft inelastic \pp collisions 
called pileup (PU).\par}

Electron candidates are reconstructed by combining clusters of energy deposits 
in the ECAL with hits in the tracker~\cite{Khachatryan:2015hwa,CMS:2020uim}. To 
increase their purity, reconstructed electrons are required to pass a multivariate 
electron identification discriminant, which combines information on track quality, 
shower shape, and kinematic quantities. For the analyses presented here, a working 
point with an identification efficiency of 90\% is used, with a misidentification 
rate of ${\approx}1\%$ from jets, in the kinematic region of interest. Muons in 
the event are reconstructed by performing a simultaneous track fit to hits in the 
tracker and in the muon chambers~\cite{CMS:2012nsv,CMS:2018rym}. The presence of 
hits in the muon chambers already leads to a strong suppression of particles 
misidentified as muons. Additional identification requirements on the track fit 
quality and the compatibility of individual track segments with the fitted track 
further reduce the misidentification rate. For the analyses presented here, muon 
identification requirements with an efficiency of ${\approx}99\%$ are chosen, 
with a misidentification rate below 0.2\% for pions. 

{\tolerance=800 
The contributions from backgrounds to the electron (muon) selection are further 
reduced by requiring the corresponding lepton to be isolated from any hadronic 
activity in the detector. This property is quantified by an isolation variable
\begin{linenomath}
\ifthenelse{\boolean{cms@external}}
{
  \begin{multline}
  \label{eq:isolation} 
    \Irelem=\frac{1}{\ptem}\Big[\sumCharged \\ + \max\left(0, 
    \sumNeutral+\sumGamma-\ptPU\right)\Big],
  \end{multline}
}
{
  \begin{equation}
    \label{eq:isolation} 
      \Irelem=\frac{1}{\ptem}\Big[\sumCharged + \max\left(0, 
      \sumNeutral+\sumGamma-\ptPU\right)\Big],
  \end{equation}
}
\end{linenomath}
where \ptem corresponds to the electron (muon) transverse momentum \pt and 
\sumCharged, \sumNeutral, and \sumGamma to the \pt (transverse energy \et) sum 
of all charged particles, neutral hadrons, and photons, in a predefined cone of 
radius $\DR = \sqrt{\smash[b]{\left(\Delta\eta\right)^{2}+\left(\Delta\phi\right
)^{2}}}$ around the lepton direction at the PV, where $\Delta\eta$ and $\Delta
\phi$ (measured in radians) correspond to the angular distances of the particle 
to the lepton in the $\eta$ and azimuthal $\phi$ directions. The chosen cone 
size is $\DR=0.3\,(0.4)$ for electrons (muons). The lepton itself is excluded 
from the calculation. To mitigate the contamination from PU, only those charged 
particles whose tracks are associated with the PV are taken into account. Since 
for neutral hadrons and photons an unambiguous association with the PV or PU is 
not possible, an estimate of the contribution from PU (\ptPU) is subtracted from 
the sum of \sumNeutral and \sumGamma. This estimate is obtained from tracks not 
associated with the PV in the case of \Irelm and from the mean energy flow per 
area unit in the case of \Irele. For negative values the result of this difference 
is set to zero. The isolation criteria given in Section~\ref{sec:selection} have 
an efficiency for isolated electrons and muons from \PGt-decays well above 95\%.
\par}

For further characterization of the event, all reconstructed PF candidates are 
clustered into jets using the anti-\kt jet clustering algorithm as implemented in 
the \FASTJET software package~\cite{Cacciari:2008gp,Cacciari:2011ma} with a distance 
parameter of 0.4. Jets resulting from the hadronization of \PQb quarks are used 
to separate signal from top anti-top quark pair (\ttbar) events. These are identified 
either exploiting the \textsc{DeepCSV} (CB- and \VHanapl) or the \textsc{DeepJet} 
(\NNana) algorithm, as described in Refs.~\cite{Sirunyan:2017ezt,Bols:2020bkb}. 
The working points chosen for the \textsc{DeepCSV} algorithm correspond to \PQb 
jet identification efficiencies of 70 and 80\% for a misidentification rate for 
jets originating from light quarks and gluons of 1 and 11\%, respectively. For 
the \textsc{DeepJet} algorithm a working point is chosen with an identification 
efficiency for \PQb jets of 80\% for a misidentification rate for jets originating 
from light quarks and gluons of 1\%~\cite{CMS-DP-2018-058}. Jets with $\pt>30\GeV$ 
and $\abs{\eta}<4.7$ and \PQb jets with $\pt>20\GeV$ and $\abs{\eta}<2.4$, (2.5 
for 2017 and afterwards) are used. In 2017, the ECAL endcaps were subject to an 
increased noise level affecting the reconstruction of jets. Jets with $\pt<50\GeV$ 
in the corresponding range of $2.65<\abs{\eta}<3.14$ have been excluded from the 
analysis resulting in an expected efficiency loss of ${\approx}4\%$ for \HTT 
events provided via VBF. 

Jets are also used as seeds for the reconstruction of \tauh candidates. This is 
done by exploiting the substructure of the jets, using the ``hadrons-plus-strips'' 
algorithm~\cite{Sirunyan:2018pgf,CMS:2022prd}. Decays into one or three charged 
hadrons with up to two neutral pions with $\pt>2.5\GeV$ are used. Neutral pions 
are reconstructed as strips with dynamic size from reconstructed electrons and 
photons contained in the seeding jet, where the strip size varies as a function 
of the \pt of the electron or photon candidates. The \tauh decay mode is then 
obtained by combining the charged hadrons with the strips. To distinguish \tauh 
candidates from jets originating from the hadronization of quarks or gluons, and 
from electrons, or muons, the \textsc{DeepTau} (DT) algorithm~\cite{CMS:2022prd} 
is used. This algorithm exploits the information of the reconstructed event 
record, comprising tracking, impact parameter, and ECAL and HCAL cluster 
information; the kinematic and object identification properties of the PF 
candidates in the vicinity of the \tauh candidate and the \tauh candidate itself; 
and several global characterizing quantities of the event. It results in a 
multiclassification output $\yDT_{\alpha}\,(\alpha=\PGt,\,\Pe,\,\PGm,\,\text{jet}
)$ equivalent to the Bayesian probability that the \tauh candidate originated 
from a \PGt lepton, an electron, muon, or the hadronization of a quark or gluon. 
From this output three discriminants are built according to 
\begin{linenomath}
  \begin{equation}
    D_{\alpha} = \frac{\yDT_{\PGt}}{\yDT_{\PGt}+\yDT_{\alpha}},\,\quad
    \alpha=\Pe,\,\PGm,\,\text{jet}.
  \end{equation}
\end{linenomath}
For the analyses presented here, predefined working points of \De, \Dm, and \Dj 
are chosen~\cite{CMS:2022prd}. The exact choice of working point depends on the 
analysis and \PGtPgt final state, and is given in Table~\ref{tab:dt-working-points}. 
For \De, the efficiencies vary from 54 (Tight) to 71\% (VVVLoose) for 
misidentification rates from 0.05--5.42\%, where the letter V in VLoose and other 
working-point labels stands for ``Very''. For \Dm, the efficiencies vary from 
70.3 (Tight) to 71.1\% (VLoose) for misidentification rates from 0.03--0.13\%. 
For \Dj, the efficiencies of the chosen working points vary from 35 (VTight) to 
49\% (Medium) for misidentification rates from 0.14--0.43\%. The misidentification 
rate of \Dj strongly depends on the \pt and quark flavor of the misidentified 
jet. The estimated value of this rate should therefore be viewed as approximate.

\begin{table*}[t]
  \centering
  \topcaption{
    Selected working points for each of the analyses presented in this paper.
    For each analysis, the working point of \Dj is the same for all final states. 
    The working points for \De and \Dm vary by final state. The labels \WH and 
    \ZH stand for the individual production modes addressed by the \VHana. More 
    details on the efficiency and misidentification rates of the individual working 
    points are given in the text and in Ref.~\cite{CMS:2022prd}.
    }
    \begin{tabular}{cccccccc}
      \hline
      & \multicolumn{3}{c}{\De} 
      & \multicolumn{3}{c}{\Dm} 
      & \Dj \\
      Analysis
      & \etau & \mutau & \tautau 
      & \etau & \mutau & \tautau & \\
      \hline
      CB  
      & \multirow{2}{*}{Tight} 
      & VVVLoose 
      & VLoose 
      & \multirow{2}{*}{VLoose} 
      & \multirow{2}{*}{Tight}  
      & \multirow{2}{*}{VLoose} 
      & Medium \\ [\cmsTabSkip]
      NN & & VVLoose & VVLoose & & & & Tight \\ [\cmsTabSkip] 
      WH  
      & Medium & Medium & Medium 
      & Medium & Medium & Medium 
      & VTight \\ [\cmsTabSkip]
      ZH & VLoose & Tight   & VLoose  & Tight  & VLoose & VLoose 
      & Medium \\ 
      \hline
    \end{tabular}
    \label{tab:dt-working-points}
\end{table*}

{\tolerance=1200
The missing transverse momentum vector \ptvecmiss is also used in event 
characterization. For the CB- and \VHanapl, it is calculated from the negative 
vector \pt sum of all PF candidates~\cite{Khachatryan:2014gga}. For the \NNana 
the pileup-per-particle identification algorithm~\cite{Bertolini:2014bba} is applied 
to reduce the dependence of \ptvecmiss on PU. This means that \ptvecmiss is computed 
from the PF candidates weighted by their probability to originate from the 
PV~\cite{Sirunyan:2019kia}. The \ptvecmiss is used for the discrimination of \PW 
boson production in association with jets (\Wjets) from signal by exploiting the 
transverse mass 
\begin{linenomath}
  \begin{equation}
    \mT^{\Pell}=\sqrt{2 \pt^{\Pell}\,\ptmiss\left(1-\cos\Delta\phi\right)},
  \end{equation}
\end{linenomath}
where \Pell stands for an electron or a muon, but might also refer to the vector 
sum of \ptvecE and \ptvecM, and $\Delta\phi$ stands for the angular difference 
of $\ptvec^{\kern1pt\Pell}$ and \ptvecmiss. The \ptvecmiss is also used for a 
likelihood-based estimate of the invariant mass \mtt of the \PGtPgt system before 
the decays of the \PGt leptons~\cite{Bianchini:2014vza}. This estimate combines 
the measurement of \ptvecmiss and its covariance matrix with the measurements of 
the visible \PGtPgt decay products, utilizing the matrix elements for unpolarized 
\PGt decays~\cite{Bullock:1992yt} for the decay into leptons and the two-body 
phase space~\cite{PDG2020} for the decay into hadrons. On average the resolution 
of this estimate amounts to 10\% in the \tautau, 15\% in the \etau and \mutau, 
and 20\% in the \emu final states, related to the number of neutrinos that escape 
detection.
\par}

\section{Event selection}
\label{sec:selection}

Common to all analyses presented in this paper is the selection of a \PGt pair. 
Depending on the final state, the online selection in the HLT step is based on 
the presence of an \emu pair, a single electron or muon, an \etau or \mutau pair, 
or a \tautau pair in the event~\cite{CMS-DP-2016-026,CMS-DP-2016-067,CMS-DP-2019-012}. 
In the \etau and \mutau final states, the presence of a lepton pair in the HLT 
step allows lower \pt thresholds on the light lepton candidate. The efficiency 
of the online selection is generally above 90\% without strong kinematic 
dependencies of the leptons selected for the offline analysis, as checked from 
independent monitor trigger setups. In the offline selection, further requirements 
on \pt, $\eta$, and \Irelem, are applied in addition to the object identification 
requirements described in Section~\ref{sec:eventreco} and summarized in 
Table~\ref{tab:selection_kin}.

In the \emu final state, an electron and a muon with $\pt>15\GeV$ and $\abs{\eta}
<2.4$ are required. Depending on the trigger path that has led to the online 
selection of an event, a stricter requirement of $\pt>24\GeV$ is imposed on one 
of the two leptons to ensure a sufficiently high efficiency of the HLT selection. 
Both leptons are required to be isolated from any hadronic activity in the 
detector according to $\Irelem<0.15\,(0.20)$.

In the \etau (\mutau) final state, an electron (muon) with $\pt>25\,(20)\GeV$ is 
required, if an event was selected by a trigger based on the presence of the 
\etau (\mutau) pair in the event. From 2017 on the threshold on the muon is 
raised to 21\GeV. If the event was selected only by a single-electron trigger, 
the \pt requirement on the electron is increased to 26, 28, or 33\GeV for the 
years 2016, 2017, or 2018, respectively. For muons, the \pt requirement is 
increased to $23\,(25)\GeV$ for 2016 (2017--2018), if selected only by a 
single-muon trigger. The electron (muon) is required to be contained in the 
central detector with $\abs{\eta}<2.1$, and to be isolated according to $\Irelem
<0.15$. The \tauh candidate is required to have $\abs{\eta}<2.3$ and $\pt>35\,
(32)\GeV$ if selected by an $\etau\,(\mutau)$ pair trigger, or $\pt>30\GeV$ if 
selected by a single-electron (single-muon) trigger. In the \tautau final state, 
both \tauh candidates are required to have $\abs{\eta}<2.1$ and $\pt>40\GeV$. 
The working points of the DT discriminants as described in Section~\ref{sec:eventreco} 
are chosen depending on the final state and are given in 
Table~\ref{tab:dt-working-points}. 

\begin{table*}[t]
  \centering
  \caption{
    Offline selection requirements applied to the electron, muon, and \tauh 
    candidates used for the selection of the \Pgt pair. The expressions first 
    and second lepton refer to the label of the final state in the first column. 
    The \pt requirements are given in \GeVns. For the \emu final state two lepton 
    pair trigger paths, with a stronger requirement on the \pt of electron 
    (muon), are used for the online selection of the event. For the \etau and 
    \mutau final states, the values (in parentheses) correspond to the lepton 
    pair (single lepton) trigger paths that have been used in the online 
    selection. A detailed discussion is given in the text. 
  }
  \begin{tabular}{lccccccc}
    \hline
    Final state & Observable 
    & \multicolumn{3}{c}{First lepton} 
    & \multicolumn{3}{c}{Second lepton} \\ 
    & & 2016 & 2017 & 2018 & 2016 & 2017 & 2018 \\
    \hline
    \emu 
    & \pt & \multicolumn{3}{c}{$>15\,(24)$} & \multicolumn{3}{c}{$>24\,(15)$} \\ 
    & $\abs{\eta}$ & \multicolumn{3}{c}{$<2.4$} & \multicolumn{3}{c}{$<2.4$} \\ 
    & \Irele & \multicolumn{3}{c}{$<0.15$} & \multicolumn{3}{c}{$<0.20$} \\ [\cmsTabSkip]
    \etau 
    & \pt & $>25\,(26)$ & $>25\,(28)$ & $>25\,(33)$ 
    & $\hphantom{>20\,(21)}$ & $>35\,(30)$ & $\hphantom{>20\,(21)}$ \\ 
    & $\abs{\eta}$ 
    & \multicolumn{3}{c}{$<2.1$}  & \multicolumn{3}{c}{$<2.3$} \\ 
    & \Irele 
    & \multicolumn{3}{c}{$<0.15$} & \multicolumn{3}{c}{--} \\ [\cmsTabSkip]
    \mutau 
    & \pt & $>20\,(23)$ & $>21\,(25)$ & $>21\,(25)$ 
    & \multicolumn{3}{c}{$>32\,(30)$} \\
    & $\abs{\eta}$ & \multicolumn{3}{c}{$<2.1$} & \multicolumn{3}{c}{$<2.3$} \\ 
    & \Irelm & \multicolumn{3}{c}{$<0.15$} & \multicolumn{3}{c}{--} \\ [\cmsTabSkip]
    \tautau 
    & \pt & \multicolumn{3}{c}{$>40$} & \multicolumn{3}{c}{$>40$} \\
    & $\abs{\eta}$ & \multicolumn{3}{c}{$<2.1$} & \multicolumn{3}{c}{$<2.1$} \\ 
    \hline
  \end{tabular}
  \label{tab:selection_kin}
\end{table*}

The selected \PGt decay candidates are required to be of opposite charges and to 
be separated by more than $\DR = 0.3$ in the $\eta$--$\phi$ plane in the \emu 
final state and 0.5 otherwise. This applies also to all selected \PGt-decay 
candidates in the \VHana, where the final states comprise a combination of one 
or more light leptons (\Pe or \PGm) and one or two \tauh candidates. The closest 
distance of the tracks to the PV is required to be $d_{z}<0.2\cm$ along the beam 
axis. For electrons and muons, an additional requirement of $d_{xy}<0.045\cm$ in 
the transverse plane is applied. In rare cases in which more than the expected 
number of \tauh candidates fulfilling all selection requirements is found in an 
event, in the CB- and \NNanapl, the candidates with the highest \Dj scores are 
chosen until the expected number of \tauh candidates is met. In subsequent steps, 
the analyses differ slightly in their selection requirements, which is a 
consequence of the different analysis strategies employed in the CB- and \NNanapl, 
and the different measurement target for the \VHana. For the CB- and \NNanapl, 
events with additional leptons fulfilling looser selection criteria are excluded 
to avoid the assignment of single events to more than one \PGtPgt final state. 
For the \NNana, which is more inclusive than the \CBana before event classification, 
a requirement of $\mT^{\Pell}<70\GeV$ is imposed in the \etau (\mutau) final state, 
to keep an orthogonal control region for the estimate of the background from events 
with quark- or gluon-induced jets, which are misidentified as \tauh leptons 
(\jettau), as discussed in Section~\ref{sec:FF-method}. In the \emu final state, 
events with at least one \PQb jet are excluded from the selection in order to 
suppress the background from \ttbar production. Finally, to prevent kinematic 
event overlap with the analysis of \HWW events, \mTemu calculated from $\ptvecE
+\ptvecM$ and \ptvecmiss is required to be less than 60\GeV in the \emu final 
state. 

Further selection details of the CB- and \VHanapl are discussed in 
Sections~\ref{sec:cut-based},~\ref{sec:WH}, and~\ref{sec:ZH}. 

\section{Background and signal modeling}
\label{sec:model}

The main backgrounds in the CB- and \NNanapl originate from \PZ boson production 
in association with jets in the \PGtPgt decay channel (\ZTT), \Wjets, \ttbar 
production, and SM events where light quark- or gluon-induced jets are produced 
through the strong interaction, referred to as quantum chromodynamics (QCD) 
multijet production. Minor backgrounds originate from the production of two \PV 
bosons (diboson), single \PQt~quark, and \PZ boson production in the \PePe and 
\PGmPGm final states (also denoted as \ZLL). We distinguish three ways in which 
these processes may contribute to the selected event samples: 

\begin{enumerate}
\item they contain two genuine \PGt leptons in their final states; 
\item at least one quark- or gluon-induced jet is misidentified as \tauh (\jettau), 
\Pe, or \PGm (\jetell); 
\item an isolated high-\pt electron or muon is misidentified as originating from 
a \PGt lepton decay or mistakenly identified as a \tauh candidate.
\end{enumerate}

Event groups 1 and 2 are estimated from data, as will be discussed in the 
following sections. Event group 3 and the signal are estimated from simulation. 
Group 1, which still relies on the simulation of the \PGt decays, group 3, 
and the signal are subject to simulation-to-data corrections, which have been 
determined from dedicated control regions, as will be discussed in 
Section~\ref{sec:corrections}.   

For the \VHana, the backgrounds generally involve the production of an additional 
\PV boson. Details of the background estimation are given in Section~\ref{sec:backgroundEstimation}

\subsection{Backgrounds with genuine \texorpdfstring{\PGt}{tau} lepton pairs}
\label{sec:tau-embedding}

For events in the CB- and \NNanapl in which, \eg, the decay of a \PZ boson results 
in two genuine \PGt leptons, the \PGt-embedding method is used, as described in 
Ref.~\cite{Sirunyan:2019drn}. For this purpose \PGmPGm events are selected in data. 
All energy deposits of the muons are removed from the event record and replaced 
by simulated \PGt lepton decays with the same kinematic properties as the selected 
muons. In this way, the method relies only on the simulation of the \PGt lepton 
decay and its energy deposits in the detector, while all other parts of the event, 
such as the reconstructed jets, their identification as originating from the PV, 
the identification of \PQb jets, or the non-\PGt related parts of \ptmiss, are 
obtained from data. This results in an improved modeling of the data compared to 
the simulation of the full process and removes the need for several of the 
simulation-to-data corrections detailed in Section~\ref{sec:corrections} for these 
events.  A detailed discussion of the selection of the original \PGmPGm events, 
the exact procedure itself, its range of validity, and related uncertainties can 
be found in Ref.~\cite{Sirunyan:2019drn}. 

Although the selected muons predominantly originate from \PZ boson decays, there
are also contributions from other processes that result in two genuine \PGt 
leptons. For example, \ttbar and diboson events where both \PW bosons decay into 
a muon and neutrino are included in the original selection of \PGmPGm events, and 
replacing the selected muons by simulated \PGt lepton decays naturally leads to 
an estimate of these processes, as well. For the selection described in 
Section~\ref{sec:selection}, 97\% of the \PGmPGm events selected for the 
\PGt-embedding method originate from \PZ boson decays, ${\approx}1\%$ from \ttbar 
production, and the rest from other processes. 

\subsection{Backgrounds with jets misidentified 
as hadronic \texorpdfstring{\PGt}{tau} lepton decays}
\label{sec:FF-method}

The main processes contributing to \jettau events in the \etau, \mutau, and \tautau 
final states are QCD multijet, \Wjets, and \ttbar production. These events are 
estimated using the ``fake factor'' or \FF-method described in 
Refs.~\cite{Sirunyan:2018qio, Sirunyan:2018zut} and adapted to each corresponding 
analysis described in this paper. For this purpose, the signal region (SR) as 
defined by the event selection given in Section~\ref{sec:selection} is complemented 
by the disjoint application region (AR) and determination regions (DR$^{i}$, where 
$i$ stands for QCD, \Wjets, or \ttbar). All other processes are estimated either 
from simulation or from the \PGt-embedding method and subtracted from the data in 
the AR, DR$^{\text{QCD}}$, and DR$^{\Wjets}$. The SR and the AR differ only in 
the working point chosen for the identification of the \tauh candidate, where for 
the AR a looser working point is chosen and the events from the SR are excluded. 

Depending on the final state, one or three independent sets of extrapolation 
factors \FFi are then derived. For the \tautau final state, where QCD multijet 
production contributes ${\approx}94\%$ of the events in the AR, only \FFQCD is 
determined, and \FFWjets and \FFttbar are assumed to be similar. In the \etau 
and \mutau final states, where the sharing is more equal, separate \FFi are used 
for QCD multijet, \Wjets, and \ttbar production. In these final states the largest 
fraction in the AR, in the range of 55--70\%, is expected to originate from \Wjets 
production; the smallest fraction in the AR, in the range of 2--5\%, is expected 
to originate from \ttbar production.

Each \FFi is determined in a dedicated DR$^{i}$, defined to enrich each 
corresponding process. The \FFi are then used to estimate the yields \NSR and 
kinematic properties of the combination of these backgrounds in the SR from the 
number of events \NAR in the AR according to 
\begin{linenomath}
  \begin{equation}
    \label{eq:FF}
    \NSR = \biggl(\sum\limits_{i}w_{i}\FFi\biggr)\NAR,
    \qquad i=\text{QCD, }\PW\text{+jets, }\ttbar.
  \end{equation}
\end{linenomath}
For this purpose, the \FFi are combined into a weighted sum, using the 
simulation-based estimate of the fractions $w_{i}$ of each process in the AR. 
A template fit to the data in the AR yields a similar result for the $w_{i}$.

For the estimation of \FFQCD, the charges of the two selected \PGt decay products 
are required to be of the same sign. For the estimation of \FFWjets, high $\mT^{
\Pell}$ and the absence \PQb jets are required. For \ttbar production a sufficiently 
pure DR with still similar kinematic properties to the SR in data can not be 
defined. Instead the \FFttbar are obtained from simulation and corrected to data 
with a selection of more than two jets, at least one \PQb jet, and more than two 
leptons in an event. 

The \FFi depend mainly on the \pt of the (leading) \tauh candidate and are derived 
on an event-by-event basis. For \FFQCD, an additional dependency on the number of 
selected jets \Njet being 0, 1, or ${\geq}2$, and for \FFWjets, a similar dependency 
on \Njet and the distance $\DR(\tauh,\Pe\,(\PGm))$ between the \tauh and the $\Pe
\,(\PGm)$ in $\eta$--$\phi$ are introduced. Subleading dependencies on $\ptvec^{
\kern1pt\Pe(\PGm)}$, \Irelem, the \pt of the second leading \tauh candidate, or 
the mass of the visible \PGtPgt decay products (\mvis) enter via bias corrections 
obtained from additional control regions in data. They are usually close to 1 but 
can range between 0.8--1.2, depending on the process and observable.

\subsection{Backgrounds with jets misidentified as an electron or muon}
\label{sec:em-background}

The number of \jetell events in the \emu final state is estimated in a way that 
is similar to the \FF-method. In this case, an AR is distinguished from the SR 
by requiring the charges of the electron and muon to be of the same sign. A DR 
is defined requiring $0.2<\Irelm<0.5$ from which a transfer factor \TF is obtained 
to extrapolate the number \NAR of events in the AR to the number \NSR of events 
in the SR according to
\begin{linenomath}
  \begin{equation}
    \label{eq:TF}
    \NSR = \TF\,\NAR.
  \end{equation}
\end{linenomath} 

{\tolerance=800 The main dependency of \TF is on the distance $\DR(\Pe,\PGm)$ 
between the \Pe and \PGm trajectories at the PV in $\eta$--$\phi$, and \Njet being 
0, 1, or ${\geq}2$. Subleading dependencies on \ptvecE and \ptvecM are introduced 
via a closure correction in the DR and a bias correction to account for the fact 
that \TF has been determined from less-isolated muons. The latter is obtained 
from another control region with $0.15\,(0.20)<\Irele<0.50$ for the (CB-) 
\NNana.\par}

\subsection{Simulated backgrounds and signal}
\label{sec:simulation}

In the \etau and \mutau final states, more than 80\% of all backgrounds, after 
the selection described in Section~\ref{sec:selection}, are obtained from one of 
the methods described in the previous sections. In the \tautau final state, this 
fraction is ${\gtrsim}95\%$. All remaining backgrounds---for example \PZ boson, 
\ttbar, or diboson production, where at least one decay of a \PV boson into 
an electron or muon is not covered by either of the methods---are obtained from 
simulation. 

The production of \PZ bosons in the \PePe and \PGmPGm final states and \PW boson 
production are simulated at leading order (LO) precision in the strong coupling 
constant \alpS, using the \MGvATNLO 2.2.2 (2.4.2) event generator~\cite{Alwall:2011uj,
MadGraph} for the simulation of the data taken in 2016 (2017--2018), exploiting 
the ``so-called'' MLM jet matching and merging scheme of the matrix element 
calculation with the parton~\cite{Alwall:2007fs}. To increase the number of 
simulated events in regions of high signal purity, supplementary samples are 
generated with up to four outgoing partons in the hard interaction. For diboson 
production \MGvATNLO is used at next-to-LO (NLO) precision in \alpS. For the 
simulation of \ttbar and single~\PQt quark production the \POWHEG 2~\cite{Nason:2004rx,
Frixione:2007vw,Alioli:2008tz,Alioli:2010xd,Alioli:2010xa,Bagnaschi:2011tu} event 
generator is used at NLO precision in \alpS. 

The signal samples are also obtained at NLO precision in \alpS using \POWHEG for 
the five main production modes, gluon fusion~\cite{Alioli:2008tz,Bagnaschi:2011tu}, 
VBF~\cite{Nason:2009ai}, \VH (split by \WH and \ZH)~\cite{Luisoni:2013cuh}, and 
\ttH~\cite{Hartanto:2015uka}. For the production via gluon fusion the distributions 
of \ptH and the jet multiplicity in the simulation are tuned to match the 
next-to-NLO (NNLO) accuracy obtained from full phase space calculations with the 
\textsc{nnlops} generator~\cite{Hamilton:2013fea,Hamilton:2015nsa}.

All signal samples have been produced assuming $\mH=125\GeV$ and rescaled to 
correspond to the cross sections and branch fraction for the decay to 
$\Pgt$ leptons as expected for $\mH=125.38\GeV$, based on the recommendations of 
the  LHC Higgs Working Group~\cite{Heinemeyer:2016spz}. This change is less than 
$-1\%$ for the cross sections for \PH production via gluon fusion and VBF, and 
$-1\%$ for the branching fraction for the decay in $\Pgt$ leptons.

For the generation of all processes, the NNPDF3.0~\cite{Ball:2014uwa} 
(NNPDF3.1~\cite{Ball:2017nwa}) set of parton distribution functions (PDFs) is 
used for the data taken in 2016 (2017--2018). All matrix element generators are 
interfaced with the \PYTHIA~8.230 event generator~\cite{Sjostrand:2014zea}, which 
is used to model the effects of parton showering, hadronization, and fragmentation, 
as well as the decay of the \PGt lepton. For this purpose, two different tunes 
(CUETP8M1~\cite{Khachatryan:2015pea} and CP5~\cite{Sirunyan:2019dfx}) are used for 
the data taken in 2016 and from 2017 onward, for the parameterization of multiparton 
interactions and the underlying event.

When comparing to data, \PZ boson, \ttbar, and single~\PQt quark events in the 
$\PQt\PW$-channel are normalized to their cross sections at NNLO precision in 
\alpS~\cite{Melnikov:2006kv,Czakon:2011xx,Kidonakis:2013zqa}. Single~\PQt quark 
production in the $t$-channel and diboson events are normalized to their cross 
sections at NLO precision in \alpS or higher~\cite{Kidonakis:2013zqa,Campbell:2011bn,
Gehrmann:2014fva}. The signal samples are normalized to their inclusive cross 
sections and branching fractions as recommended by the LHC Higgs Working 
Group~\cite{deFlorian:2016spz}, assuming an \PH mass of 125.38\GeV~\cite{CMS:2020xrn}. 

For all simulated events, additional inclusive inelastic \pp collisions generated 
with \PYTHIA are added according to the expected PU profile in data to take the 
effect of the observed PU into account. All events generated are passed through 
a \GEANTfour-based~\cite{Agostinelli:2002hh} simulation of the CMS detector and 
reconstructed using the same version of the CMS event reconstruction software as 
used for the data.

\section{Simplified template cross section schemes}
\label{sec:stxs-scheme}

The analyses presented in this paper aim for measurements of inclusive and
differential production cross sections for the signal at increasing levels of 
granularity following the STXS scheme specified by the LHC Higgs Working 
Group~\cite{deFlorian:2016spz}. In this scheme two stages are defined. Stage-0 
assumes the separation by production modes, of which \ggH, \qqH, and \VH are of 
relevance for this paper. For \ggH, which subsumes gluon fusion and $\Pg\Pg\to
\PZ(\PQq\PQq)\PH$ production, and \qqH, which comprises VBF and $\PQq\PQq\to\PV
(\PQq\PQq)\PH$ production, the simulated signals of each corresponding process 
are combined for signal extraction, assuming relations between the processes as 
expected from the SM. 

At stage-1.1~\cite{Berger:2019wnu} and -1.2, these production modes are further 
split into STXS bins according to the jet multiplicity at stable-particle level, 
the invariant mass of the two leading jets \mjj, if present in an event, and \ptH. 
The \VH process is further split by \PV type and \pt (\ptV). For all cross 
section measurements, the absolute value of the \PH rapidity is required to be 
less than 2.5. The only difference of stage-1.2 with respect to stage-1.1 that 
is of relevance for this paper is the further splitting of the \ggH bin with 
$\ptH>200\GeV$ into a bin with $200<\ptH\leq300\GeV$ and a bin with $300\GeV<\ptH$. 

\begin{figure*}[htbp]
  \centering
  \includegraphics[width=0.9\textwidth]{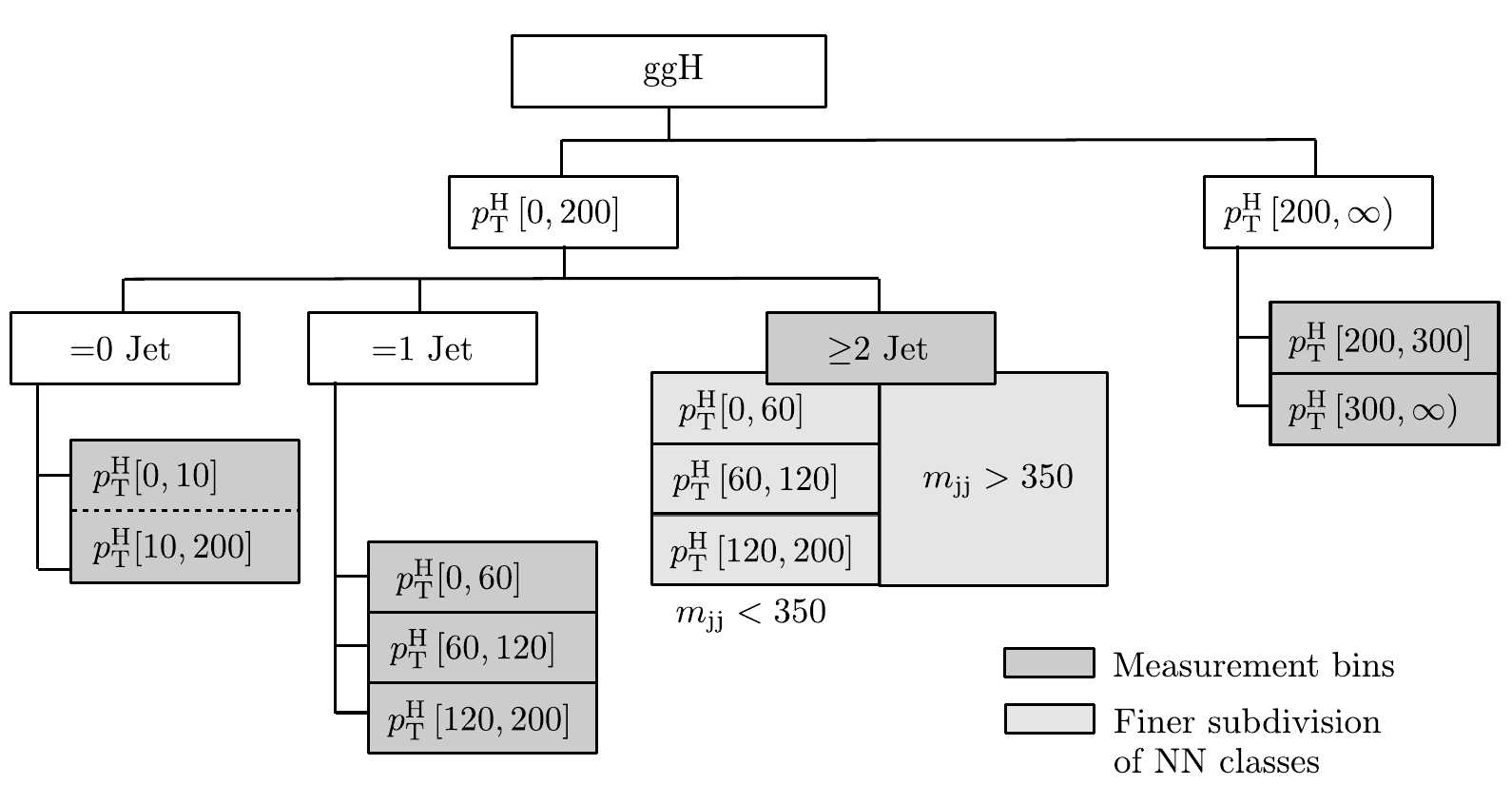}
  \caption{
    Binning for \ggH production in the reduced STXS stage-1.2 scheme as exploited 
    by the analyses presented in this paper. The dark gray boxes indicate the STXS 
    bins measured by the analyses. Depending on the analysis, a split is applied 
    to the 0-jet bin, as is indicated by the dashed line. The boxes colored in 
    light gray indicate a finer subdivision of the NN signal classes for $\Njet
    \geq2$ used for the classification task, as explained in Section~\ref{sec:NN-based}. 
    Thresholds on \ptH are given in \GeVns.
  }
  \label{fig:stxs_ggh}
\end{figure*}

\begin{figure*}[htbp]
  \centering
  \includegraphics[width=0.9\textwidth]{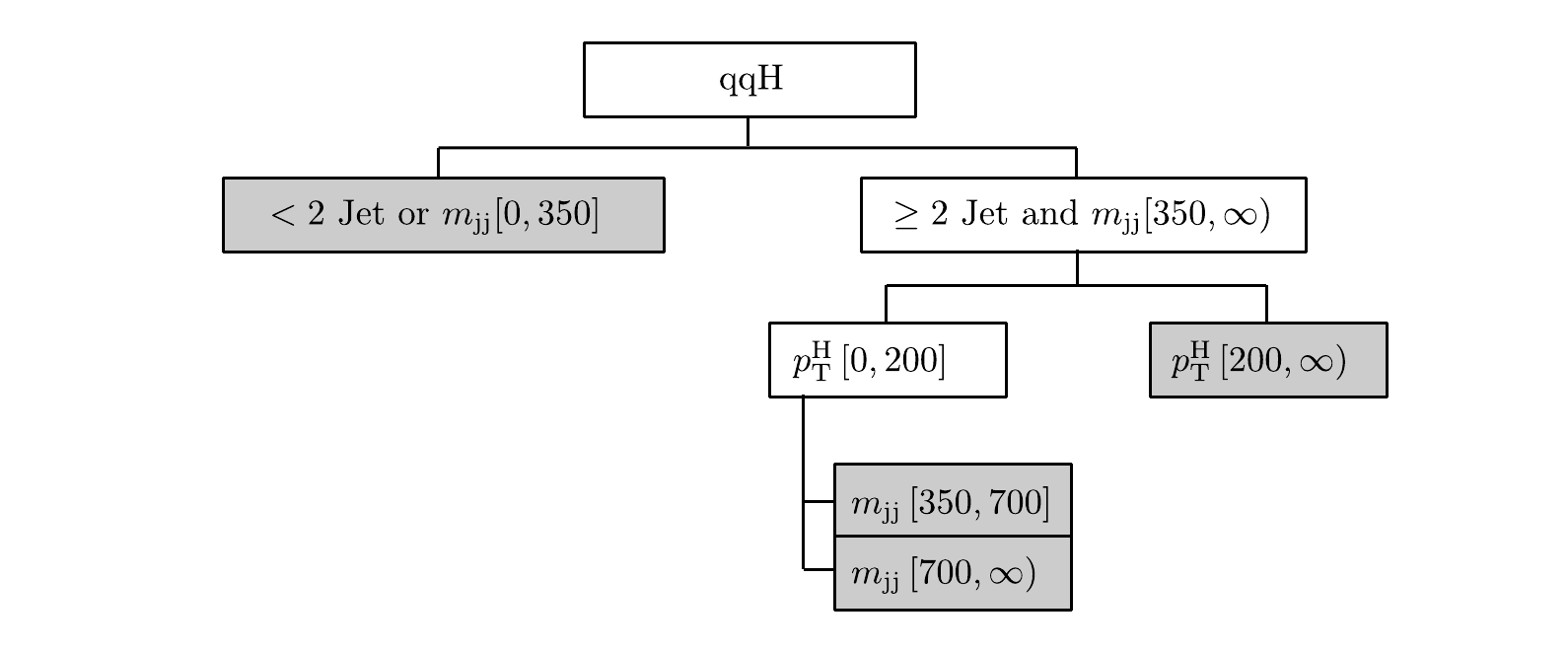}
  \caption{
    Binning for \qqH production in the reduced STXS stage-1.2 scheme as exploited 
    by the analyses presented in this paper. The gray boxes indicate the STXS 
    bins measured by the analyses. Thresholds on \ptH and \mjj are given in \GeVns.
  }
  \label{fig:stxs_qqh}
\end{figure*}

\begin{figure*}[htbp]
  \centering
  \includegraphics[width=0.9\textwidth]{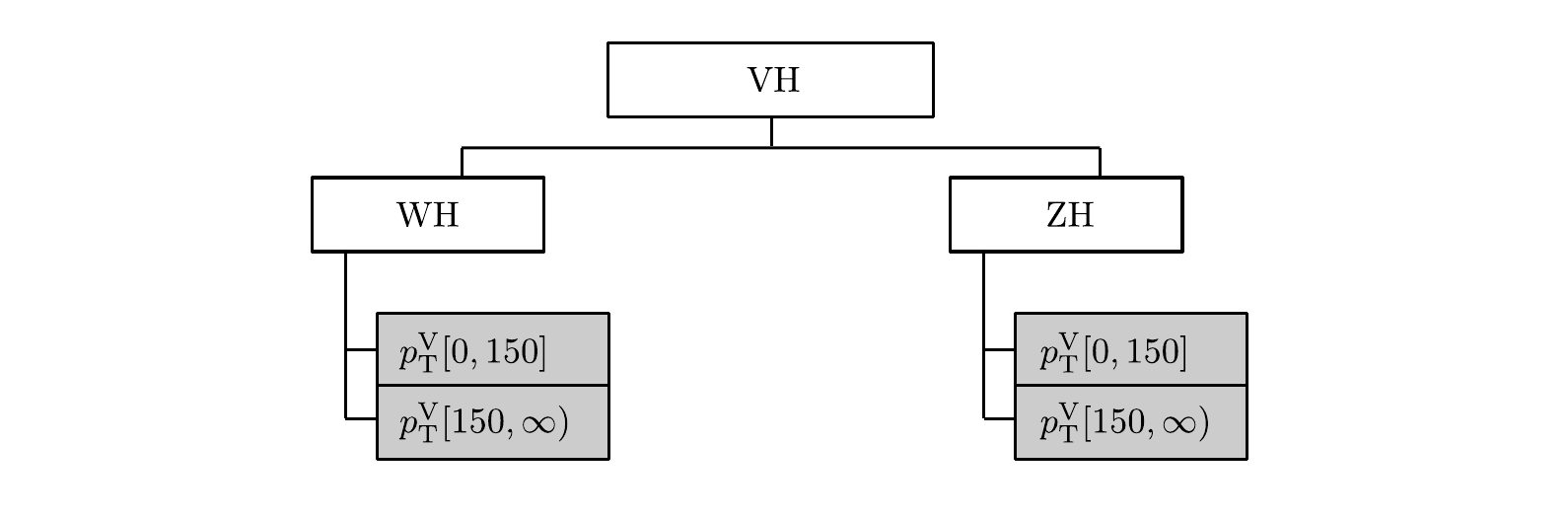}
  \caption{
    Binning for \VH production in the reduced STXS stage-1.2 scheme as exploited 
    by the analysis presented in this paper. The gray boxes indicate the STXS 
    bins measured by the analysis. Thresholds on \ptV are given in \GeVns.
  }
  \label{fig:stxs_vh}
\end{figure*}

Since the analyses under consideration do not have enough events to exploit all 
STXS bins defined in Ref.~\cite{Berger:2019wnu}, the bins have been combined as 
shown in Figs.~\ref{fig:stxs_ggh}--\ref{fig:stxs_vh} resulting in 7--8 
measurements for \ggH, 4 measurements for \qqH, and 4 measurements for \VH 
production. In the figures, the gray boxes correspond to the measured STXS bins. 
For the \CBana, the \ggH 0-jet STXS bin is measured inclusively in \ptH. For the 
\NNana this bin is split in \ptH. Throughout the text the \qqH bin with ${<}2$ 
jets or $0<\mjj<350\GeV$ is also labeled as ``non-VBF-topo''. For all STXS bins, 
histogram template distributions are obtained from the simulated signal processes 
discussed in Section~\ref{sec:simulation}. These are fitted to the data together 
with corresponding template distributions from each considered background process, 
for signal extraction in each given STXS bin, as discussed in 
Section~\ref{sec:results}.

\section{Neural network based analysis}
\label{sec:NN-based}

For the \NNana, all selected events are provided as input to a set of 
multiclassification NNs. The outputs of these NNs are used to distribute the 
events into background classes depending on the \PGtPgt final state, as shown in 
Table~\ref{tab:nn-event-classes}, and a number of signal classes. 

\begin{table*}[htbp]
  \centering
  \topcaption{
    Background processes and event classes for each \PGtPgt final state for the 
    \NNana. All event classes enter the NN trainings for the event multiclassification 
    with the same statistical weight (\ie, with uniform prevalence). The label 
    $\ttbar(\Pe/\PGm+X)$ refers to the fraction of \ttbar production that is not 
    covered by any of the estimation methods from data, as listed in the first 
    two rows. The classes tt, misc, zll, and db are defined in the introduction 
    of Section~\ref{sec:NN-based}.
  }
  \begin{tabular}{ccccc}
    \hline
    & \multicolumn{4}{c}{Classes per final state} \\
    Process & \emu & \etau & \mutau & \tautau \\
    \hline
    \PGt-embedding & genuine \PGt & genuine \PGt 
    & genuine \PGt & genuine \PGt \\ [\cmsTabSkip]
    QCD/\FF-method & \jetell   & \jettau  
    & \jettau  & \jettau  \\ [\cmsTabSkip]
    $\ttbar(\Pe/\PGm+X)$ & tt   & tt   & tt 
    & misc \\ [\cmsTabSkip]
    \ZLL  & misc & zll  & zll 
    & misc \\ [\cmsTabSkip]
    Diboson/single~\PQt    & db   & misc & misc 
    & misc \\
    \hline
  \end{tabular}
  \label{tab:nn-event-classes}
\end{table*}

The background classes, which are based on the experimental signatures of groups 
of processes rather than individual processes, closely resemble the background 
model discussed in Section~\ref{sec:model}. The genuine \PGt, \jettau, and 
\jetell classes are trained on data. The \ttbar (tt) and diboson (db) classes 
are trained on simulated events from \ttbar and diboson production excluding 
those parts of the processes which are already covered by the \PGt-embedding 
method. The db class in addition subsumes single~\PQt quark production. The \ZLL 
(zll) classes in the \etau and \mutau final states, are trained on simulated 
\ZEE and \ZMM events. Finally the miscellaneous (misc) classes comprise those 
processes that are either difficult to isolate or too small to be treated as 
single templates for signal extraction. These are \ZLL events in the \emu final 
state; diboson and single~\PQt quark production in the \etau and \mutau final 
states; and \ttbar, \ZLL, diboson, and single~\PQt quark production in the 
\tautau final state. 

Depending on the stage of the STXS measurement that the analysis is targeting, 
two different sets of NNs are used that differ by the number of signal classes. 
Two signal classes are defined for the stage-0 measurement corresponding to the 
\ggH and \qqH processes. For the stage-1.2 measurement, the stage-0 processes are 
divided into 15 subclassess, splitting the signal events for training by their 
jet multiplicity and kinematic properties at the stable-particle level. This 
subdivision follows the STXS stage-1.2 scheme as shown in Figs.~\ref{fig:stxs_ggh} 
and~\ref{fig:stxs_qqh}, with the exception that the signal class for \ggH events 
with $\Njet\geq2$ is subdivided into four additional STXS bins, according to \ptH 
and \mjj to accommodate future combined coupling measurements. 

The goal of this strategy is to achieve not only the best possible separation 
between the signal and each of the most relevant backgrounds, but also across all 
individual STXS stage-1.2 bins. 

\subsection{Neural network layout}
\label{sec:nn-layout}

For each measurement, a distinct NN is trained for each of the four \PGtPgt final 
states. All NNs have a fully connected feed-forward architecture with two hidden 
layers of 200 nodes each. The activation function for the hidden nodes is the 
hyperbolic tangent. For the output layers the nodes are defined by the classes 
discussed in the previous section and the activation function is chosen to be 
the softmax function~\cite{GoodBengCour16}. The NN output function \yl of each 
output node $l$ can be interpreted as a Bayesian conditional probability for an 
event to be associated with event class $l$, given its input features $\vec{x}$. 
This conditional probability interpretation does not take the prior of the production 
rate of each corresponding process into account. 

In the \etau, \mutau, and \tautau final states each NN has the same 14 input 
features comprising the \pt of both \PGt candidates and their vector sum; the \pt 
of the two leading jets, their vector sum, their difference in $\eta$, and \mjj; 
\Njet; the number of \PQb jets \Nbtag; \mtt; \mvis; and the estimates of the 
momentum transfer of each exchanged vector boson under the VBF hypothesis, as 
used in Ref.~\cite{Gritsan:2016hjl}. In the \emu final state, where events 
containing \PQb jets have been excluded from the analysis, \Nbtag carries no 
discriminating information. Instead, \mTemu has been added as an input feature 
with some separating power. These variables have been selected from a larger 
feature space based on their importance for the classification task derived from 
the metric defined in Ref.~\cite{Wunsch:2018oxb}. Moreover, to account for 
differences in data-taking conditions, the data-taking year is an additional 
input to each NN that is provided through one-hot-encoding, such that the correct 
data-taking year obtains the value 1, while all other data-taking years obtain 
the value 0.

Before entering the NNs, all input features are standardized for their distributions 
to have mean 0 and s.d.\ 1. Potentially missing features in a given event, such 
as \mjj for events with less than two selected jets, are assigned a default value 
that is close enough to the transformed value space to not influence the NN 
decision. 

\subsection{Neural network training}
\label{sec:nn-training}

The samples discussed in Section~\ref{sec:model} are used for the training of the 
NNs. Those processes, which are part of the misc event class, are weighted according 
to their expected production rates to represent the event mixture as expected in 
the test samples. 

The parameters to be optimized during training are the weights ($\{w_{a}\}$) and 
biases ($\{b_{b}\}$) of \yl. The classification task is encoded in the NN loss 
function, chosen to be the categorical cross entropy
\begin{linenomath}
\ifthenelse{\boolean{cms@external}}
{
  \begin{multline}
      L^{(k)}\left(\{\yl^{(k)}\},\{\yprimepk\}\right) = \\
      -\yprimepk
      \ln\left(\yl^{(k)}(\{w_{a}\},\{b_{b}\},\{\vec{x}_{p}^{(k)}\})\right);
      \, \yprimepk=\delta_{lp}, \\
    \label{eq:nn-loss-function}
  \end{multline}
}
{
  \begin{equation}
    L^{(k)}\left(\{\yl^{(k)}\},\{\yprimepk\}\right) = 
    -\yprimepk
    \ln\left(\yl^{(k)}(\{w_{a}\},\{b_{b}\},\{\vec{x}_{p}^{(k)}\})\right)\,;
    \qquad \yprimepk=\delta_{lp}, \\
  \label{eq:nn-loss-function}
\end{equation}
}
\end{linenomath}
where $k$ indicates the event, on which $L$ is evaluated. All training events are 
implicitly labeled by the true process $p$ to which they belong. The NN output 
function for event $k$ to belong to event class $l$ is given by $\yl^{(k)}$. The 
function \yprimepk encodes the prior knowledge of the training. It is 1 if the 
predicted class $l$ of event $k$ coincides with $p$, and is 0 otherwise. The 
$\yl^{(k)}$ depend on the weights, biases, and input features $\{\vec{x}_{p}^{(k)}
\}$ of event $k$ to the corresponding NN. Before training, the weights are 
initialized with random numbers using the Glorot initialization 
technique~\cite{glorot2010} with values drawn from a uniform distribution. The 
biases are initialized with zero. The training is then performed as a minimization 
task on the empirical risk functional
\begin{linenomath}
\begin{equation}
      R\left[\{\yl^{(k)}\},\{\yprimepk\}\right] = \frac{1}{N}
      \sum\limits_{k=1}^{N}\,L^{(k)}\left(\{\yl^{(k)}\},\{\yprimepk\}\right) \\
    \label{eq:empirical-risk}
  \end{equation}
\end{linenomath}
in the space of $\{w_{a}\}$ and $\{b_{b}\}$ using randomly sampled mini-batches 
of 30 events per signal and background class and data-taking year, drawn from the 
training data set using a balanced batch approach~\cite{Shimizu2018}. This approach 
has shown improved convergence properties on training samples with highly imbalanced 
lengths. The batch definition guarantees that all true event classes enter the 
training with equal weight in the evaluation of $R$, \ie, uniform prevalence. On 
each mini-batch a gradient step is applied defined by the partial derivatives of 
$L$ in each weight, $w_{a}$, and bias, $b_{b}$, using the Adam minimization 
algorithm~\cite{Kingma:2014vow}, with a constant learning rate of $10^{-4}$. 

To guarantee statistical independence, events that are used for training are 
not used for any other step of the analysis. The performance of the NN during 
training is monitored evaluating $R$ on a validation subset that contains a fraction 
of 25\% of randomly chosen events from the training sample, which are excluded 
from the gradient computation. The training is stopped if the evaluation of $R$ 
on the validation data set does not indicate any further decrease for a sequence 
of 50 epochs, where an epoch is defined by 1000 mini-batches. The NNs used for 
the analysis are then defined by the weights and biases of the epoch with the 
minimal value of $R$ on the validation sample. To improve the generalization 
property of the NNs, two regularization techniques are introduced. First, after 
each hidden layer, a layer with a dropout probability of 30\% is added. Second, 
the weights of the NNs are subject to an L2 (Tikhonov) 
regularization~\cite{Tikhonov:1963} with a regularization factor of $10^{-5}$.

The NNs draw their power not only from the isolated values of each corresponding 
feature, but also from correlations across features. To provide an objective 
statistical measure of an adequate modeling of the inclusive NN feature space, 
goodness-of-fit (GoF) tests have been performed on 1D and 2D histograms of all 
features and their pairwise combinations. This has been done prior to the 
application of the NNs to the test data set and the data. These GoF tests are 
based on a saturated likelihood model, as described in Ref.~\cite{Baker:1983tu}, 
exploiting the data model as described in Section~\ref{sec:model} and including 
all systematic uncertainties of the model and their correlations as used for 
signal extraction. During these tests we generally observe an agreement of the 
model with the data within 5--10\% of the event density, well contained in the 
combined systematic variations of the uncertainty model typically ranging between 
10--15\%, prior to the maximum likelihood fit used for signal extraction.

\subsection{Characterization of the classification task}
\label{sec:nn-characterization}

\begin{figure*}[htbp]
  \centering
  \includegraphics[width=0.60\textwidth]{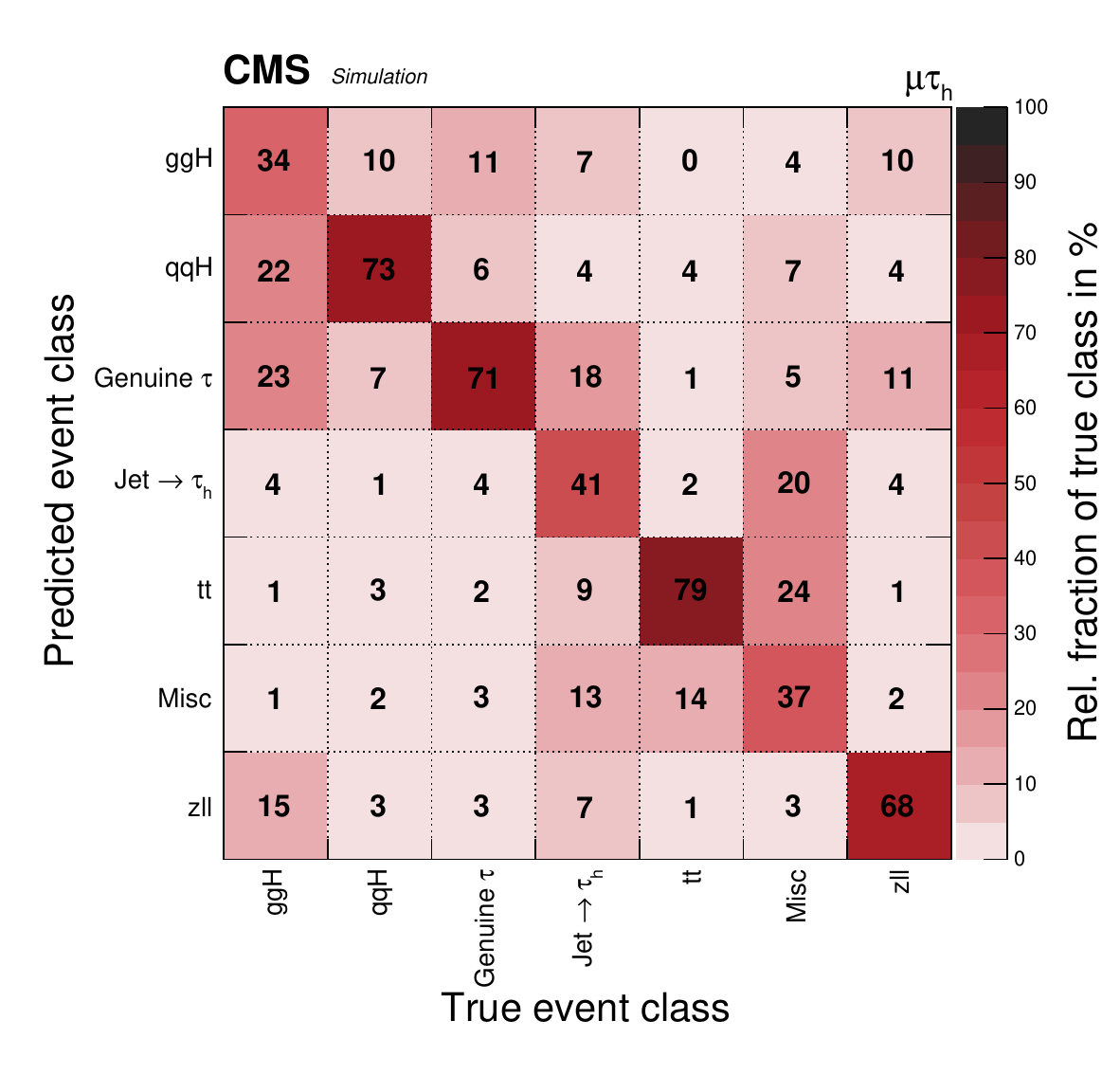}
  \includegraphics[width=0.75\textwidth]{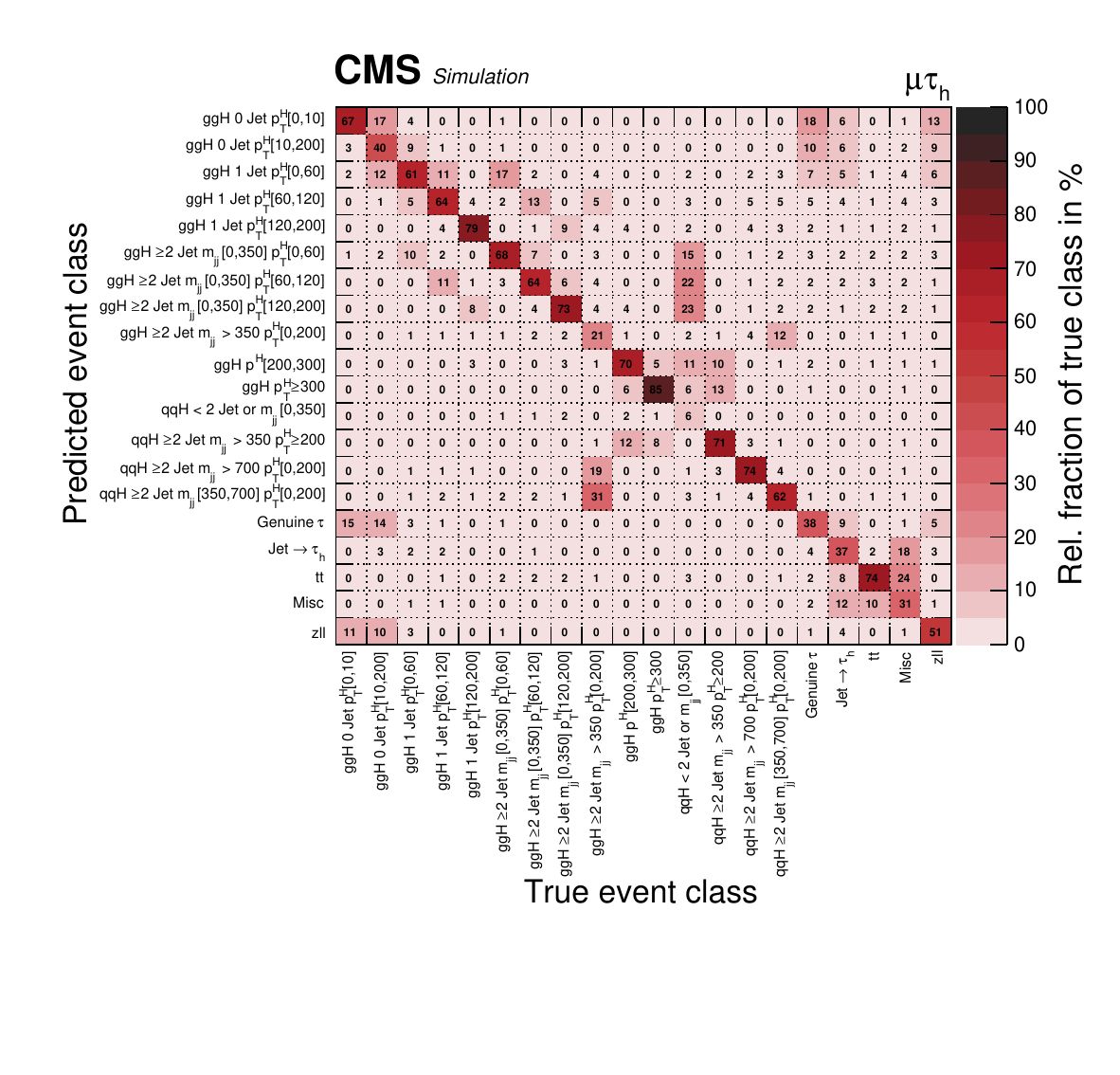}
  \caption{
    Confusion matrices for the NN classification tasks used for the (upper)
    STXS stage-0/inclusive and (lower) STXS stage-1.2 cross section measurements 
    described in Section~\ref{sec:NN-based}. These confusion matrices have 
    been evaluated on the full test data set, comprising all data-taking years 
    in the \mutau final state. They are normalized such that all entries in 
    each column, corresponding to a given true event class, sum to unity. 
  }
  \label{fig:nn-confusion-matrices}
\end{figure*}

The overall success of the NNs in adapting to the given classification task is 
monitored with the help of confusion matrices. Examples of such confusion 
matrices for the STXS stage-0 and -1.2 measurements, evaluated on the full 
analyzed data set in the \mutau final state, are shown in 
Fig.~\ref{fig:nn-confusion-matrices}. For the given representation the matrices 
have been normalized such that all entries in each column, corresponding to a 
given true event class, sum to unity. The values on the diagonals in the figure 
thus represent the sensitivity of the NN to each given true event class. Random 
association would lead to a uniform distribution across predicted event classes 
with a weight of one out of seven (twenty) for the stage-0 (-1.2) training. 

For the stage-0 classification, sensitivities of 70\% and larger can be observed 
for the \qqH, genuine \PGt, tt, and zll classes. The \jettau and misc classes 
have less prominent features to identify them, which partially relates to the fact 
that they comprise several different processes. The \ggH class also reveals 
ambiguities, especially with respect to the \qqH and genuine \PGt classes. 
Adding the fractions of true \ggH events in these three NN output classes results 
in a sensitivity of 79\% to distinguish \ggH events from events without genuine 
\PGt leptons in the final state, giving hint to the importance of \PGt-related 
features for the NN response, in this case. The distinction of \ggH from genuine 
\PGt events mostly relies on features related to \ptH, such as the vector sum \pt 
of the two \PGt candidates. On the other hand \ggH events with high \ptH tend to 
higher jet multiplicities, which makes them harder to distinguish from \qqH events. 
These trends are confirmed by the confusion matrix of the stage-1.2 classification 
that reveals larger off-diagonal elements relating \ggH events with $\Njet=0$ and 
low \ptH with genuine \PGt events and \ggH events with $\Njet\geq2$ and $\mjj>350
\GeV$ with \qqH events. 

The stage-1.2 classification reveals high sensitivities to the \qqH signal classes 
with high \mjj and the \ggH signal classes with high \ptH. Larger confusion, 
sticking out from the general trend, is observed for \qqH events with $\Njet<2$ 
or $\mjj<350\GeV$. We observe that 70\% of these events migrate from their original 
truth labeled class into one of three \ggH NN output classes with $\Njet\geq2$ and 
$\mjj<350\GeV$. This can be explained by the similarity of the observable 
signatures. For \ggH events with $\Njet\geq2$ and $\mjj<350\GeV$ the reconstructed 
jets may originate from the matrix element calculation, but most probably they 
emerge from the parton-shower model. In the case of initial-state radiation, \mjj 
can take large values and thus mimic the VBF signature in the detector. The fact 
that the NN associates 76\% of the \qqH events originating from the bin in 
discussion with NN output classes with $\Njet\geq2$ and $\mjj<350\GeV$ indicates 
the experimental signature that the NN has identified to be decisive for this 
classification. The fact that no migrations from the corresponding \ggH bins into 
the \qqH bin are observed can be explained by the more specific signatures in the 
\ggH bins, which are additionally split in \ptH. The \qqH bin, which is inclusive 
in \ptH, acts like a superclass to these \ggH bins in this respect. An event that 
is compatible with a certain \ptH hypothesis will be associated with one of the 
\ggH classes rather than the \qqH class. The background processes are identified 
with a sensitivity comparable to the stage-0 training. The slightly larger confusion 
can be understood by the increased number of classes and therefore increased 
variety of signatures that a given event can be associated with. 

We visualize the influence of single features and their pairwise linear correlations 
on the NN response exploiting the metric \talpha based on Taylor expansions of 
the \yl after training, with respect to the features $\vec{x}$ up to second order, 
as described in Ref.~\cite{Wunsch:2018oxb}. For a classic gradient descent training, 
\talpha corresponds to the mean of the absolute values of the given Taylor 
coefficient obtained from the whole sampled input space. In this way we identify 
\mtt, \mvis, \mjj, and especially correlations across these features as most 
influential for the NN classification. For the stage-0 \ggH event class we 
identify \mtt, \mvis, and corresponding (self-) correlations of \mtt and \mvis 
as the most important characteristics for identification; the term self-correlation 
corresponds to the second derivative in the Taylor expansion and reveals, \eg, 
that the signal is peaking in the \mtt and \mvis distributions. The fact that 
these characteristics are shared across the \ggH, \qqH, and genuine \PGt event 
classes explains the relatively high degree of confusion across these categories 
for the stage-0 training. The separation of \ggH from \qqH events mostly relies 
on characteristics related to \mjj. As previously discussed this is more difficult 
for \ggH events with $\Njet\geq2$. The distinction of genuine \PGt events mostly 
relies on the vector \pt sum of the two \PGt candidates and on \mtt. For the tt 
event class, we find that the information that \ttbar events are nonpeaking in 
\mvis and \mtt contributes as much as \Nbtag and the jet properties to the observed 
high sensitivity. These findings, which similarly apply to all final states, 
demonstrate that the NNs have indeed captured the features that are expected to 
provide the best discrimination between event classes.

\subsection{Classification and discriminants for signal extraction}
\label{sec:nn-application}

For each event, the maximum of the \yl obtained from all classes $l$ defines the 
class to which the event is assigned. This maximum takes values ranging from one 
over the number of classes, for events that cannot be determined without ambiguity, 
to one, for events that can clearly be associated with a corresponding class. 
Histogrammed distributions of \yl for each corresponding class are also used as 
input for signal extraction. 

\begin{figure*}[htbp]
  \centering
  \includegraphics[width=0.45\textwidth]{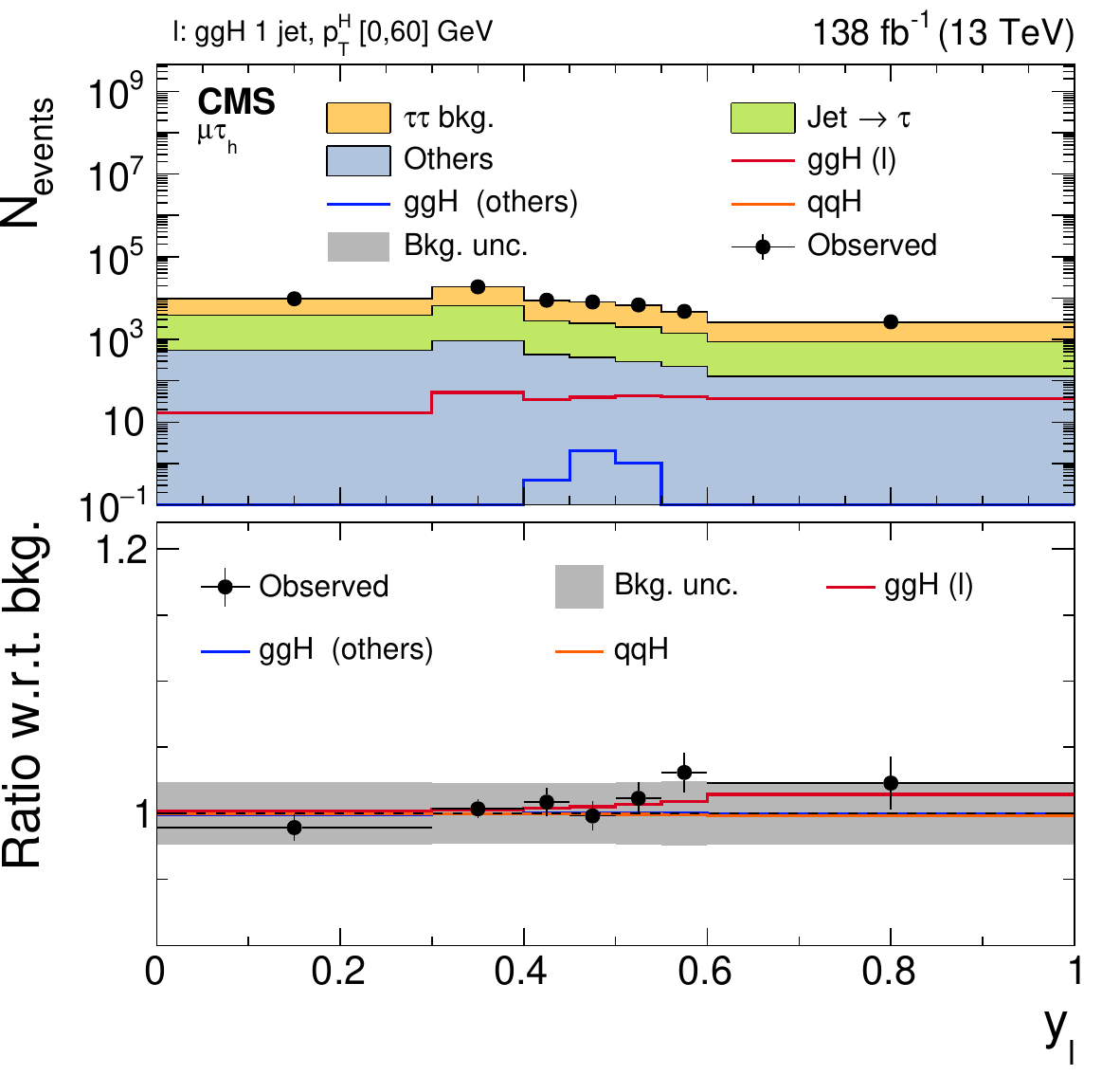}
  \includegraphics[width=0.45\textwidth]{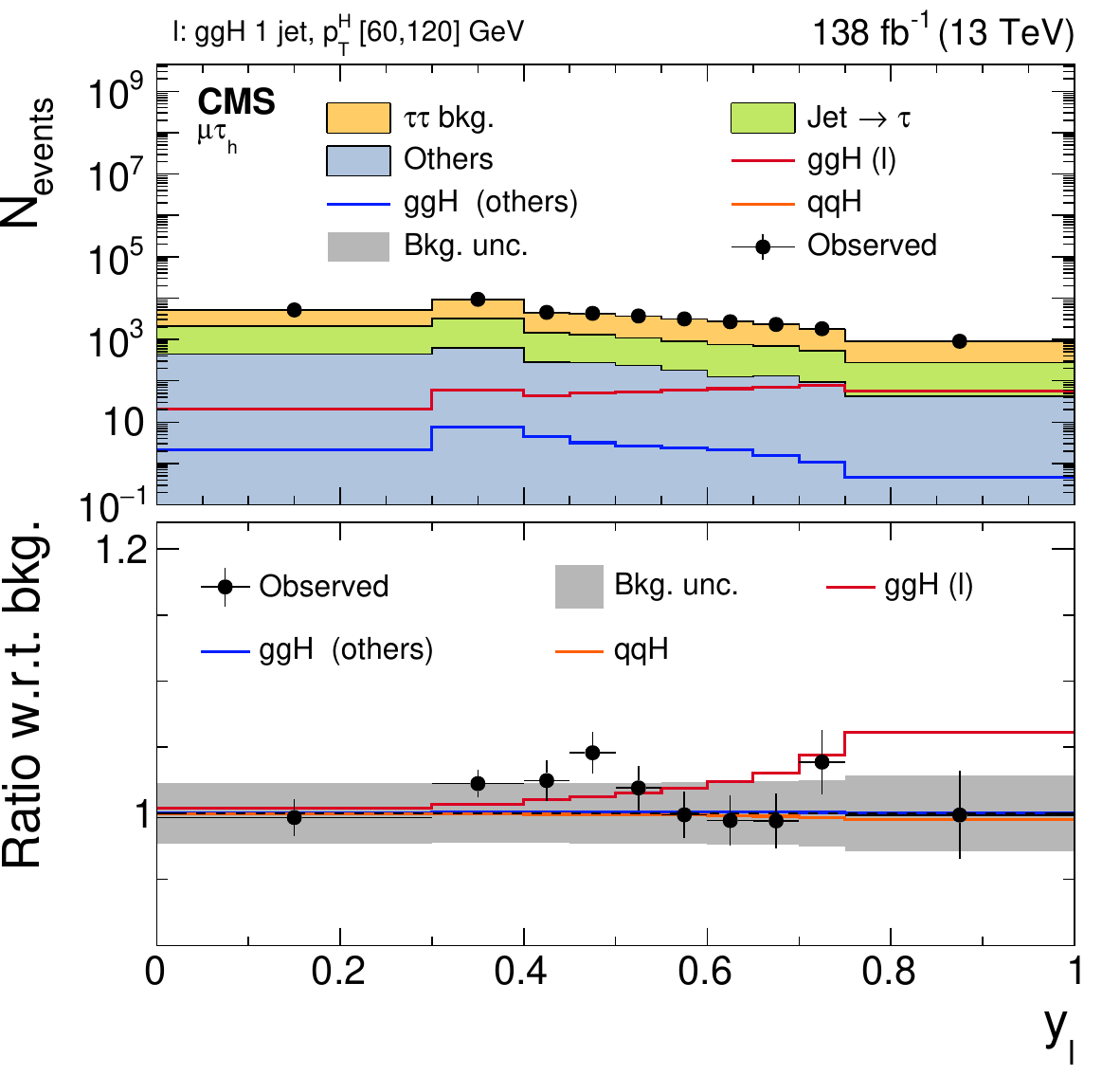}
  \includegraphics[width=0.45\textwidth]{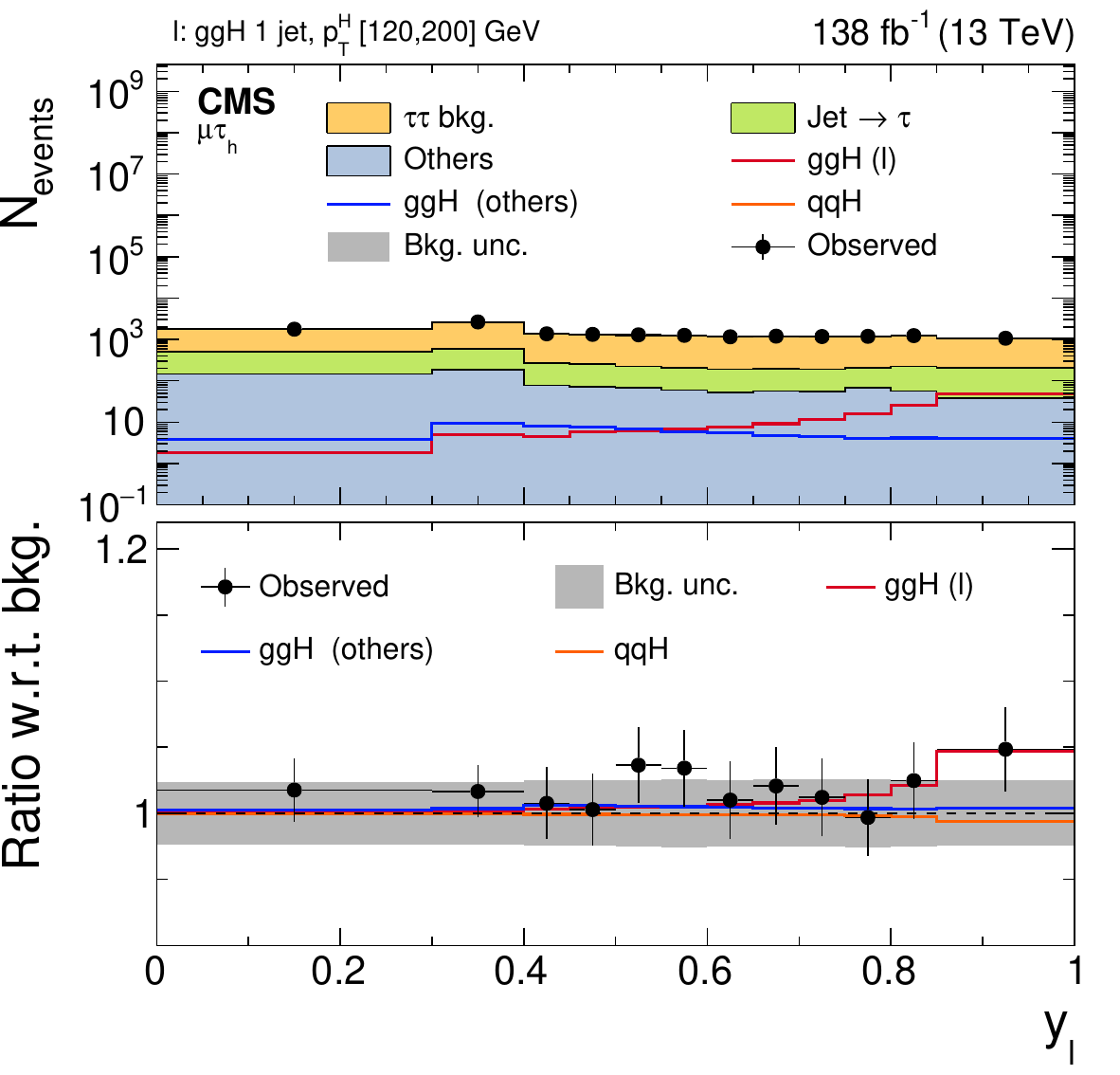}
  \includegraphics[width=0.45\textwidth]{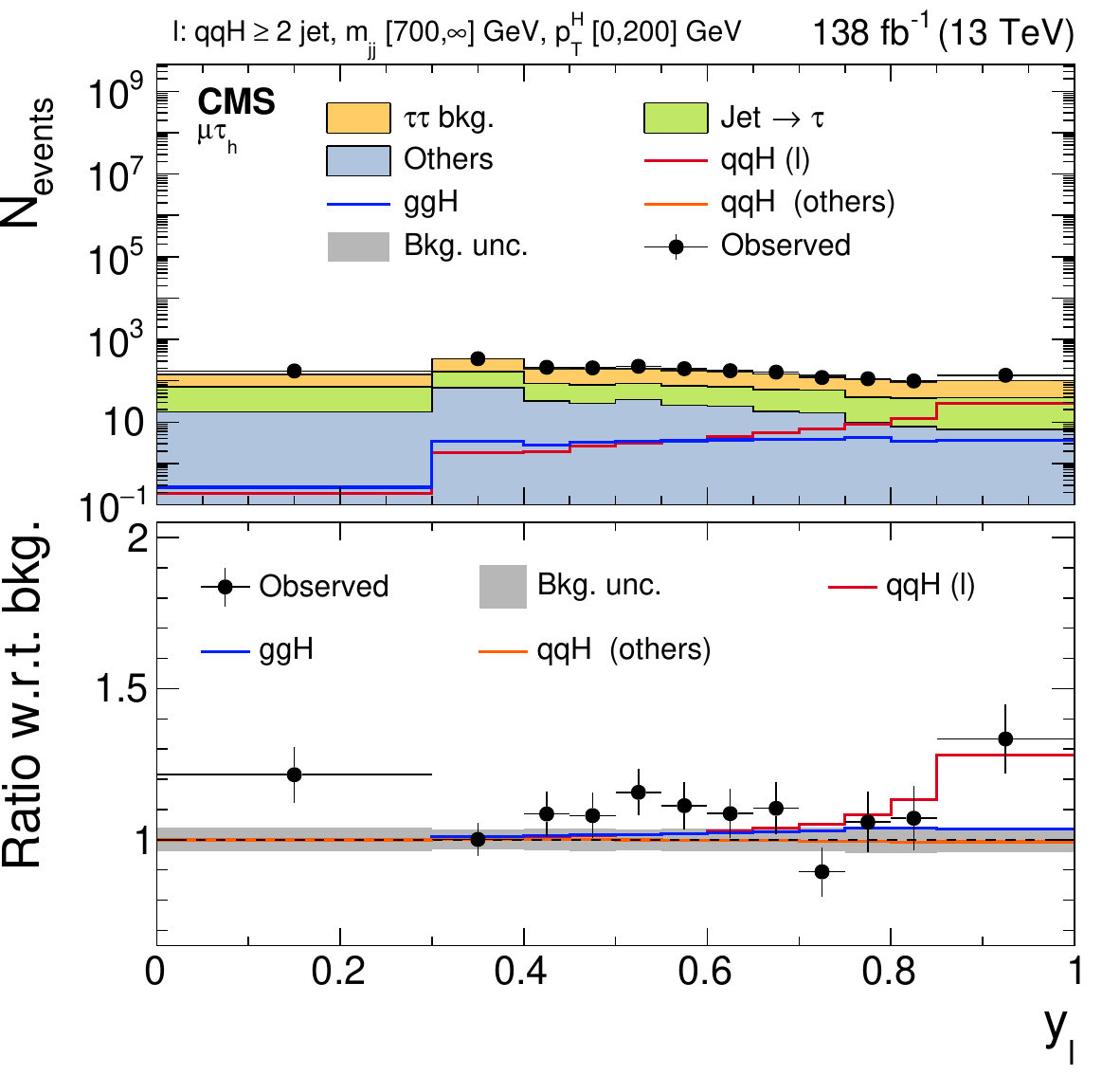}
  \caption{
    Observed and predicted distributions of \yl for the signal classes for the 
    \ggH $\Njet=1$ subspace in three increasing STXS stage-1.2 bins in \ptH and 
    the \qqH $\Njet\geq2$ bin with $\mjj\geq700\GeV$, $\ptH [0, 200]\GeV$, in 
    the \mutau final state of the \NNana. The distributions show all data-taking 
    years combined. The signal contributions for (red) each STXS bin corresponding 
    to the given event class, and the remaining inclusive (blue) \ggH and (orange) 
    \qqH processes, excluding the STXS bin in consideration, are also shown as 
    unstacked open histograms. All distributions are shown after the fit of the 
    model, used for the extraction of the STXS stage-1.2 signals, to the data 
    from all final states and data-taking years. Signal contributions in particular 
    have been scaled according to the obtained fit results. In the lower panel 
    of each plot the differences either of the corresponding additional signals 
    or the data relative to the background expectation after fit are shown. 
  }
  \label{fig:nn-discriminator-stage-1}
\end{figure*}

For the stage-1.2 measurement, this leads to 18 input distributions for all signal 
and background classes in the \tautau final state and 20 input distributions in 
the \emu, \etau, and \mutau final states, for each data-taking year. Sample 
distributions of \yl, combined for all data-taking years, are shown 
in Fig.~\ref{fig:nn-discriminator-stage-1}. Here and in all figures of that kind 
the data are represented by the black points, where the error bars indicate the 
statistical uncertainty of the event count in each corresponding histogram bin. 
The processes of the background model are represented by the stacked filled 
histograms.

\begin{figure*}[htbp]
  \centering
  \includegraphics[width=0.45\textwidth]{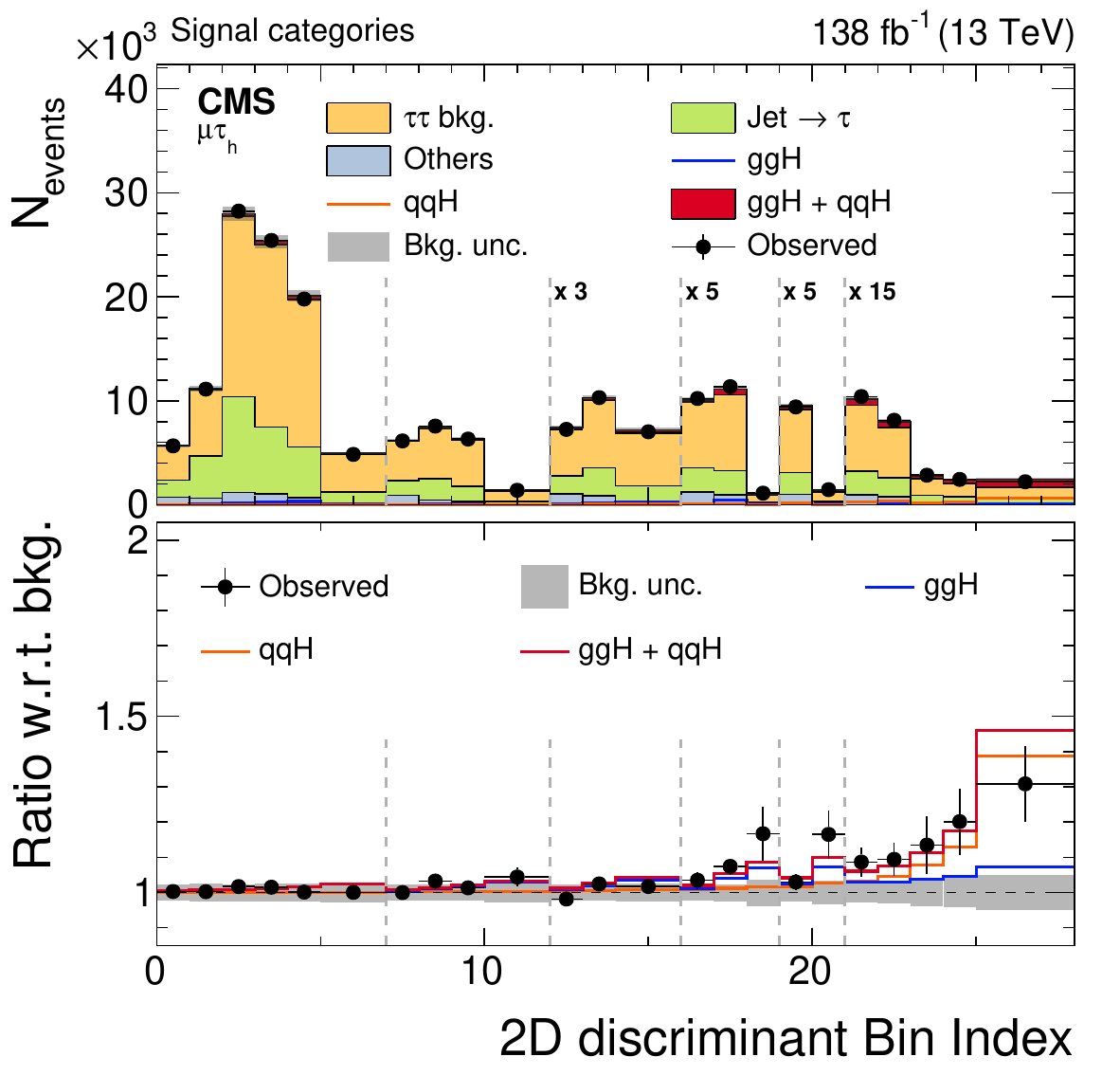}
  \includegraphics[width=0.45\textwidth]{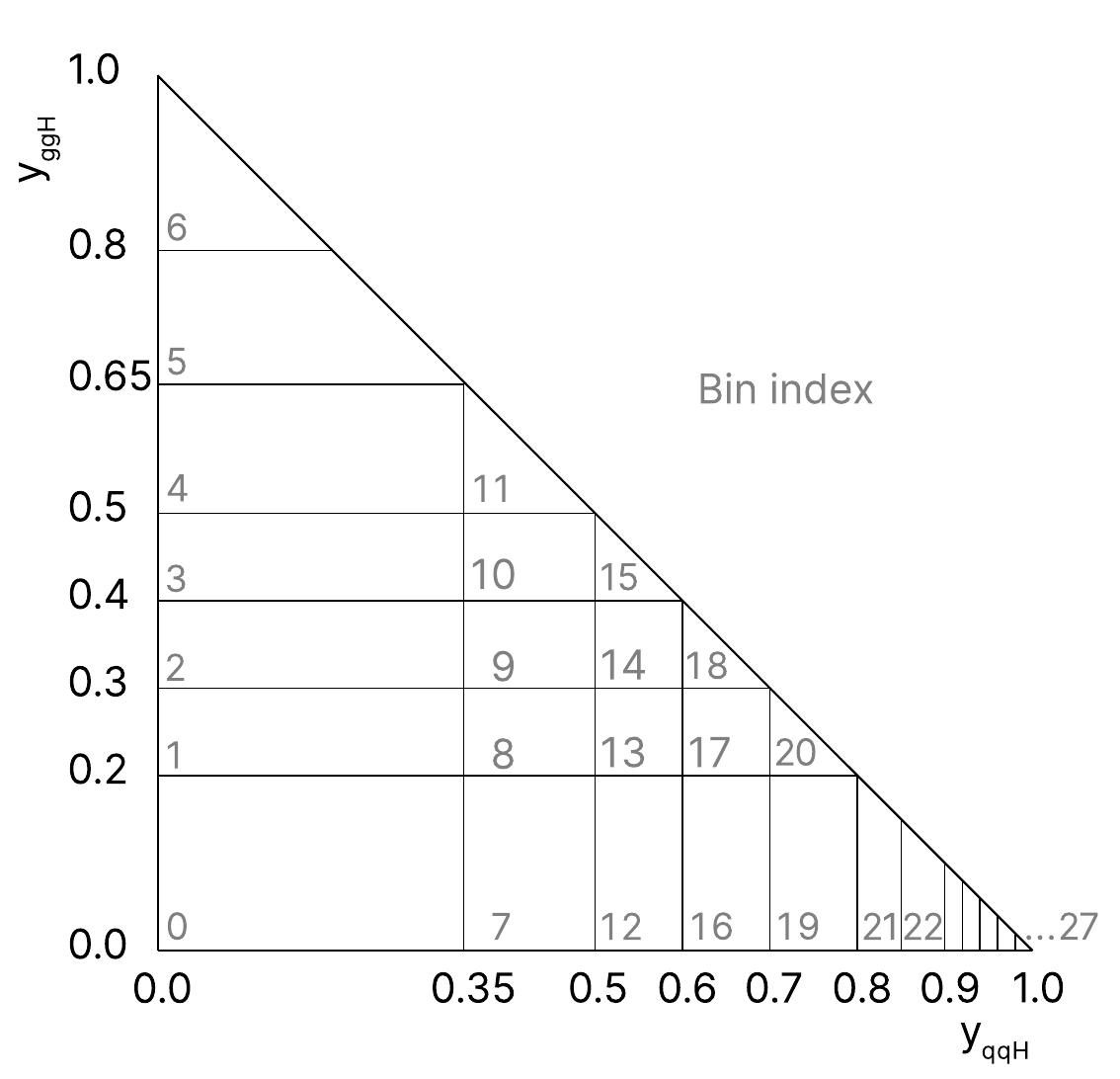}
  \caption{
    Observed and predicted distributions of the 2D discriminant as used for the 
    extraction of the STXS stage-0 and inclusive signals, for all data-taking 
    years in the \mutau final state of the \NNana. Also shown is the definition 
    of each individual bin, on the right. The background distributions are shown 
    after the fit of the model, used for the extraction of the inclusive signal 
    to the data from all final states and data-taking years. For this fit some 
    of the bins have been merged to ensure a sufficient population of each bin. 
    The distributions are shown with the finest common binning across all data-taking 
    years. The signal contributions for the (red) inclusive, (blue) \ggH, and 
    (orange) \qqH signals are also shown as unstacked open histograms, scaled 
    according to the correspondingly obtained fit results. The vertical gray 
    dashed lines indicate six primary bins in \yqqH, the last four of which to 
    the right have been enhanced by factors ranging from 3 to 15 for improved 
    visibility. In the lower panel of the plot on the left the differences either 
    of the corresponding additional signals or the data relative to the background 
    expectation after fit are shown.
  }
  \label{fig:nn-discriminator-stage-0}
\end{figure*}

For the stage-0 measurement, the training with two signal categories is used, 
resulting in five event classes in the \tautau final state and seven event 
classes in the \emu, \etau, and \mutau final states, for each data-taking year. 
For this measurement, the signal classes are combined into a 2D discriminant 
depending on \yggH and \yqqH. The binning scheme used for this discriminant and 
the distribution of the unrolled discriminant for all data-taking years combined, 
in the \mutau final state, are shown in Fig.~\ref{fig:nn-discriminator-stage-0}. 
The binning is grouped in up to 11 bins in \yqqH and up to seven bins in \yggH. 
Vertical gray dashed lines in Fig.~\ref{fig:nn-discriminator-stage-0} (left) 
indicate the main groups of bins in \yqqH in this scheme. Each main group 
corresponds to increasing values in \yqqH from left to right. From bin 0--20, 
the binning within each main group indicates increasing values in \yggH. The 
last main group on the right of the figure is ordered by \yqqH only. The same 
input distributions are also used for a measurement of the inclusive \PH production 
cross section. In both figures, the differences either of the corresponding 
additional signals or the data relative to the background expectation after the 
fit used for signal extraction are shown in the lower panel of each plot. 

\section{Cut-based analysis}
\label{sec:cut-based}

For the \CBana the event selection as described in Section~\ref{sec:selection} 
is modified and extended in a few places. To reduce the contamination from \Wjets 
production in the \etau and \mutau final states in the \CBana, a requirement of 
$\mT^{\Pell}<50\GeV$ is imposed. To suppress the background from \ttbar production, 
events that contain at least one \PQb jet are excluded from the selection not 
only in the \emu, but also in the \etau, and \mutau final states. For similar 
reasons, in the \emu final state an additional requirement of $\Dzeta>-30\GeV$ 
is imposed on the event variable \Dzeta defined as 
\begin{linenomath}
\ifthenelse{\boolean{cms@external}}
{
  \begin{equation}
    \label{eqn:Dzeta}
    \begin{aligned}
      \Dzeta &= \pzetamiss - 0.85\,\pzetavis ; \\
      \pzetamiss &= \ptvecmiss\cdot\hat{\zeta} ; \\
      \pzetavis &= \left(\ptvecE + 
      \ptvecM\right)\cdot\hat{\zeta},
    \end{aligned}
  \end{equation}
}
{
  \begin{equation}
    \label{eqn:Dzeta}
      \Dzeta = \pzetamiss - 0.85\,\pzetavis ; \qquad
      \pzetamiss = \ptvecmiss\cdot\hat{\zeta} ; \qquad
      \pzetavis = \left(\ptvecE + 
      \ptvecM\right)\cdot\hat{\zeta},
  \end{equation}  
}
\end{linenomath}
where $\hat{\zeta}$ corresponds to the bisectional direction between the electron 
and muon trajectories at the PV in the transverse plane~\cite{Abulencia:2005kq}. 
The variables \pzetamiss and \pzetavis can each take positive or negative values. 
Their linear combination has been chosen to optimize the sensitivity of the 
analysis in the \emu final state. A more detailed discussion of the variable 
\Dzeta is given in Ref.~\cite{Sirunyan:2018qio}. Finally, the requirement on 
\Irelm is tightened from 0.20 to 0.15 in the \emu final state. 

After selection, event categories are designed to increase the sensitivity to 
the signal by isolating regions with large signal-to-background ratios, and 
provide sensitivity to the stage-0 and -1.2 \ggH and \qqH STXS bins.

Events are distributed in different categories, which separate the different 
\PH production modes, corresponding to the stage-0 processes in the STXS scheme. 
A 0-jet category is used to collect events with no reconstructed jet. This 
category predominantly contains background, but also some signal events, which 
mainly originate from \ggH production. It therefore mainly acts as a control 
region for backgrounds. Two event categories target \qqH production. Depending 
on the \PGtPgt final state, these are defined by the presence of more than 2 
jets with $\mjj>350\GeV$ or a large separation in $\eta$ (\detajj), and an 
estimate of \ptH (\ptHhat) larger or smaller than 200\GeV, where \ptHhat is 
obtained from the sum of the \ptvec of the two \PGt candidates and \ptvecmiss. 
All other events enter two so-called ``boosted'' categories, which are 
distinguished by the presence of exactly one or at least two jets in an event. 
The boosted categories are supposed to contain mostly \ggH events with \PH 
recoiling against one or several jets, but they also contain contributions from 
\qqH events that did not pass the VBF category selection. This leads to five 
categories for each \PGtPgt final state.

In each of these categories, 2D distributions are then built to provide more 
granularity for the analysis. The observables for these distributions are chosen 
to separate the signal from the backgrounds, but also to provide additional 
sensitivity to the individual STXS stage-1.2 bins. One of the observables is 
always chosen to be \mtt. In the 0-jet category of the \etau and \mutau final 
states, the \pt of the \tauh candidate is taken as a second observable, as the 
contribution from backgrounds with misidentified \tauh candidates significantly 
decreases with \pt. In the \emu and \tautau final states, where the sensitivity 
to 0-jet signal events is low, no second observable is chosen, and 1D distributions 
are used. In the VBF categories, the second observable is \mjj. In addition to 
aligning with the definition of the STXS stage-1.2 \qqH bins, using this variable 
as an observable increases the analysis sensitivity to the \qqH process as a 
whole, since the signal-to-background ratio quickly increases with increasing 
values of \mjj. In the boosted categories, the second observable is chosen to be 
\ptHhat. 

\begin{figure*}[htbp]
  \centering
  \includegraphics[width=0.90\textwidth]{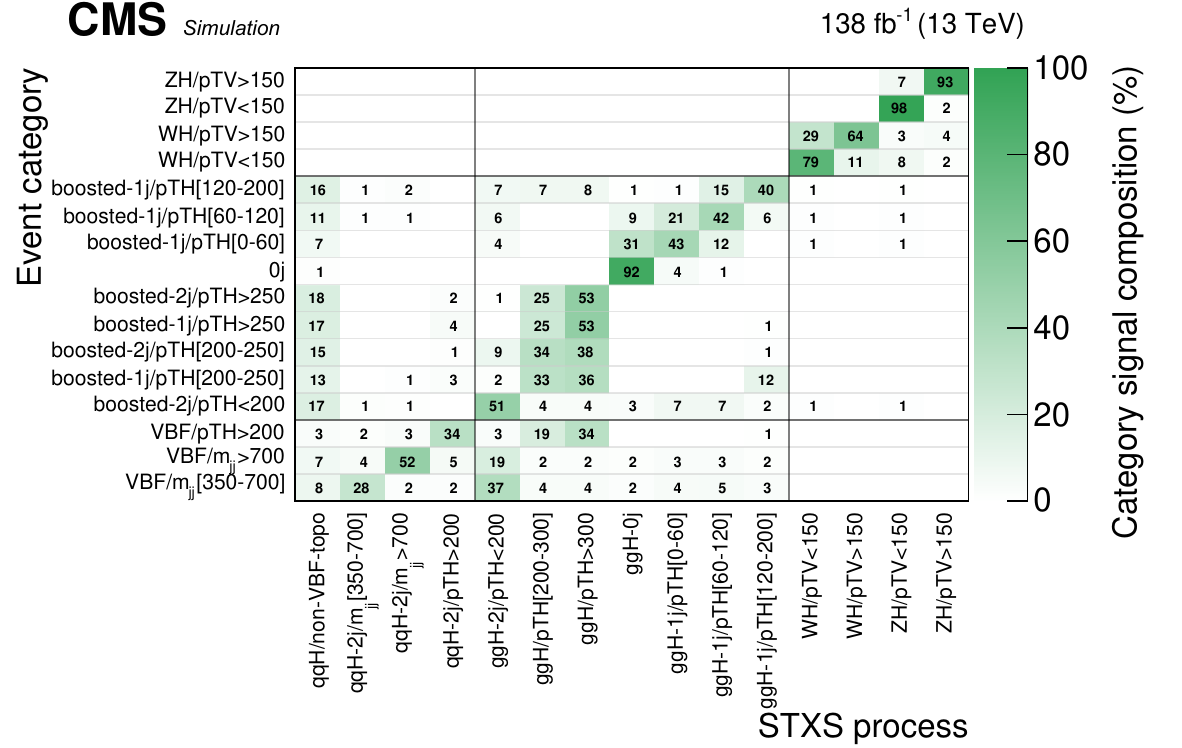}
  \caption{
    Signal composition of the subcategories of the \CBana, in terms of the STXS 
    stage-1.2 bins (in \%). The rows correspond to the signal categories described 
    in the text including the signal categories of the \VHana as described in 
    Section~\ref{sec:VH}. The columns refer to the STXS bins specified in 
    Figs.~\ref{fig:stxs_ggh}--\ref{fig:stxs_vh}. The STXS stage-1.2 \qqH bin 
    with $\Njet<2$ or $\mjj<350\GeV$ is labeled by ``qqH/non-VBF-topo''. This 
    figure is based on all \PGtPgt final states. 
  }
  \label{fig:cb-purity}
\end{figure*}

\begin{figure*}[htbp]
  \centering
  \includegraphics[width=1\textwidth]{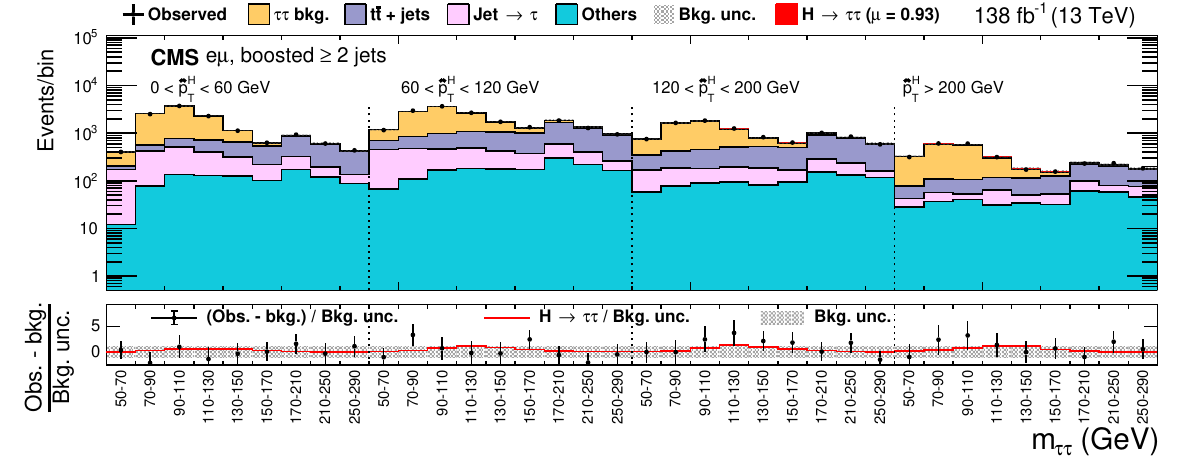}
  \includegraphics[width=1\textwidth]{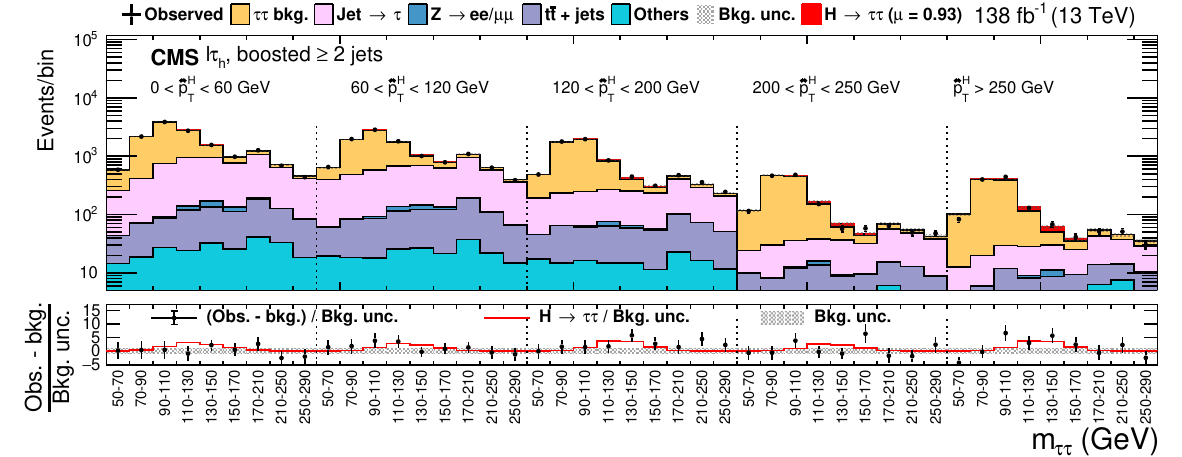}
  \includegraphics[width=1\textwidth]{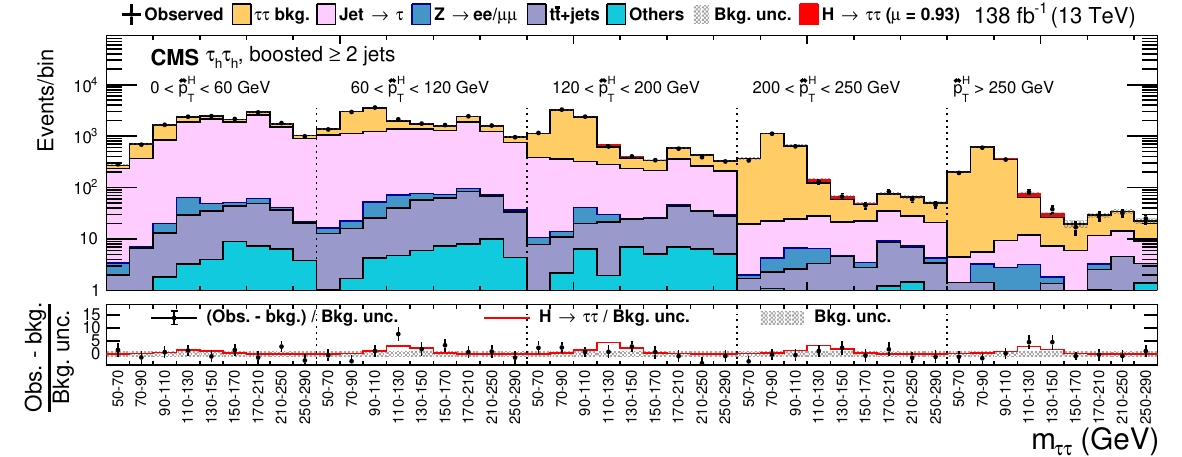}
  \caption{
    Observed and predicted 2D distributions in the ${\geq}2$ jet category of the 
    (upper) \emu, (middle) \ltau, and (lower) \tautau final states of the \CBana. 
    The predicted signal and background distributions are shown after the fit 
    used for the extraction of the inclusive signal. The ``Others'' background 
    contribution includes events from diboson and single~\PQt quark production, 
    as well as \HWW decays. The uncertainty bands account for all systematic and 
    statistical sources of uncertainty, after the fit to the data. 
  }
  \label{fig:cb-discriminator}
\end{figure*}

The category definitions, as well as the observables per category, are summarized 
in Table~\ref{tab:cb-categories}. Figure~\ref{fig:cb-purity} shows the composition 
of the categories in terms of signal in the individual STXS stage-1.2 bins 
integrated over all \PGtPgt final states. The subcategorization on the vertical 
axis of the figure is given by the categorization given in Table~\ref{tab:cb-categories} 
and the binning for \mjj and \ptHhat of the corresponding 2D distributions. In 
the boosted categories, the signal is generally composed of at least 50\% of the 
signal in the corresponding STXS bin, but there are migrations between adjacent 
\ptH bins because of the limited resolution of \ptHhat, as well as contributions 
from \ggH events with 0 jets in the low \ptHhat subcategories, and from \qqH 
events with 1 or 2 jets with low \mjj in the boosted subcategories with high 
\ptHhat. In the VBF categories, there is a mixture of \ggH and \qqH events with 
high \mjj, as well as limited contributions from \ggH events with lower \mjj but 
high \ptH. In the 2D distributions, the VBF categories with $\mjj>700\GeV$ are 
subdivided with additional \mjj thresholds going up to 1800\GeV depending on the 
category and final state. This binning provides an additional separation between 
the \ggH and \qqH events in the VBF categories. The 2D distributions for the 
boosted ${\geq}2$-jets category for the \emu, \ltau, and \tautau final states, 
for all data-taking years combined, are shown in Fig.~\ref{fig:cb-discriminator}. 

\begin{table*}[htbp]
  \centering
  \topcaption{
    Event categories of the \CBana. The STXS stage-0 and -1.2 measurements are 
    extracted by performing a maximum likelihood fit of 1D and 2D distributions 
    in these categories using the observables listed in the last column.
  }
  \begin{tabular}{llll}
    \hline
    Final state & Category & Selection & Observables \\
    \hline
    \multirow{8}{*}{\ltau, \emu} & 0-jet & 0 jet & \mtt, $\pt^{\tauh}$ (\ltau) \\
    & & & \mtt (\emu) \\[\cmsTabSkip]
    & VBF low \ptH & ${\ge}2$ jets, $\mjj>350\GeV$, $\ptHhat<200\GeV$ & \mtt, \mjj \\
    & VBF high \ptH & ${\ge}2$ jets, $\mjj>350\GeV$, $\ptHhat>200\GeV$ & \mtt, \mjj \\
    & Boosted 1 jet & 1 jet & \mtt, \ptHhat \\
    & Boosted ${\ge}2$ jets & Not in VBF, ${\ge}2$ jets & \mtt, \ptHhat \\
    \hline
    \multirow{5}{*}{\tautau} & 0-jet & 0 jet & \mtt \\
    & VBF low \ptH & ${\ge}2$ jets, $\Delta\eta_{jj}>2.5$\,(2.0 for 2016), & \mtt, \mjj \\
    & & $100<\ptHhat<200\GeV$ & \\
    & VBF high \ptH & ${\ge}2$ jets, $\Delta\eta_{jj}>2.5$\,(2.0 for 2016),  & \mtt, \mjj \\
    & & $\ptHhat>200\GeV$ & \\
    & Boosted 1 jet & 1 jet & \mtt, \ptHhat \\
    & Boosted ${\ge}2$ jets & Not in VBF, ${\ge}2$ jets & \mtt, \ptHhat \\
    \hline
  \end{tabular}
  \label{tab:cb-categories}
\end{table*}

\section{VH production modes}
\label{sec:VH}

Higgs boson production in association with a \PW or \PZ boson has a cross section 
that is much lower than the cross section for \ggH or \qqH production, but it 
provides an additional check of the predictions of the SM. The best sensitivity 
for \VH production is obtained using \PH decay modes with large branching 
fractions, such as \HTT and \HBB.

Four final states are considered for \WH production, corresponding to the \PW 
boson decays into $\Pe\Pgn$ and $\PGm\Pgn$, combined with the \etau, \mutau, and 
\tautau final states of the \HTT decay, and discarding the final state with two 
electrons and one \tauh because of the large background from electron charge 
misidentification. The \emu final state of the \HTT decay is not studied because 
of its small branching fraction and overlap with the \WH analysis with \HWW 
decays. For \ZH production, six final states are studied, corresponding to the 
\PZ boson decays into \PePe and \PGmPGm, combined with the \etau, \mutau, and 
\tautau final states of the \HTT decay. The \emu decay of the Higgs boson is not 
studied here as it is already included in the \ZH analysis targeting \HWW decays. 
This analysis supersedes a previous analysis that was performed using 2016 data 
only, which had found signal strengths relative to the SM prediction for \WH and 
\ZH production of $3.39^{+1.68}_{-1.54}$ and $1.23^{+1.62}_{-1.35}$, 
respectively~\cite{Sirunyan:vh}.

\subsection{WH final states}
\label{sec:WH}

After the trigger selection detailed in Section~\ref{sec:selection}, \Pe, \PGm, and 
\tauh candidates comprising the \emutau, \etautau, \mumutau, and \mutautau final 
states are required to satisfy the additional selection criteria listed in 
Table~\ref{tab:WHcuts}. Final states with a \PGm benefit from a lower \pt threshold 
at the trigger level. Beyond that, events with additional electrons, muons, or 
\PQb jets are rejected for all \WH final states.   

\begin{table}[htbp]
  \centering
  \topcaption{
    Selection requirements for \Pe, \PGm and \tauh candidates used in the \WH 
    analysis. The three \pt thresholds specified for the \Pe in the \etautau 
    final state refer to the years 2016, 2017, and 2018, respectively. The column 
    labeled \Irelem refers to the lepton isolation variable, defined in 
    Eq.~(\ref{eq:isolation}).
  }
  \begin{tabular}{cccc}
    \hline
    Candidate & \pt  (\GeVns) & $\abs{\eta}$ & \Irelem  \\
    \hline
    \Pe (\emutau)  & ${>}15$  &  ${<}2.1$ & ${<}0.15$ \\
    \Pe (\etautau) & ${>}(26,\,28,\,33)$  &  ${<}2.5$ & ${<}0.15$ \\
    \PGm           & ${>}10$  &  ${<}2.4$ & ${<}0.15$ \\ 
    \tauh          & ${>}20$  &  ${<}2.3$ & \NA{}     \\
    \hline
  \end{tabular}
  \label{tab:WHcuts}
\end{table}

In the \emutau final state, the \Pe and \PGm can come from either the \PW boson 
or from one of the \PGt leptons of the \HTT decay. The light lepton with the 
largest \pt is assigned to the \PW boson, and the other to \PH, which gives the 
correct pairing more than 75\% of the time. The \Pe and \PGm are required to 
have the same charges to reduce backgrounds with prompt and isolated electrons 
and muons, \eg, from \ZTT events in the \emu final state. The \tauh candidate is 
required to have the opposite charge relative to the light leptons, to give the 
proper charge for \PH. The \pt threshold for both the electron and muon is set 
to 15\GeV, instead of the baseline value of 10\GeV for the muon, to reduce 
backgrounds where jets are misidentified as electrons or muons. Three additional 
selection requirements are imposed in this final state to increase the purity of 
the expected signal selection. Specifically, these are: 

\begin{enumerate}
\item the scalar \pt sum of the two light leptons and the \tauh (\LT) is required 
to be larger than 100\GeV; 
\item the \abs{\Delta\eta} between the lepton associated with the \PW boson and 
the other two leptons is required to be less than 2; 
\item  the \abs{\Delta\phi} between the light lepton associated with the \PW 
boson and the other two leptons is required to be larger than 2. 
\end{enumerate}

The same selection requirements are applied for the \mumutau final state, where 
the same charge requirement on the muons already significantly reduces the 
background from \ZMM events. 

For the \etautau and \mutautau final states, the \tauh that has the same sign as 
the electron (muon) is required to have a $\pt>30\GeV$, \LT is required to be 
larger than 130\GeV, and the magnitude of the \ptvec sum of the \Pe (\PGm), the 
\tauh candidates, and \ptvecmiss (\ST) is required to be $\ST<70\GeV$.

\subsection{ZH final states}
\label{sec:ZH}

The \ZH analysis proceeds by identifying a pair of light leptons from either a 
\ZEE or \ZMM decay, and a pair of \PGt candidates comprising the \etau, \mutau, 
or \tautau final states. At least one of the light leptons from the \PZ boson, 
referred to here as the triggering lepton, is required to satisfy the online 
selection. The selection criteria for the light leptons are summarized in 
Table~\ref{tab:ZH_Zcuts}. The selection criteria for the \PGtPgt final state are 
summarized in Table~\ref{tab:ZHcuts}. The charges for the light leptons associated 
with the \ZLL decay, as well as the charges of the individual \PGt decays 
associated with the \PGtPgt final state are required to be of opposite sign. As 
is the case for the \WH final states, events with additional electrons, muons, 
or \PQb jets are rejected for all \ZH final states. Moreover, to increase the 
signal purity and to reject \jettau background events, the scalar \pt sum of the 
\PGt decay products in the \tautau final state is required to be larger than 
60\GeV.

\begin{table*}[htbp]
  \centering
  \topcaption{
    Selection requirements for \Pe and \PGm candidates that are associated with 
    the \ZLL decay in the \ZH analysis. The three \pt thresholds and the three 
    $\eta$ restrictions specified for the triggering leptons refer to the years 
    2016, 2017, and 2018, respectively. The column labeled \Irelem refers to the 
    lepton isolation variable, defined in Eq.~(\ref{eq:isolation}). 
  }
  \begin{tabular}{c@{\hspace{0.6cm}}ccc@{\hspace{0.6cm}}ccc}
    \hline 
    & \multicolumn{3}{c}{Triggering} & \multicolumn{3}{c}{Other} \\
    Object             & \pt  (\GeVns) & $\abs{\eta}$       & \Irelem  & \pt  
    (\GeVns) & $\abs{\eta}$ & \Irelem  \\ \hline
    \Pe (\ZEE)  & ${>}26,\,28,\,33$ & ${<}2.1$             & ${<}0.15$  & ${>}10$ & ${<}2.5$ & ${<}0.15$ \\ 
    \PGm (\ZMM) & ${>}23,\,25,\,25$ & ${<}2.1,\,2.4,\,2.4$ & ${<}0.15$  & ${>}10$ & ${<}2.4$ & ${<}0.15$ \\ 
    \hline
  \end{tabular}
  \label{tab:ZH_Zcuts}
\end{table*}

\begin{table}[htbp]
  \centering
  \topcaption{
    Selection requirements for \Pe, \PGm, and \tauh candidates associated with 
    the \PGtPgt decay in the \ZH analysis. First and second lepton refers to the 
    label of the final state in the first column. The column labeled \Irelem 
    refers to the lepton isolation variable, defined in Eq.~(\ref{eq:isolation}).
  }
  \begin{tabular}{c@{\hspace{0.6cm}}ccc@{\hspace{0.6cm}}cc}
    \hline
    & \multicolumn{3}{c}{First lepton} & \multicolumn{2}{c}{Second lepton} \\
    Mode & \pt (\GeVns)  & $\abs{\eta}$ & \Irelem  & 
    \pt (\GeVns) & $\abs{\eta}$  \\\hline
    \etau   & ${>}10$ & ${<}2.5$ & ${<}0.15$ & ${>}20$ & ${<}2.3$ \\
    \mutau & ${>}10$ & ${<}2.4$ & ${<}0.15$  & ${>}20$ & ${<}2.3$ \\
    \tautau & ${>}20$ & ${<}2.3$ & \NA   & ${>}20$ & ${<}2.3$ \\
    \hline
  \end{tabular}
  \label{tab:ZHcuts}
\end{table}

\subsection{VH observables}
\label{sec:VHobservables}

\begin{figure*}[htbp]
  \centering
  \includegraphics[width=0.75\textwidth]{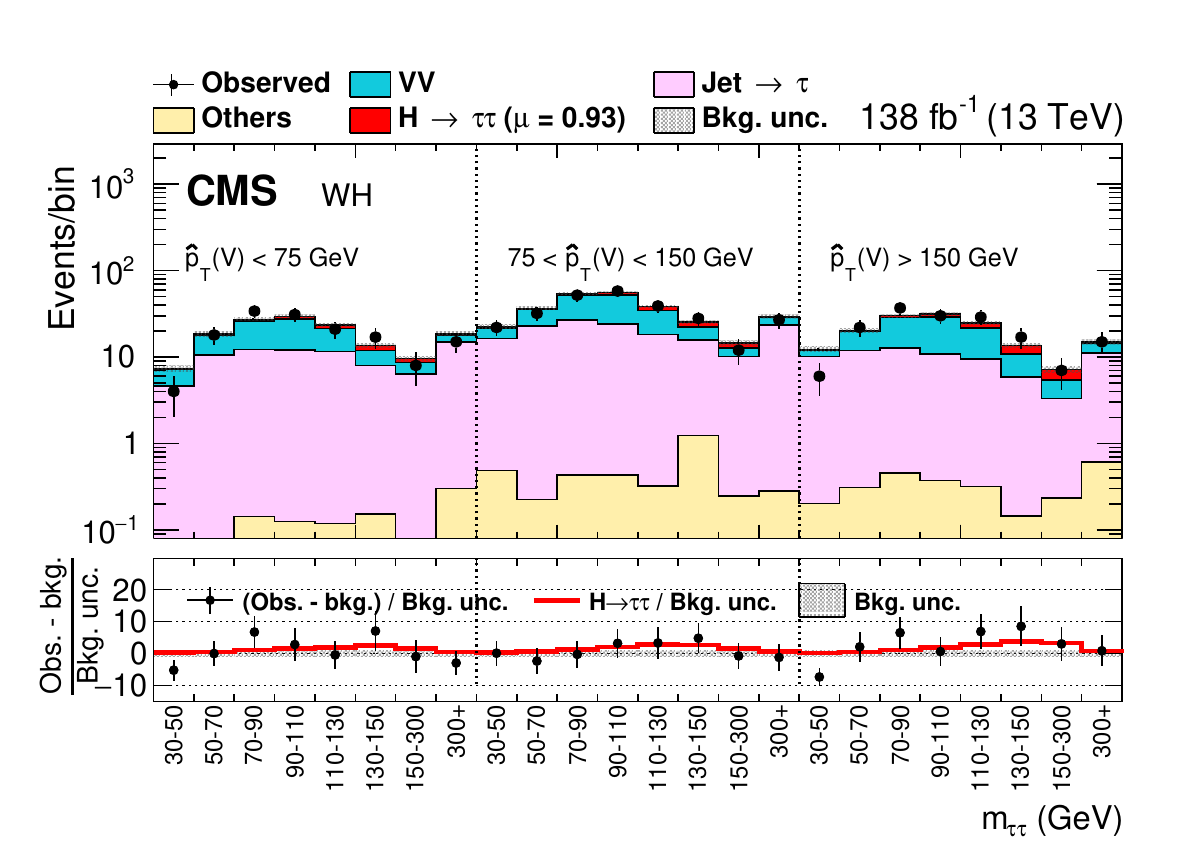}\\
  \includegraphics[width=0.75\textwidth]{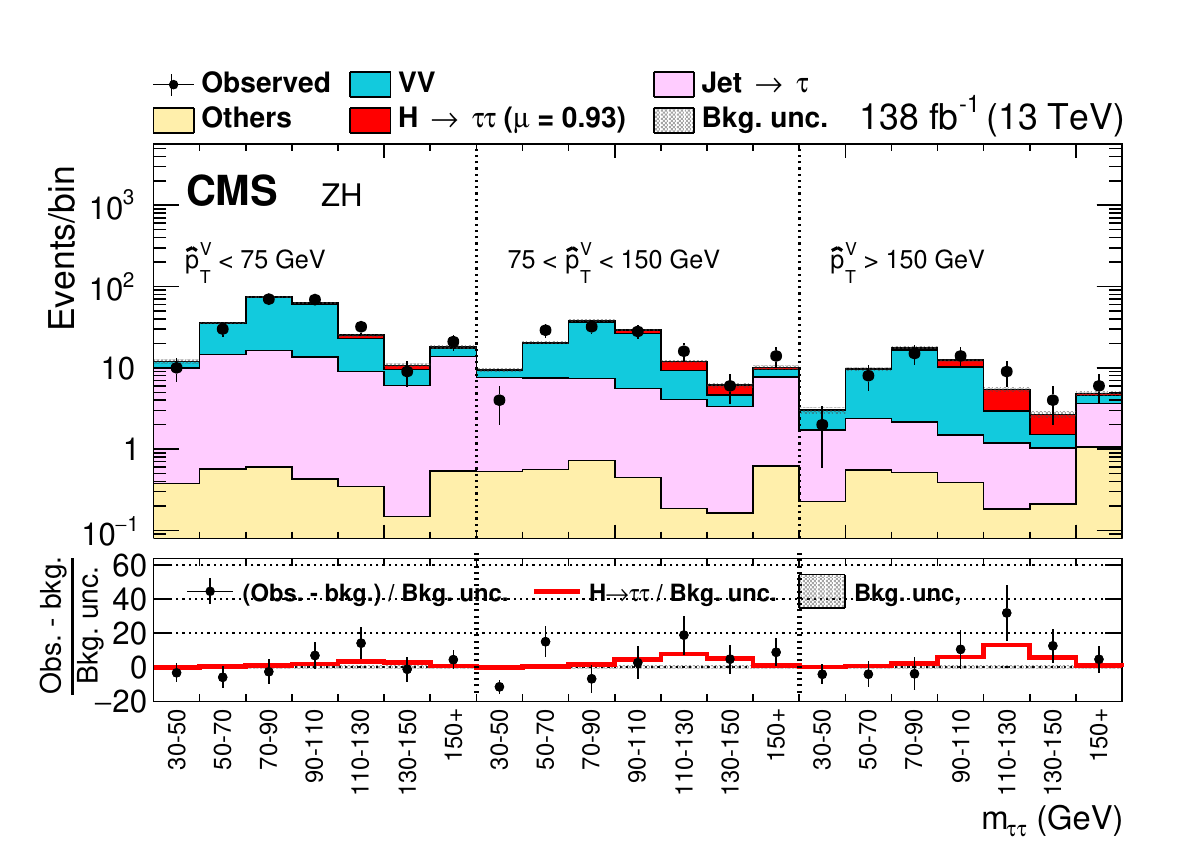}
  \caption{
    Observed and predicted distribution of \mtt in three bins of \ptVhat, 
    combined for all final states and all data-taking years for the (upper)
    \WH and (lower) \ZH analyses.
  }
  \label{fig:vh-signal-extraction}
\end{figure*}

The results are extracted by fitting 2D distributions of \mtt and the \pt of the 
reconstructed vector boson (\ptVhat), where \mtt is chosen because it provides 
the best discrimination between signal and background, and \ptVhat is chosen to 
separate the different \VH STXS stage-1.2 bins. For the \WH process, \ptmiss 
originates from both the \PW and \PH decays, resulting in a total of three or 
four neutrinos that escape detection. The system is underconstrained and the 
individual neutrinos cannot be fully reconstructed. However, an estimate of \ptVhat 
can still be made: Simulation studies indicate that the invisible \pt of the \HTT 
decay is on average 47 (69)\% of the visible \pt of the \PGtPgt system in the 
\tautau (\ltau) final state. Assuming that the sum of the neutrino momenta in 
the \PGtPgt final state points in the direction of the visible \pt, the neutrino 
system from the \HTT decay is assumed to have the corresponding proportions of 
the visible \PGtPgt four-momentum, this is known as the collinear approximation. 
The \ptvecmiss associated with the \PW decay is taken to be the difference between 
the reconstructed \ptvecmiss and the assumed \ptvec of the neutrino system from 
the \HTT decay. No significant dependence of this assumption on the \pt of the 
visible \PGtPgt decay products is observed. This estimate of \ptVhat provides a 
better separation between the STXS stage-1.2 bins than using other estimates of 
\ptV. For the \ZH process, the \pt of the \PZ boson, \ptVhat, is well reconstructed 
from the selected \ZLL final state. The only source of \ptmiss are the \PGt decays, 
and \mtt is obtained from the maximum likelihood estimate, as indicated in 
Section~\ref{sec:eventreco}. Observed and predicted distributions of \mtt in 
three bins of \ptVhat, combined for all final states and all data-taking years 
for the \WH and \ZH analyses are shown in Fig.~\ref{fig:vh-signal-extraction}.

\subsection{VH Background estimation}
\label{sec:backgroundEstimation}

Backgrounds to the \VH final states are estimated using methods similar to those 
described in Sections~\ref{sec:FF-method} and~\ref{sec:simulation}. There are, 
however, some notable differences for the estimates, which are obtained from data, 
as described below.   

\subsubsection{Backgrounds with the same final state as the signal}

In the \WH analysis, the dominant irreducible background is 
$\PW(\Pell\Pgn)\PZ(\PGtPgt)$, which can only be separated from the signal by 
using \mtt. Other small irreducible backgrounds are $\PZ\PZ\to4\Pell$, $\ttbar\PZ$, 
$\ttbar\PW$, $\PW\PW\PW$, $\PW\PW\PZ$, $\PW\PZ\PZ$, and $\PZ\PZ\PZ$. In the \ZH 
analysis, the dominant irreducible background is $\PZ\PZ\to4\Pell$, with small 
contributions from $\PW\PW\PZ$, $\PW\PZ\PZ$, $\PZ\PZ\PZ$, and $\ttbar\PZ$. The 
\HWW process is treated as an irreducible background to both \WH and \ZH production. 
All of the above processes are estimated from simulation. Simulated events where 
one of the reconstructed \Pe, \PGm, or \tauh candidates corresponds to a jet at 
the generator level are vetoed, because they are estimated as part of the backgrounds 
described below. 

\subsubsection{\texorpdfstring{Backgrounds with jets misidentified 
as an electron, muon, or hadronic \PGt lepton decay}{Backgrounds 
with jets misidentified as an electron, muon, or hadronic tau 
lepton decay}}

Reducible backgrounds have at least one jet that is misidentified as one of the 
objects in the final state (\Pe, \PGm, or \tauh). In the \WH analysis, they mostly 
consist of \PZ and \ttbar production, with small contributions from \Wjets, and 
diboson production, where one boson subsequently decays leptonically and the other 
hadronically. In the \ZH analysis, the reducible backgrounds mostly include \ZLL 
decays with one or two jets that are misidentified as \PGt decays.    

These backgrounds are estimated from data using a method that is similar to the 
one described in Section~\ref{sec:FF-method}, but differing in detail, as 
described below. The \FFi, with $i=\Pe,\,\PGm,\,\tauh$, denote the probabilities 
that a loosely selected \Pe, \PGm, or \tauh candidate, which is assumed to 
originate from a jet, will satisfy a tighter selection criterion, where in this 
context ``tight'' and ``loose'' refer to the object identification criteria of 
each corresponding object used in the analyses. The \FFi are measured in $\Pell
\Pell{+}\text{jets}$ events and are common for all \VH final states. They are 
applied in slightly different ways for the $\Pell\Pell\tauh$ and $\Pell\tauh\tauh$ 
final states in the \WH analysis, because of the different background compositions. 
The applications of the \FFi in the \WH and \ZH analyses differ because the number 
of final state objects subject to misidentification is not the same.

\noindent\textit{$\text{Jet}\to\tauh$ misidentification measurement}

The probability that a jet is misidentified as a \tauh candidate is measured 
using \ZMM events with additional jets. These events are selected by requiring 
two muons of opposite charges that satisfy the muon identification criteria, as 
given in Section~\ref{sec:eventreco}, and have $\Irelm<0.15$, $\pt>10\GeV$, and 
$\abs{\eta}<2.4$. The leading muon must have $\pt>23\,(25)\GeV$ in 2016 (2017--2018) 
data and the event must be selected through the single-muon trigger path in the 
online selection. The muons must be separated by $\DR>0.3$ from each other. In 
addition the event must have a \tauh candidate that satisfies the VVVLoose 
working point of \Dj, as defined in Ref.~\cite{CMS:2022prd}. The \tauh candidate 
must have $\pt>20\GeV$ and $\abs{\eta}<2.3$. Almost all \tauh candidates selected 
in this way are expected to originate from misidentified quark- or gluon-induced 
jets. The \FFjet are then measured as function of the \pt of the \tauh candidate 
for different \tauh decay modes. They are taken as the ratio between the number 
of \tauh candidates, selected as described above and passing the tighter DT working 
point used in the analysis, and the number of \tauh candidates selected as above, 
without any additional requirements. Genuine \tauh decays contribute to the \FFjet 
measurement at the percent level, mainly originating from $\PW\PZ$ and \ZZ production 
with leptonic \PV boson decays. Their contribution is estimated from simulation 
and subtracted from the data before taking the ratio. Depending on the decay mode 
and \pt of the \tauh candidate, \FFjet values lie in the range 0.03--0.15. 

\noindent\textit{$\text{Jet}\to\Pe$ misidentification measurement}

The probability that a jet is misidentified as an electron is also measured in 
\ZMM events with additional jets. These events are selected in the same way as 
described above, but with the loosely selected \tauh candidate replaced by a 
reconstructed electron with no specific identification or isolation requirements. 
This electron must satisfy $\pt>10\GeV$ and $\abs{\eta}<2.5$, and be separated 
by $\DR>0.3$ from the muons. In addition, we veto all events with $\mT^{\Pe}>40
\GeV$ to increase the purity of the measurement sample. The \FFe are then measured 
as a function of the electron candidate \pt. They are taken as the ratio between 
the number of electron candidates selected as described above, fulfilling in 
addition the identification requirement given in Section~\ref{sec:eventreco} and 
$\Irele<0.15$ and the number of electron candidates selected as described above 
without any additional requirements. The fraction of events selected in data with 
a genuine electron, which ranges between 20--30\%, is estimated from simulation 
and subtracted from the data before taking the ratio. The values of \FFe thus 
obtained fall in the range 0.004--0.013.

\noindent\textit{$\text{Jet}\to\PGm$ misidentification measurement}

Similarly to the $\text{jet}\to\Pe$ misidentification measurement, the probability 
that a jet is misidentified as a muon is measured using \ZEE events with additional 
jets. These events are selected in the same way as described above, but with the 
muons replaced by electrons. The electrons must satisfy the criteria outlined in 
Section~\ref{sec:eventreco}, $\pt>10\GeV$, and $\abs{\eta}<2.5$. The leading 
electron must have $\pt>26$, 28, and 33\GeV in 2016, 2017, and 2018, respectively, 
and the event must be selected through the single-electron trigger path in the 
online selection. The electrons are required to be separated by $\DR>0.3$ from 
each other. In addition, we veto all events with $\mT^{\PGm}>40\GeV$ to increase 
the purity of the measurement sample, and select a muon with no specific identification 
or isolation criteria. This muon must satisfy $\pt>10\GeV$, $\abs{\eta}<2.5$, and 
be separated by $\DR>0.3$ from the electrons. The \FFmu are then measured as 
function of the muon candidate \pt. They are taken as the ratio between the number 
of muon candidates selected as described above, with the additional requirement 
that the muon must satisfy the muon identification criteria as given in 
Section~\ref{sec:eventreco} and $\Irelm<0.15$ and the number of muon candidates 
selected as described above without any additional requirements. The fraction of 
events selected in data with a genuine muon, which ranges between 20--40\%, is 
estimated from simulation and subtracted from the data before taking the ratio. 
The values of \FFmu thus obtained fall in the range 0.01--0.03. 

\section{Systematic uncertainties}
\label{sec:uncertainties}

Control regions comprising event samples, that are not used to carry out the 
measurements, are used to ascertain how well the model describes the data, and to 
derive corrections and their corresponding uncertainties as needed.

A summary of all systematic uncertainties considered in the analyses is given in 
Table~\ref{tab:uncertainties}. They mainly comprise uncertainties in the object 
reconstruction and identification, in the signal and background modelings, and 
due to the limited population of template distributions available for the signal 
and background model. The last group of uncertainties is incorporated for each 
bin of each corresponding template individually following the approach proposed in 
Ref.~\cite{Barlow:1993dm}. All other uncertainties lead to correlated changes 
across bins taking the form of either normalization changes or general nontrivial 
shape-altering variations. Depending on the way they are derived, correlations 
may also arise across data-taking years, individual signal and background samples, 
or individual uncertainties.  

\begin{table*}[htbp]
  \centering
  \topcaption{
    Summary of the most important systematic uncertainties discussed in the text. 
    The columns indicate the source of uncertainty, the process class that it 
    applies to, the variation, and how it is correlated. A checkmark is given 
    also for partial correlations. More details are given in the text.
  }
  \begin{tabular}{llcccr@{--}lcc}
    \hline
    \multicolumn{2}{l}{Uncertainty} & \multicolumn{3}{c}{Process} 
    & \multicolumn{2}{c}{Variation} & \multicolumn{2}{c}{Correlated across} \\
    \multicolumn{2}{l}{} & Sim. & \PGt-emb. & \FF & \multicolumn{2}{c}{} 
    & Years & Processes \\
    \hline 
    \multirow{2}{*}{\PGt-emb.} & Acceptance & $\NA$ & $\checkmark$ & $\NA$ 
    & \multicolumn{2}{c}{4\%} & $\NA$ & $\NA$ \\
    & \ttbar fraction & $\NA$ & $\checkmark$ & $\NA$ 
    & 0.1 & 10\% & $\NA$ & $\NA$ \\[\cmsTabSkip]
    \multirow{3}{*}{\PGm} 
    & Id      & $\checkmark$ & $\checkmark$ & $\NA$ 
    & \multicolumn{2}{c}{2\%} & $\checkmark$ & $\checkmark$ \\
    & Trigger & $\checkmark$ & $\checkmark$ & $\NA$ 
    & \multicolumn{2}{c}{2\%} & $\NA$ & $\checkmark$ \\
    & $p_{\PGm}$ scale & $\checkmark$ & $\checkmark$ & $\NA$ 
    &  $0.4$ & $2.7\%$ & $\checkmark$ & $\checkmark$ \\[\cmsTabSkip]
    \multirow{3}{*}{$\Pe$} 
    & Id      & $\checkmark$ & $\checkmark$ & $\NA$ 
    & \multicolumn{2}{c}{2\%} & $\checkmark$ & $\checkmark$ \\
    & Trigger & $\checkmark$ & $\checkmark$ & $\NA$ 
    & \multicolumn{2}{c}{2\%} & $\NA$ & $\checkmark$ \\
    & $E_{\Pe}$ scale & $\checkmark$ & $\checkmark$ & $\NA$ 
    & \multicolumn{2}{c}{See text} & $\checkmark$ & $\checkmark$ \\[\cmsTabSkip]
    \multirow{3}{*}{\tauh} 
    & Id      & $\checkmark$ & $\checkmark$ & $\NA$ 
    & 3 & 5\% & $\NA$ & $\checkmark$ \\
    & Trigger & $\checkmark$ & $\checkmark$ & $\NA$ 
    & 5 & 10\% & $\NA$ & $\checkmark$ \\
    & $E_{\tauh}$ scale & $\checkmark$ & $\checkmark$ & $\NA$ 
    & 0.2 & 1.1\% & $\NA$ & $\checkmark$ \\[\cmsTabSkip]
    \multirow{2}{*}{$\PGm\to\tauh$} & Mis-Id & $\checkmark$ & $\NA$ & $\NA$ 
    & 10 & 70\% & $\NA$ & $\NA$ \\
    & $E_{\tauh}$ scale & $\checkmark$ & $\NA$ & $\NA$ 
    & \multicolumn{2}{c}{2\%} & $\NA$ & $\NA$ \\[\cmsTabSkip]
    \multirow{2}{*}{$\Pe\to\tauh$} & Mis-Id & $\checkmark$ & $\NA$ & $\NA$ 
    & \multicolumn{2}{c}{40\%} & $\NA$ & $\NA$ \\
    & $E_{\tauh}$ scale & $\checkmark$ & $\NA$ & $\NA$ 
    & $1.0$ & $2.5\%$ & $\NA$ & $\NA$ \\[\cmsTabSkip]
    \multicolumn{2}{l}{$\PZ$ boson $\pt$ reweighting} & $\checkmark$ & $\NA$ & $\NA$
    & $10$ & $20\%$ & $\checkmark$ & $\NA$ \\[\cmsTabSkip]
    \multicolumn{2}{l}{$E_{\text{Jet}}$ scale and resolution} & $\checkmark$ & $\NA$ & $\NA$
    & $0.1$ & $10\%$ & $\checkmark$ & $\checkmark$ \\[\cmsTabSkip]
    \multicolumn{2}{l}{$\PQb$-jet (mis-)Id} & $\checkmark$ & $\NA$ & $\NA$
    & \multicolumn{2}{c}{$(1)\,10\%$} & $\NA$ & $\checkmark$ \\[\cmsTabSkip]
    \multicolumn{2}{l}{$\ptmiss$ calibration} & $\checkmark$ & $\NA$ & $\NA$
    & \multicolumn{2}{c}{See text} & $\checkmark$ & $\checkmark$ \\[\cmsTabSkip]
    \multicolumn{2}{l}{ECAL timing shift} & $\checkmark$ & $\NA$ & $\NA$
    & 2 & 3\% & $\checkmark$ & $\checkmark$ \\[\cmsTabSkip]
    \multicolumn{2}{l}{\PQt quark \pt reweighting} & $\checkmark$ & $\NA$ & $\NA$
    & \multicolumn{2}{c}{See text} & $\checkmark$ & $\NA$ \\[\cmsTabSkip]
    \multicolumn{2}{l}{Integrated luminosity} & $\checkmark$ & $\NA$ & $\NA$
    & \multicolumn{2}{c}{$1.6\%$} & $\checkmark$ & $\checkmark$ \\[\cmsTabSkip]
    \multicolumn{2}{l}{Process normalizations} & $\checkmark$ & $\NA$ & $\NA$
    & \multicolumn{2}{c}{See text} & $\checkmark$ & $\NA$ \\[\cmsTabSkip]
    \multicolumn{2}{l}{Signal acceptance} & $\checkmark$ & $\NA$ & $\NA$
    & 1 & 10\% & $\checkmark$ & $\NA$ \\[\cmsTabSkip]
    \multirow{4}{*}{\FF} & Sample size & $\NA$ & $\NA$ & $\checkmark$
    & 3 & 5\% & $\NA$ & $\NA$ \\
    & Nonclosure & $\NA$ & $\NA$ & $\checkmark$
    & \multicolumn{2}{c}{10\%} & $\NA$ & $\NA$ \\
    & Non-\FF processes & $\NA$ & $\NA$ & $\checkmark$
    & \multicolumn{2}{c}{7\%} & $\NA$ & $\NA$ \\
    & \FF proc. composition & $\NA$ & $\NA$ & $\checkmark$
    & \multicolumn{2}{c}{7\%} & $\NA$ & $\NA$ \\
    \hline 
  \end{tabular}
  \label{tab:uncertainties}
\end{table*}

\subsection{Corrections to the model}
\label{sec:corrections}

{\tolerance=1200
The following corrections equally apply to simulated and \PGt-embedded events, 
where the \PGt lepton decay is also simulated. Since the simulation of \PGt-embedded
events happens under different detector conditions, corrections and related 
uncertainties may differ, as detailed in Ref.~\cite{Sirunyan:2019drn}. Corrections 
are derived for residual differences in the efficiency of the selected triggers, 
differences in the electron and muon tracking efficiencies, and in the efficiency 
of the identification and isolation requirements for electrons and muons. These 
corrections are obtained in bins of \pt and $\eta$ of the corresponding lepton, 
using the ``tag-and-probe'' method described in Ref.~\cite{Khachatryan:2010xn} 
with \ZEE and \ZMM events. They usually amount to not more than a few percent. 
In a similar way, corrections are obtained for the efficiency of triggering on 
the \tauh decays and for the \tauh identification efficiency, following the 
procedures described in Ref.~\cite{Sirunyan:2018pgf}. The latter is derived as a 
function of the \pt of the \tauh candidate in four bins below 40\GeV and one bin 
above. For $\pt^{\tauh}>40\GeV$ a correction is derived also for each \tauh decay 
mode, which is used only in the \tautau final state. Corrections to the energy 
scale of the \tauh decays and for electrons misidentified as \tauh candidates are 
derived for each data-taking year and each \tauh decay mode individually, from 
likelihood scans of discriminating observables, like the mass of the visible decay 
products of the \tauh candidate, as detailed in Ref.~\cite{Sirunyan:2018pgf}. For 
the trigger efficiency, corrections are obtained from parametric fits to the 
trigger efficiency curves derived for each corresponding sample and data.
\par}

The following corrections only apply to fully simulated events. During the 
2016--2017 data taking, a gradual shift in the timing of the inputs of the ECAL 
L1 trigger in the region of $\abs{\eta}>2.0$ caused a specific 
inefficiency~\cite{Sirunyan:2020zal}. For events containing an electron (a jet) 
with $\pt\gtrsim 50(100)\GeV$, in the region of $2.5<\abs{\eta}<3.0$ the efficiency 
loss is 10--20\%, depending on \pt, $\eta$, and time. Corresponding corrections 
depend on the multiplicity, $\eta$, and \pt distributions of the jets in an event. 
They are at the level of 1\% for \ttbar events and up to 5\% for signal events in 
VBF. In the \qqH STXS stage-1.2 bins they can be up to 15\%. Uncertainties range 
between 0.2 and 3\% depending on the size of the correction. 

The energies of jets are corrected to the expected response of the jet at the 
stable-hadron level, using corrections measured in bins of the jet \pt and 
$\eta$, as described in Ref.~\cite{Khachatryan:2016kdb}. These corrections are 
usually not larger than 10--15\%. Residual data-to-simulation corrections are 
applied to the simulated event samples, which are also propagated to \ptvecmiss. 
They usually range from less than 1\% at high jet \pt in the central part of 
the detector to a few percent in the forward region. An explicit correction to 
the direction and magnitude of \ptvecmiss is obtained from differences between 
estimates of the hadronic recoil in \ZMM events in data and simulation, as 
described in Ref.~\cite{Sirunyan:2019kia}. This correction is applied to the 
simulated \ZLL, \Wjets, and signal events, where a hadronic recoil against a 
single particle is well defined, and replaces the propagation of jet energy scale 
corrections to the \ptvecmiss. The efficiencies for genuine and misidentified 
\PQb jets to pass the working points of the \PQb jet identification discriminants, 
as given in Section~\ref{sec:eventreco}, are determined from data, using \ttbar 
events for genuine \PQb jets and \PZ boson production in association with jets 
for jets originating from light quarks or gluons. Data-to-simulation corrections 
are obtained for these efficiencies and used to correct the number of \PQb jets 
in the simulation, as described in Ref.~\cite{Sirunyan:2017ezt}. 

Data-to-simulation corrections are further applied to \ZEE (\PGmPGm) events 
in the \etau (\mutau) and \tautau final states, in which an electron (muon) is 
reconstructed as a \tauh candidate, to account for residual differences in the 
$\Pe(\PGm)\to\tauh$ misidentification rate between data and simulation. 
Deficiencies in the modeling of \ZLL events, which have been simulated only at 
LO precision in $\alpS$, are corrected for by a weighting of the simulated \ZMM 
events to data in bins of $\pt^{\PGmPGm}$ and $m_{\PGmPGm}$. In addition, all 
simulated \ttbar events are weighted to better match the top quark \pt distribution 
observed in data~\cite{Khachatryan:2015oqa}. 

In the \NNana the background classes described in Section~\ref{sec:NN-based} and 
summarized in Table~\ref{tab:nn-event-classes} are included in the likelihood 
model discussed in Section~\ref{sec:results} to further constrain these backgrounds 
during the signal extraction process. 

\subsection{Uncertainties related to the \texorpdfstring{\PGt}
{tau}-embedding method or the simulation}

The following uncertainties related to the level of control of the reconstruction 
of electrons, muons, and \tauh decays after selection refer to simulated and 
\PGt-embedded events. Unless stated otherwise, they are partially correlated 
across \PGt-embedded and simulated events.

Uncertainties in the electron and muon tracking, reconstruction, identification, 
and isolation efficiencies amount to 2\% each for electrons and 
muons~\cite{CMS:2020uim,CMS:2018rym}. They are introduced as normalization 
uncertainties. Uncertainties in the electron or muon trigger efficiencies contribute 
an additional normalization uncertainty of 2\% for events selected with each 
corresponding trigger. Due to differences in the trigger leg definitions they 
are treated as uncorrelated for single-lepton and lepton-pair triggers, which 
may result in shape altering effects in the overall model, since both triggers 
act on different regimes in lepton \pt.

Uncertainties in the muon momentum scale range between 0.4--2.7\%, depending on 
the muon $\eta$~\cite{CMS:2018rym}. For fully simulated events an uncertainty in 
the electron energy scale is derived from the calibration of ECAL crystals, and 
applied on an event-by-event basis~\cite{Khachatryan:2015hwa}. It usually amounts 
to less than 1.5\% with a dependence on the 0.1\% level in magnitude on the 
electron \pt and $\eta$. For \PGt-embedded events uncertainties of 0.50--1.25\%, 
split by the ECAL barrel and endcap regions, are derived for the corrections 
described in Section~\ref{sec:corrections}. Due to the different ways, the 
uncertainties are determined and differences in detector conditions, they are 
treated as uncorrelated across simulated and \PGt-embedded events. They lead to 
shape-altering variations and are treated as correlated across years.

Uncertainties in the \tauh identification range from 3--5\% in its different 
bins of \pt and decay mode. They are obtained from the corrections described in 
Section~\ref{sec:corrections} following procedures as described 
in Ref.~\cite{Sirunyan:2018pgf}. Due to the nature of the way they are derived, 
these uncertainties are statistically dominated and therefore treated as uncorrelated 
across decay modes, \pt bins, and data-taking years. An additional normalization 
uncertainty of 3\% is quadratically added in the \mutau and \tautau final states 
to account for the use of different working points of \De and \Dm compared to 
the selection that has been used for the determination of the corrections. 
Uncertainties in the \tauh trigger efficiency usually range between 5--10\% 
depending on the \tauh \pt. They are obtained from parametric fits to data and 
simulation, and lead to shape-altering effects. They are treated as uncorrelated 
across trigger paths and data-taking years. Uncertainties in the \tauh momentum 
scale range between 0.2--1.1\% depending on the \tauh \pt and decay mode. They 
are treated as uncorrelated across decay modes, \pt bins, and data-taking years 
for the same reason as given in the case of the uncertainties in the $\tauh$ 
identification efficiency. 

Two further sources of uncertainty are considered for \PGt-embedded events, 
as discussed in Section~\ref{sec:tau-embedding}. A 4\% normalization uncertainty 
accounts for the level of control in the efficiency of the \PGmPGm selection 
in data, which is unfolded during the \PGt-embedding procedure. The dominant 
part of this uncertainty originates from the trigger used for selection. Since 
these trigger setups differed across data-taking years, this uncertainty is treated 
as uncorrelated in this respect. Another shape- and normalization-altering 
uncertainty relates to the normalization of $\ttbar\to\PGmPGm+\text{X}$ decays, 
which are part of the \PGt-embedded event samples. It ranges from less than 1 to 
3\%, depending on the event composition of the model. For this uncertainty the 
number and shape of \ttbar events contained in the \PGt-embedded event samples 
are estimated from simulation, for which the corresponding decay has been selected 
at the parton level. This estimate is then varied by ${\pm}10\%$. A more detailed 
discussion of these uncertainties is given in Ref.~\cite{Sirunyan:2019drn}.

For fully simulated events, as discussed in Section~\ref{sec:simulation}, the 
following additional uncertainties apply: For the rare cases of electrons or 
muons misidentified as \tauh candidates, an uncertainty is derived in bins of 
\pt, $\eta$, and decay mode of the misidentified \tauh candidate. It amounts 
to up to 40\% for electrons and ranges 10--70\% for muons. The relatively large 
size of these uncertainties originates from the rareness of such cases in the 
control regions that are used to measure the rates. These uncertainties influence 
the measurements at the sub percent level. Uncertainties in the momentum scale 
of these misidentified leptons range 0.8--6.6\% (amount to 1\%) for 
electrons~\cite{CMS:2020uim} (muons~\cite{CMS:2018rym}). Uncertainties in the 
energy calibration and resolution of jets are applied with different correlations 
depending on their sources, comprising the limited sample size for the measurements 
used for calibration, the time-dependence of the energy measurements in data due 
to detector aging, and corrections introduced to cover residual differences 
between simulation and data~\cite{Khachatryan:2016kdb}. They range between sub 
percent level and $\mathcal{O}(10\%)$ depending on the kinematic properties of 
the jets in an event. They have the largest impact in the \mtt distribution for 
simulated events, since they are propagated to \ptvecmiss, and lead to migration 
effects in \Njet or \mjj. Uncertainties in the jet energy resolution typically 
have a lower impact than the sum of the jet energy scale uncertainties. They only 
have relevance due to migrations in \mjj.

Depending on the process under consideration, two independent uncertainties in 
\ptmiss are applied. For processes that are subject to recoil corrections, \ie, 
\PZ boson, \Wjets production, and signal, uncertainties in the calibration and 
resolution of the hadronic recoil are applied, ranging 1--5\%. For all other 
processes an uncertainty in \ptmiss results from the uncertainties in the jet 
energy calibration, which is propagated to \ptvecmiss. An additional uncertainty 
is obtained from the amount of unclustered energy in the event, which is not 
subject to the jet energy calibration~\cite{Sirunyan:2019kia}. These uncertainties 
mostly affect the event selections based on $\mT^{\Pell}$ and \Dzeta, the 
distribution of \mtt, and \ptHhat used for event classification in the \CBana.

{\tolerance=800
Uncertainties in the misidentification rate for light quark- or gluon-induced 
jets as \PQb jets amount to 1\%. The uncertainty in the efficiency to identify 
\PQb jets amounts to 10\%~\cite{Sirunyan:2017ezt,Bols:2020bkb}. 
\par}

Additional uncertainties account for the timing shift of the inputs to the ECAL 
L1 trigger described in Section~\ref{sec:corrections}. These result in a 
normalization variation of 1--3\% for the inclusive VBF signal sample. A 
shape-altering uncertainty is derived in the reweighting of the top quark \pt 
distribution, described in Section~\ref{sec:corrections}, applying the 
correction twice or not applying it at all. It usually has only a small effect 
on the final discriminants. The integrated luminosities of the 2016--2018 
data-taking periods are individually known with uncertainties in the 1.2--2.5\%
range~\cite{CMS:2021xjt,CMS-PAS-LUM-17-004,CMS-PAS-LUM-18-002}, while the 
total Run-2 (2016--2018) integrated luminosity has an uncertainty of 1.6\%, the 
improvement in precision reflecting the (uncorrelated) time evolution of some 
systematic effects. Finally, uncertainties in the predictions of the normalizations 
of all simulated processes amount to 4\% for \ZLL and \Wjets 
production~\cite{Melnikov:2006kv}, 6\% for \ttbar production~\cite{Czakon:2011xx,
Kidonakis:2013zqa}, and 5\% for diboson and single~\PQt quark 
production~\cite{Kidonakis:2013zqa,Campbell:2011bn,Gehrmann:2014fva}. These 
uncertainties have been applied in all places, where the corresponding simulated 
samples have been used, in the analyses. They are correlated across years.

\subsection{Uncertainties related to jets misidentified as hadronic 
\texorpdfstring{\PGt}{tau} lepton decays, electrons, or muons}

The \FFi described in Sections~\ref{sec:FF-method} and the \FFjet described 
in Sections~\ref{sec:backgroundEstimation}, and their corrections are subject to 
statistical fluctuations in each corresponding DR$^{i}$. These uncertainties are 
split into normalization- and shape-altering parts and propagated to the final 
discriminants for each analysis. They usually amount to a few percent and are 
treated as uncorrelated across the kinematic and topological bins they are derived 
in. Additional uncertainties are applied to capture the needs and magnitudes of 
bias corrections and extrapolation factors, varying from a few percent to 
$\mathcal{O}(10\%)$, depending on the kinematic properties of the \tauh candidate 
and the topology of the event. These are both normalization and shape-altering 
uncertainties. An additional source of uncertainty concerns the subtraction of 
processes other than the enriched process in each corresponding DR$^{i}$. These 
are subtracted from the data using simulated or \PGt-embedded events. The combined 
shape of the events to be removed is varied by 7\%, and the measurements are 
repeated. The impacts of these variations are then propagated to the final 
discriminants as shape-altering uncertainties. An uncertainty in the estimate of 
the three main background fractions in the AR, as described in 
Section~\ref{sec:FF-method}, is estimated from a variation of each individual 
contribution also by 7\%, increasing or decreasing the remaining fractions such 
that the sum of all contributions remains unchanged. The amount of variation is 
motivated by the uncertainty in the production cross sections and acceptances of 
the involved processes, bounded by the constraint on the process composition that 
can be clearly obtained from the AR. The effect of this variation is observed to 
be very small, since usually one of the contributions dominates the event composition 
in the AR. Due to their mostly statistical nature, and differences across years, 
all uncertainties related to the \FF-method are treated as uncorrelated across 
years. 

Several shape-altering uncertainties affect the estimation of the backgrounds 
where a jet is misidentified as an electron or muon, as described in 
Sections~\ref{sec:em-background} and~\ref{sec:backgroundEstimation}. The first 
group covers statistical uncertainties in the determination of the \TF, \FFe, 
and \FFmu. The second set covers statistical uncertainties in the bias corrections 
in \ptvecE and \ptvecM. Due to the way they are derived these uncertainties are 
treated as uncorrelated across individual kinematic bins. The last uncertainty 
covers the bias correction due to the determination of the \TF from a control 
region with nonisolated muons, for the method described in 
Section~\ref{sec:em-background}. The combined effect of these uncertainties is 
usually not larger than 10\%.

\subsection{Uncertainties related only to the signal}
\label{sec:signal-uncertainties}

For the measurement of the signal strengths $\mu_{s}$ of the \PH production cross 
sections with respect to the SM expectations, the following uncertainties are taken 
into account, following the recommendations of the LHC Higgs Working 
Group~\cite{deFlorian:2016spz}: For $\mathcal{B}(\HTT)$ an uncertainty of 1.2\% 
is assumed. For $\mathcal{B}(\HWW)$, which is considered as background in the 
\emu final state of the CB- and \NNanapl and in the \VHana, an uncertainty of 
1.5\% is assumed. The uncertainties due to PDF variations and the uncertainty 
in \alpS are obtained following the PDF4LHC recommendations, taking the root 
mean square of the variation of the results when using different replicas of the 
default NNPDF sets, as described, \eg, in Ref~\cite{Butterworth:2015oua}. 
Uncertainties due to the choice of the renormalization ($\mu_{r}$) and factorization 
($\mu_{f}$) scales in the calculation of the matrix elements are obtained from 
an independent variation of these scales by factors of 0.5 and 2, omitting the 
variations where one scale is multiplied by 2 and the corresponding other scale 
by 0.5. The uncertainties are then obtained from an envelope of these variations.   

For the inclusive and STXS stage-0 measurements the uncertainties due to PDF 
variations and the uncertainty in \alpS amount to 3.2, 2.1, 1.9, and 1.6\% for 
the gluon fusion, VBF, \WH, and \ZH production modes, respectively. The 
uncertainties from the variations of $\mu_{r}$ and $\mu_{f}$ amount to 3.9, 0.4, 
0.7, and 3.8\%, for each corresponding process.

{\tolerance=800
For the STXS stage-1.2 measurements, additional uncertainties are obtained varying 
$\mu_{r}$ and $\mu_{f}$ within each STXS bin using the simulation of signal with 
\POWHEG as described in Section~\ref{sec:simulation}. These uncertainties account 
for relative variations across STXS stage-1.2 bins and migration effects. Acceptance 
effects, within a given STXS stage-1.2 bin are also taken into account. Their 
combined effect is typically below 1\%. An uncertainty in the parton-shower model 
of \PYTHIA is obtained by varying the scales in the initial- and final-state 
radiation models; the observed effect typically ranges 1--3\%, but can become as 
large as 10\% for gluon fusion production with VBF-like topologies. For the 
measurements of \PH production cross sections, the effects of theoretical 
uncertainties in the normalizations of the signal templates within each STXS bin 
are removed from the uncertainty model. 
\par}

\section{Results}
\label{sec:results}

The model used to extract the signal from the data is defined by an extended binned 
likelihood of the form
\begin{linenomath}
\ifthenelse{\boolean{cms@external}}
{
  \begin{multline}
    \label{eq:likelihood}
    \mathcal{L}\left(\{k_{i}\},\,\{\mu_{s}\},\,\{\theta_{j}\}\right) = \\
    \prod\limits_{i}\mathcal{P}\bigl(k_{i}|\sum\limits_{s}\mu_{s}\,S_{s}(\{\theta_{j}\})
     +\sum\limits_{b}B_{b}(\{\theta_{j}\})\bigr)\,
    \prod\limits_{j}\mathcal{C}(\tilde{\theta}_{j}|\theta_{j}),
  \end{multline}
}
{
  \begin{equation}
    \label{eq:likelihood}
    \mathcal{L}\left(\{k_{i}\},\,\{\mu_{s}\},\,\{\theta_{j}\}\right) =
    \prod\limits_{i}\mathcal{P}\bigl(k_{i}|\sum\limits_{s}\mu_{s}\,S_{s}(\{\theta_{j}\})
    +\sum\limits_{b}B_{b}(\{\theta_{j}\})\bigr)\,
    \prod\limits_{j}\mathcal{C}(\tilde{\theta}_{j}|\theta_{j}),
  \end{equation}
}
\end{linenomath}
where $i$ labels all bins of the input distributions for each signal class, with 
index $s$, and background class, with index $b$, defined for each of the analyses 
as described in Sections~\ref{sec:NN-based}--\ref{sec:VH}. Signal and background 
templates are obtained from the data model as discussed in Section~\ref{sec:model}, 
with further specifications for the \VHana in Section~\ref{sec:VH}. The function 
$\mathcal{P}(k_{i}|\sum \mu_{s}\,S_{s}(\{\theta_{j}\})+\sum B_{b}(\{\theta_{j}\}
))$ in Eq.~(\ref{eq:likelihood}) corresponds to the Poisson probability to observe 
$k_{i}$ events in bin $i$ for a prediction of $\sum\mu_{s}\,S_{s}$ signal and 
$\sum B_{b}$ background events in that given bin. The scaling parameters $\mu_{s}$ 
of the signal contributions $S_{s}$ with respect to the SM expectation are the 
parameters of interest (POIs). Their number varies between 16 and 1, depending 
on the measurement: 16 POIs for the STXS stage-1.2 measurement with a one-to-one 
correspondence to the targeted STXS stage-1.2 bins; three POIs (\muggh, \muqqh, 
and \muvh) for the STXS stage-0 measurements; and a single POI (\muinc) for the 
inclusive signal-strength measurement. For the inclusive and STXS stage-0 
measurements, starting from the STXS stage-1.2 set of POIs, the POIs are 
successively combined, always assuming relations across the signal templates of 
the combined STXS bins as for the SM. For each individual measurement all POIs 
are obtained from one single fit of the model derived in this way to the data.
For each measurement, the corresponding POIs are not constrained. In particular 
they are not bound to be ${\geq}0$ in the absence of signal, but are allowed to 
take also negative values, to obtain unbiased confidence intervals and facilitate 
combinations with independent data.

For the \NNana the input distributions are obtained from the \yl, where $l$ 
comprises the signal and background classes. Depending on the measurement, two 
different sets of input distributions are used for the signal. For the stage-0 
measurement, single 2D histograms for signal are used, binned in \yqqH and 
\yggH. An example distribution in the \mutau final state is shown in 
Fig.~\ref{fig:nn-discriminator-stage-0}. Together with the input distributions 
for the background classes this results in 66 input distributions split by 
\PGtPgt final state and data-taking year. For the stage-1.2 measurement, 15 
1D histograms are used for each individual STXS stage-1.2 bin. We note that the 
event categories targeting the $\ggH\,\Njet\geq2$ bin have been subdivided in 
anticipation of future measurements, as discussed in Section~\ref{sec:NN-based}. 
Therefore, the number of categories is slightly larger than the number of POIs 
used for signal extraction. Together with the input distributions for the background 
classes, this results in 234 input distributions split by \PGtPgt final states 
and data-taking years. Example distributions for three adjacent bins in \ptH in 
the $\ggH\,\Njet=1$ subspace, and for the \qqH STXS bin with $\Njet\geq2$ and 
$\mjj>350\GeV$ are shown in Fig.~\ref{fig:nn-discriminator-stage-1}.  

For the \CBana, the signal extraction is based on 60 input distributions provided 
for five event categories split by \PGtPgt final state and data-taking year. The 
inputs are usually provided as 2D histograms binned in two discriminating 
observables, with the exception of the 0-jet categories in the \emu and \tautau 
final states, for which 1D distributions are used. An overview of the discriminating 
observables, from which the input distributions are obtained, is given in 
Table~\ref{tab:cb-categories}. Example distributions are shown in 
Fig.~\ref{fig:cb-discriminator}. 

For the \VHana, 30 input distributions are used, split by four (six) final states 
in the \WH (\ZH) analysis and data-taking years. The inputs are provided as 2D 
histograms binned in \mtt and \ptVhat, as defined in Section~\ref{sec:VH}. Input 
distributions of all final states and all data-taking years combined are shown in 
Fig.~\ref{fig:vh-signal-extraction}.
 
Systematic uncertainties are incorporated in the form of penalty terms for additional 
nuisance parameters $\{\theta_{j}\}$ in the likelihood, appearing as a product 
with predefined probability density functions $\mathcal{C}(\tilde{\theta}_{j}|
\theta_{j})$, where $\tilde{\theta}_{j}$ corresponds to the nominal value for 
$\theta_{j}$. The predefined uncertainties in the $\tilde{\theta}_{j}$, as 
discussed in Section~\ref{sec:uncertainties}, may be constrained by the fit 
to the data. We do not observe any strong constraints nor shifts of any of the 
nuisance parameters in the fits from which the most probable values of the POIs 
and their uncertainties are obtained.

Maximum likelihood estimates for the POIs are obtained from the combination of 
the CB- and \VHanapl (labeled by CB in the figures) on the one hand, and the NN- and 
\VHanapl (labeled by NN in the figures) on the other hand. In all cases the \ggH, 
\qqH, and \VH STXS stage-0 processes are treated as signals. A 1\% contribution 
of \PQb quark associated \PH production is additionally subsumed into the \ggH 
process, assuming relations between the production modes as expected from the SM. 
The \PH production in association with a \ttbar pair, which contributes at the 
1\% level, is treated as background with a production cross section fixed to the 
SM expectation. The same is true for the contamination from \HWW events, mostly 
contributing to the \emu final state in the CB- and \NNanapl, and in the \VHana, 
with the only exception of the likelihood scans discussed in the end of this 
section, where this contribution is also treated as signal. 

\begin{figure*}[htbp]
  \centering
  \includegraphics[width=0.75\textwidth]{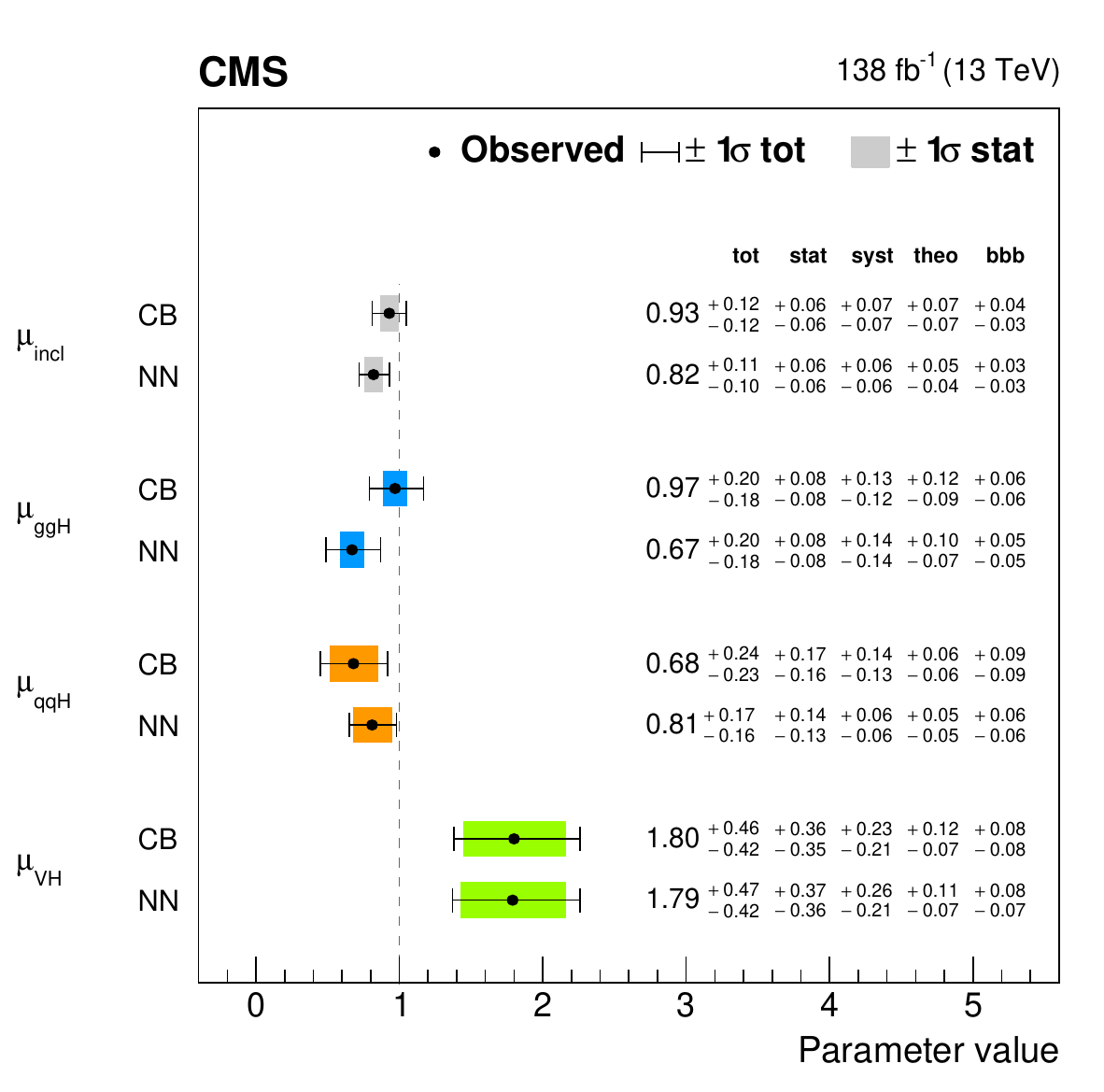}
  \caption{
    Measurements of the signal strength parameters for inclusive \PH production 
    (\muinc) and the \ggH (\muggh), \qqH (\muqqh), and \VH (\muvh) STXS stage-0 
    processes. The combination of the CB- and \VHanapl is labeled by CB. The 
    combination of the NN- and \VHanapl is labeled by NN. Central values maximizing 
    the likelihood and a split of uncertainties as explained in the text are 
    provided with each result.
  }
  \label{fig:stxs-stage-0}
\end{figure*}

The measurements of the inclusive signal strength parameter \muinc and the STXS 
stage-0 processes \muggh, \muqqh, and \muvh are shown in Fig.~\ref{fig:stxs-stage-0}. 
Here and in the following, in addition to the central value that maximizes the 
likelihood the combined systematic and statistical uncertainties (tot), as well 
as a split of each uncertainty into its purely statistical component (stat), 
experimental (syst) and theoretical (theo) systematic uncertainties, and 
uncertainties due to the finite sample sizes used for template production (bbb) 
are given; the expression bbb stands for ``bin-by-bin'', indicating that these 
uncertainties are incorporated for each template bin individually, as discussed 
in the introduction of Section~\ref{sec:uncertainties}. This split of 
uncertainties is obtained by successively fixing the nuisance parameters in each 
group to their maximum likelihood estimates and subtracting the resulting 
uncertainty in quadrature from the previous total. Correlations of uncertainties 
across these groups are assumed to be small, so that this procedure is justified.

For the combination of the NN- and \VHanapl an inclusive signal strength of 
$\muinc=0.82\pm 0.11$ is obtained, compatible within two s.d.\ with the SM 
expectation. The $p$-value for such an outcome of the measurement under the 
assumption of the SM hypothesis is found to be 0.1. For the \ggH process, for 
which the overall uncertainty is already dominated by systematic uncertainties, 
a signal strength of $\muggh=0.67\pm 0.19$ is obtained. For the $\qqH$ and $\VH$ 
processes, which are still dominated by statistical uncertainties, the signal 
strengths are found to be $\muqqh=0.81\pm 0.17$ and $\muvh=1.79\pm 0.45$. The 
result for $\muvh$ significantly supersedes a previous measurement that has been 
obtained from a smaller data set~\cite{Sirunyan:vh}. 

The correlation between $\muggh$ and $\muqqh$ is found to be $-0.35$, compatible 
with the observed migration effects of \ggH events with $\Njet\geq2$ into the 
\qqH category. As a result of the vetoes of additional leptons in the CB- or 
\NNanapl, very little or no correlation is expected between \muvh and any of the 
other POIs, which is confirmed by the observation. 

For the combination of the CB- and \VHanapl, similar results are obtained with a 
$p$-value for the outcome of the inclusive measurement under the assumption of 
the SM hypothesis of 0.6. The CB- and the \NNanapl obtain compatible results also 
in terms of the constraints on \muinc and \muggh, which are a measure for the 
sensitivity of the analyses to these quantities. This can be understood in the 
following way: 

\begin{enumerate}
\item For the \ggH process systematic uncertainties start to dominate over 
statistical uncertainties. The NNs are trained based on the cross entropy loss 
function, as defined in Eq.~(\ref{eq:nn-loss-function}), which is a maximum 
likelihood estimate for the separation of signals and backgrounds in the 
presence of only statistical uncertainties. Taking systematic uncertainties into 
account during the training of the NNs can help to improve the overall 
constraining power in future versions of the \NNana; 
\item We observe that the separation of \ggH from \qqH and \ZTT is challenging. 
This is visible from Fig.~\ref{fig:nn-confusion-matrices} (upper), and discussed 
in Section~\ref{sec:nn-characterization}. The challenging phase space regions 
are \ggH with $\Njet\geq2$, where the \ggH process becomes similar to the \qqH 
process, contributing roughly 60\% to the expected inclusive \ggH template 
cross section; and \ggH with low \ptH, where the signal is difficult to separate 
from the \ZTT background. The template cross section for \ggH is expected to be 
roughly 10 times larger than for the \qqH process. For this reason we anticipate 
the findings for the inclusive result to be similar to those for the stage-0 
\ggH result. 
\end{enumerate}

The constraint on \muqqh obtained from the \NNana on the other hand is 30\% 
stronger than for the \CBana. This can be understood from the fact that 

\begin{enumerate}
\item the \qqH process is easier to distinguish from \ggH and \ZTT events; 
\item the uncertainty in \muqqh is still dominated by the statistical variance 
of the measurement. 
\end{enumerate}

Another observation that can be made is that the systematic uncertainties 
especially in \muqqh are smaller for the \NNana compared to the \CBana. This 
trend is even more visible for some STXS stage-1.2 bins, as will be discussed 
below. Here the effect can be understood by the following means: 

\begin{enumerate}
\item The \CBana relies on one or two input distributions and a small number of 
inputs for primary categorization, to discriminate signal from background. The 
\NNana uses a 14-dimensional (14D) feature space to distinguish signals from 
backgrounds. Usually more than one feature, as well as correlations across features, 
considerably contribute to the separation task, as discussed in 
Section~\ref{sec:nn-characterization}. Because of the fact that more features, 
which are subject to complementary systematic variations, contribute to the NN 
response with roughly equal weight, systematic variations in a single feature 
may have less impact on the NN response. We find this best illustrated by the 
classification of the \qqH processes, where both the leading dijet system and the 
\PGtPgt system contribute to the separation of signals from backgrounds, and for 
the classification of \ttbar events, to which not only \PQb-tagging information, 
but a large variety of kinematic observables equally contribute to the NN 
response;
\item The likelihood of the \NNana includes the discriminating distributions not 
only for signal, but also for the background classes, as given in 
Table~\ref{tab:nn-event-classes}. All discriminating distributions are ordered 
by purity, in each corresponding event class, for increasing discriminant values. 
This results in high-purity control regions for each corresponding background 
class, and a transition from these control regions to the signal regions. This 
helps to constrain the nuisance parameters related to the main background 
processes of the analysis, in the maximum likelihood estimate, within the given 
uncertainties; 
\item Finally, the smaller uncertainties due to the finite sample sizes used for 
template production are explained by the use of larger sets of simulated signal 
samples, also used for training of the NNs. In addition the \NNana uses roughly 
a factor of five fewer bins for the input distributions to the likelihood, compared 
to the \CBana.
\end{enumerate}

\begin{figure*}[htbp]
  \centering
  \includegraphics[width=0.45\textwidth]{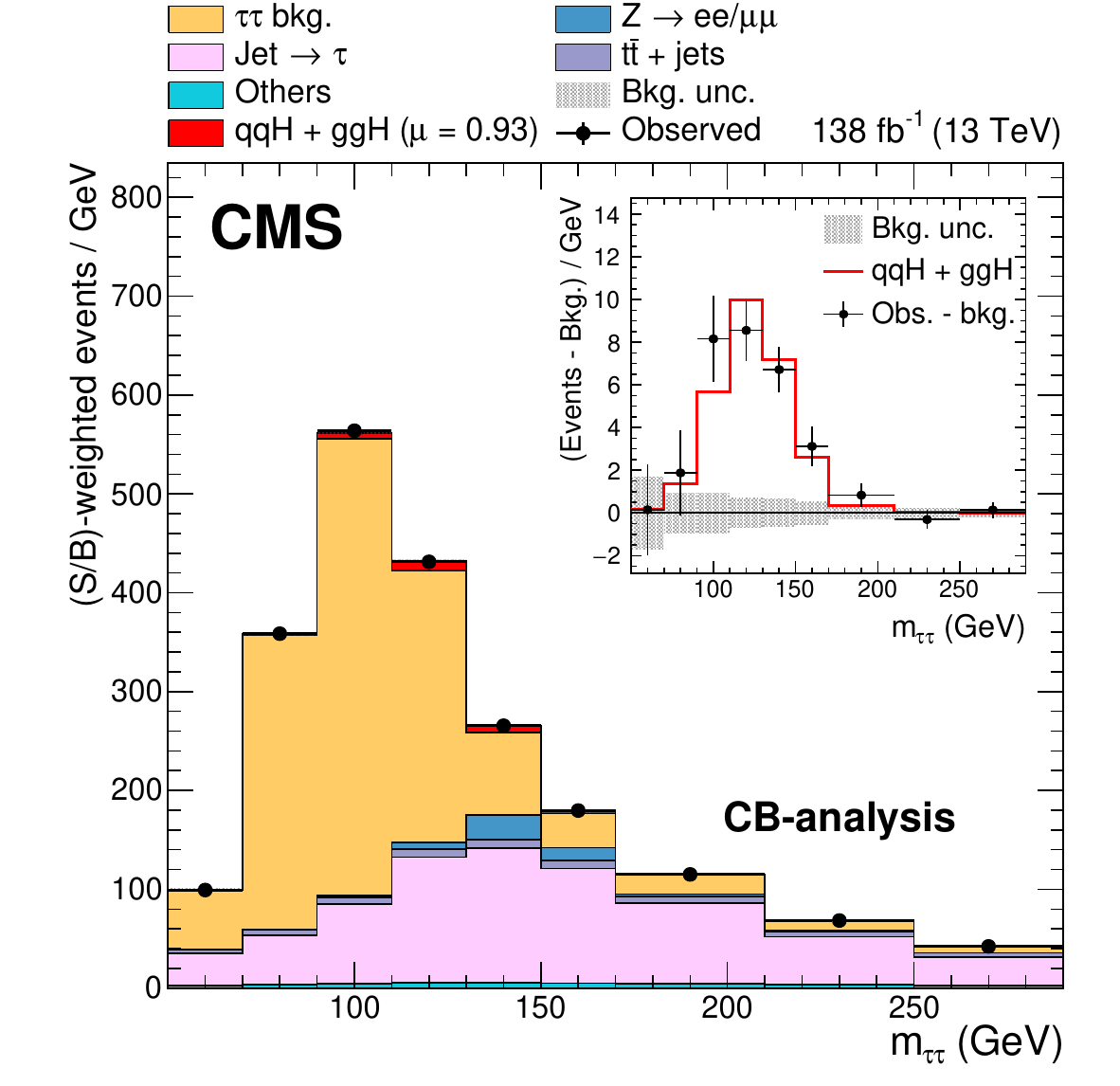}
  \includegraphics[width=0.45\textwidth]{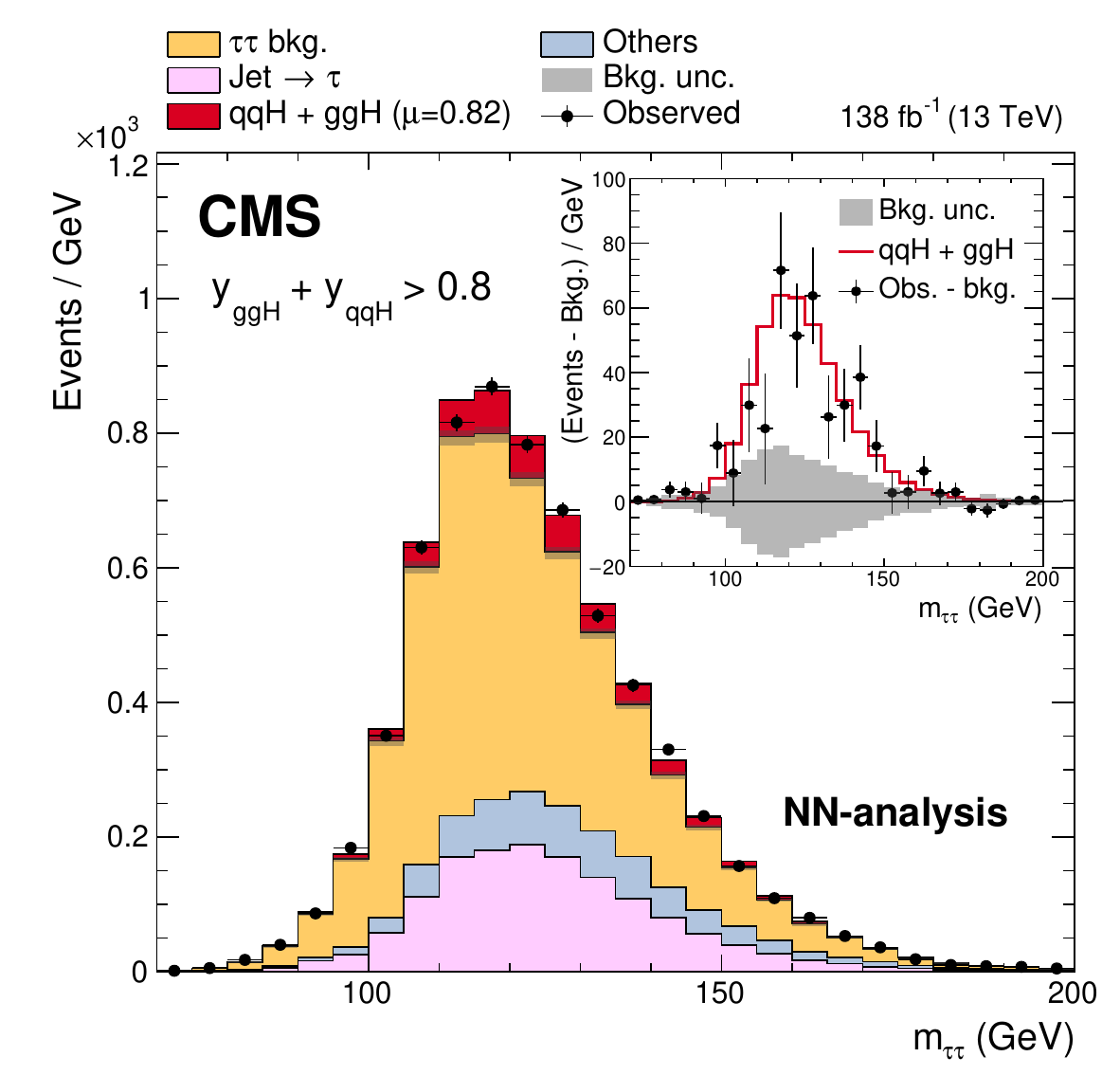}
  \includegraphics[width=0.45\textwidth]{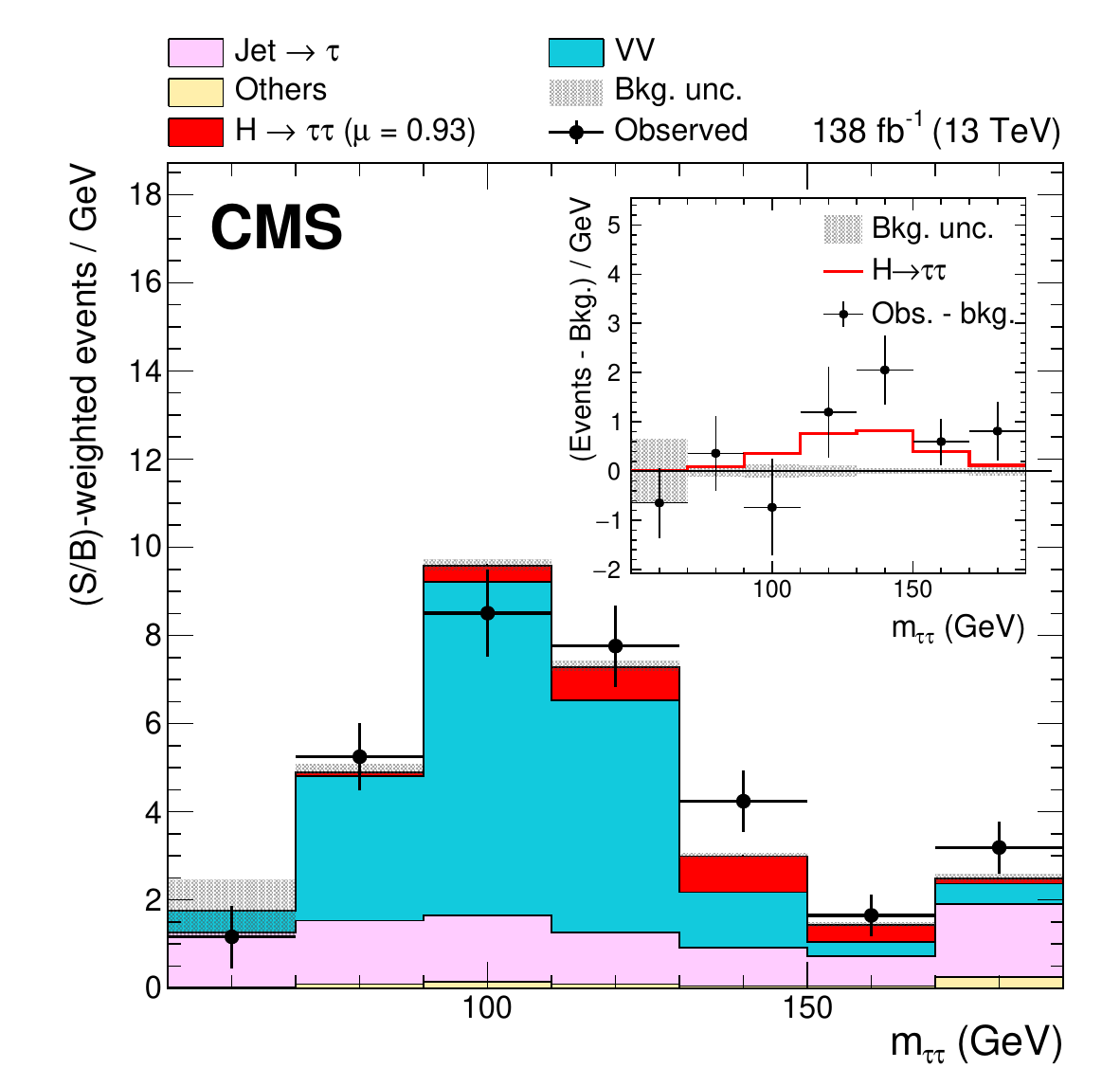}
  \includegraphics[width=0.45\textwidth]{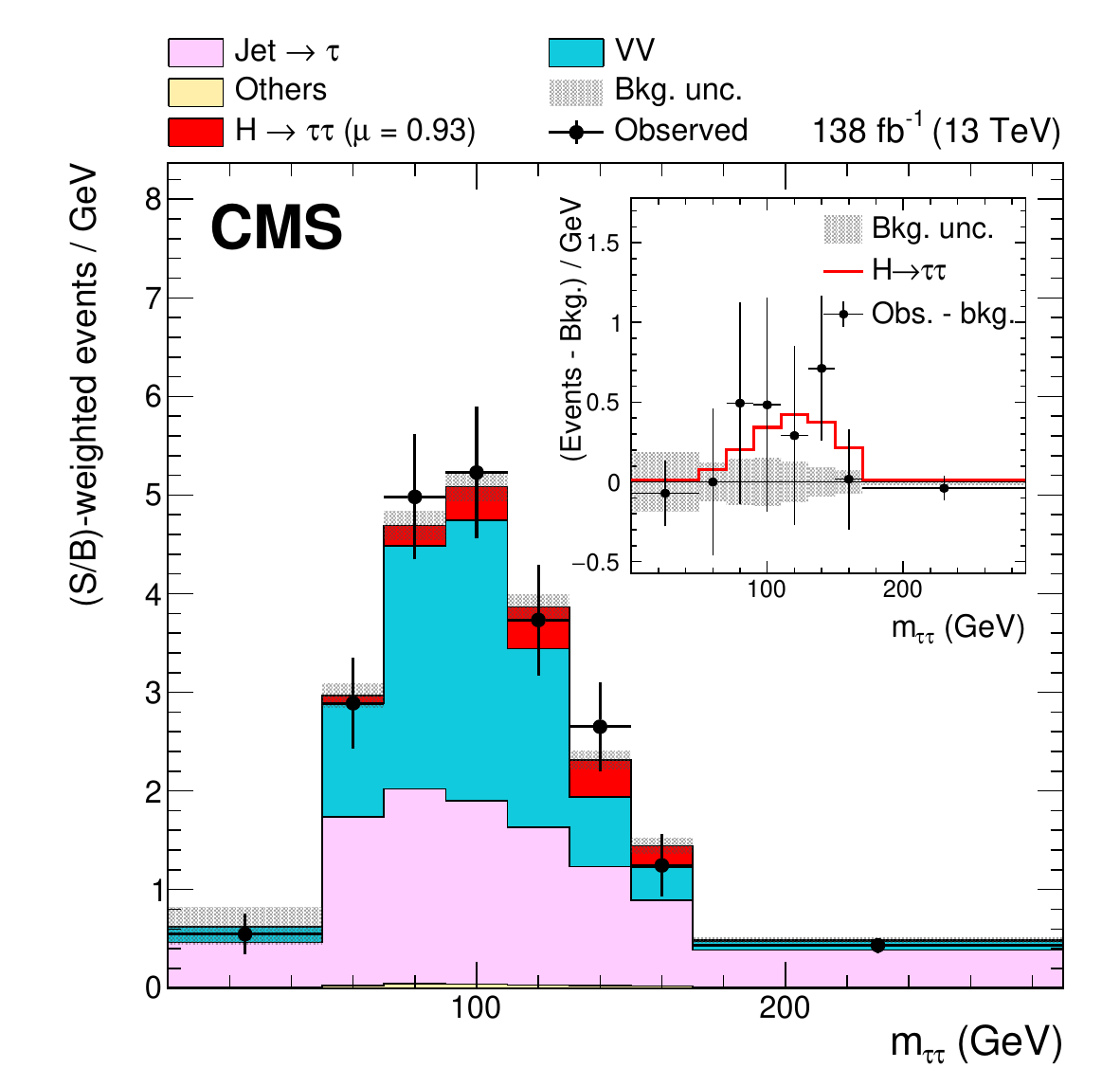}
  \caption{
    Observed and expected \mtt distributions of the analyzed data in all \PGtPgt 
    final states. For the (upper left) CB-, (lower left) \WH-, and (lower right) 
    \ZH-analyses, \mtt from all event categories and data-taking years in a 
    predefined window around the assumed \PH mass~\cite{CMS:2020xrn} is 
    histogrammed, weighted by the signal (S) over combined background (B) ratio 
    in each event category. For the (upper right) \NNana, the events fulfilling 
    $\yggH+\yqqH>0.8$ are shown without further weight applied. The inclusive 
    contribution of the \PH signal with \muinc as obtained from the maximum 
    likelihood estimates is also shown, in each subfigure. For the upper right 
    figure \muinc is obtained from the combination of the VH- and the \NNanapl, 
    for the other figures \muinc is obtained from the combination of the VH- and 
    the CB-analyses.
  }
  \label{fig:sbweighted} 
\end{figure*}

A visualization of the manifesting signal in the \mtt distribution is shown 
in Fig.~\ref{fig:sbweighted}. For the CB-, \WH-, and \ZH-analyses these 
distributions are obtained from histogramming \mtt from all event categories 
in all \PGtPgt final states and data-taking years in a predefined window around 
the assumed \PH mass~\cite{CMS:2020xrn}, weighted by the signal (S) over combined 
background (B) ratio in each event category. For the \NNana, the events fulfilling 
$\yggH+\yqqH>0.8$ are shown without further weight. In all cases, the points 
with the error bars correspond to the observation. The stacked filled histograms 
correspond to the expectation from the background model. The inclusive contribution 
of the \PH signal with \muinc as obtained from the maximum likelihood fit to the 
data is also shown in each of the figures. A clear signal over the background is 
visible, especially for the CB- and \NNanapl. 

\begin{figure*}[htbp]
  \centering
  \includegraphics[width=0.48\textwidth]{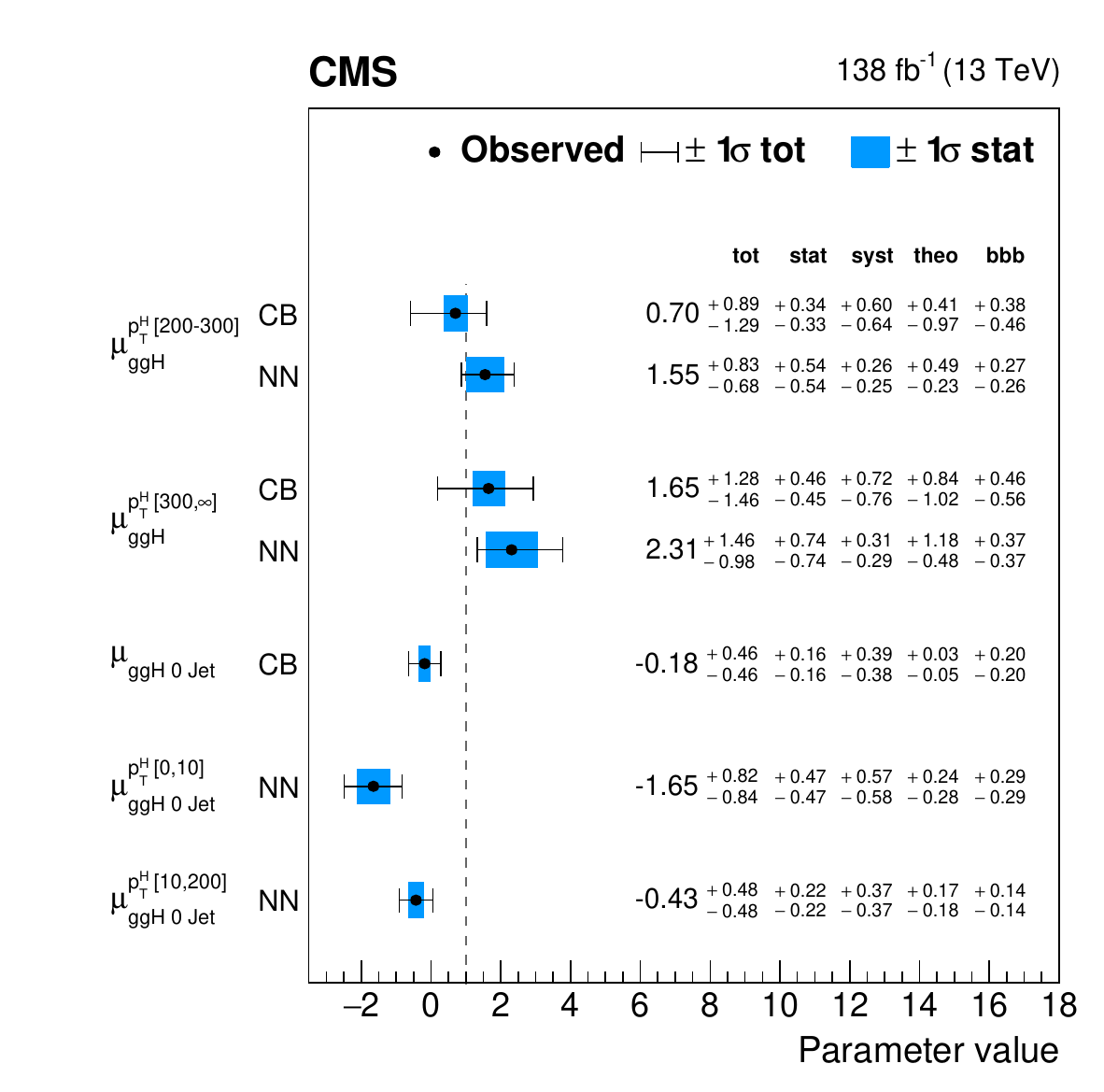}
  \includegraphics[width=0.48\textwidth]{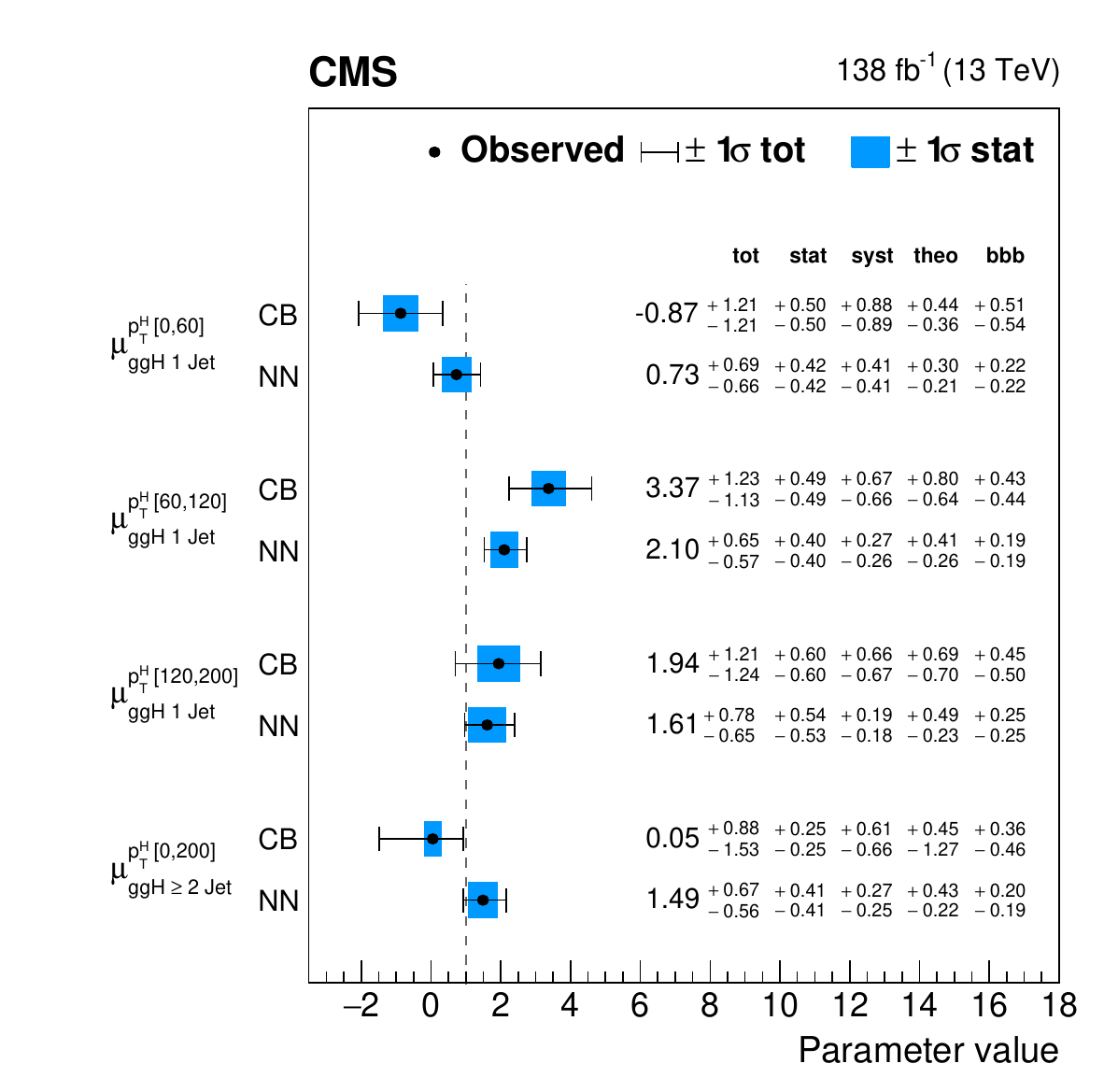}
  \includegraphics[width=0.48\textwidth]{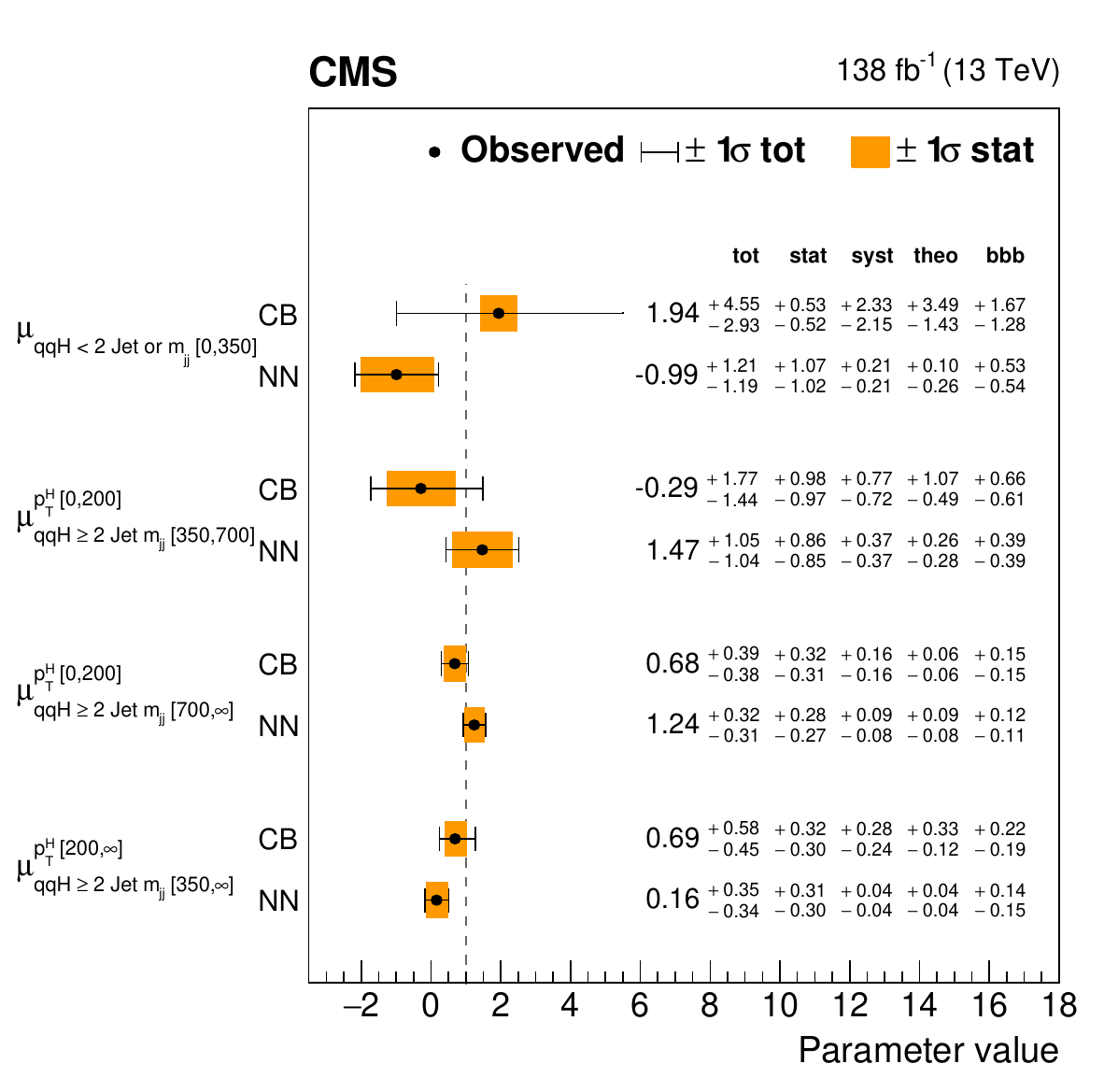}
  \includegraphics[width=0.48\textwidth]{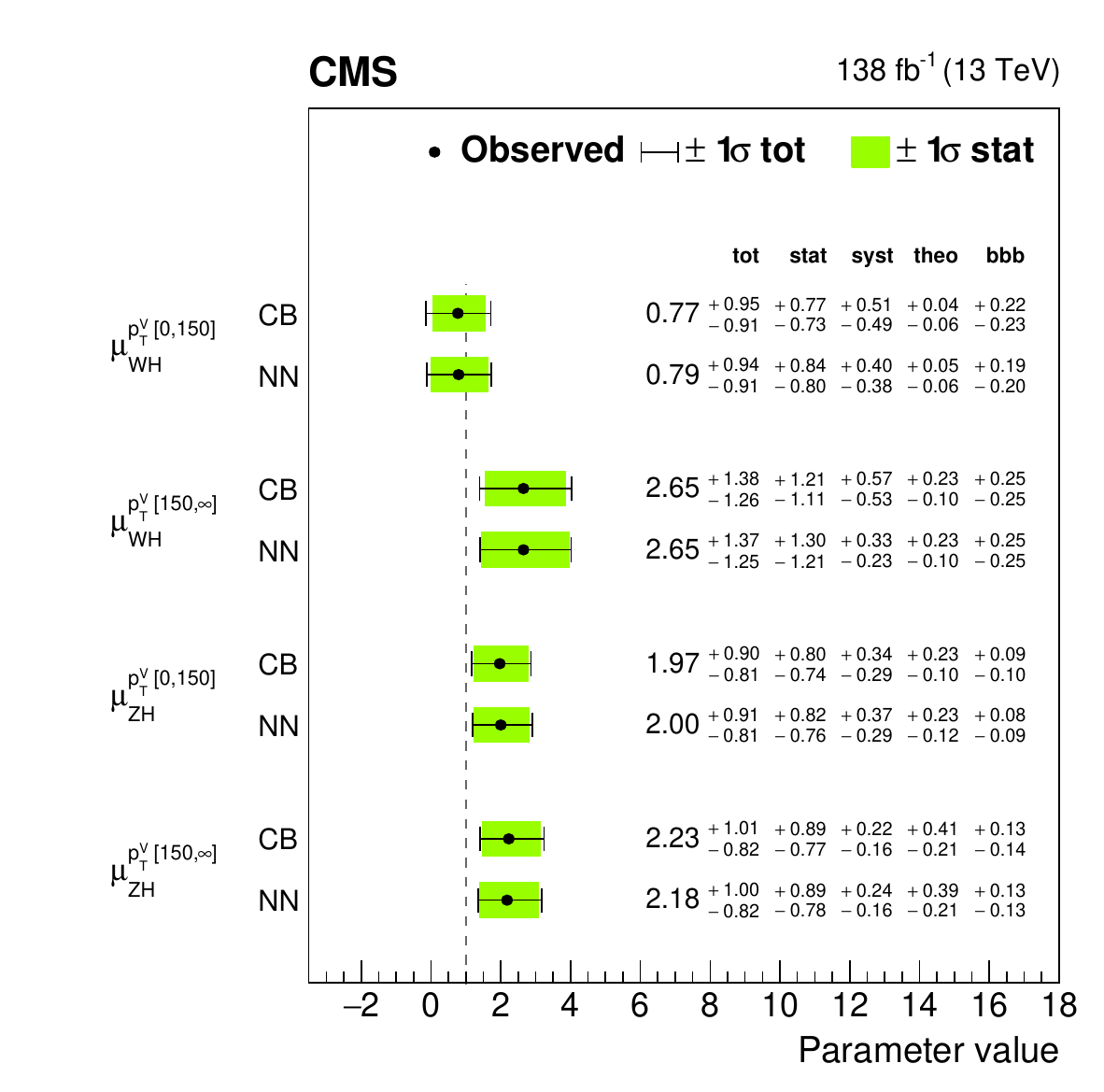}
  \caption{
    Measurements of the signal strength parameters $\mu_{s}$ in the STXS stage-1.2 
    bins for the (upper row) \ggH, (lower left) \qqH, and (lower right) \VH 
    processes. The combination of the CB- and \VHanapl is labeled by CB. The 
    combination of the NN- and \VHanapl is labeled by NN. Central values maximizing 
    the likelihood and a split of uncertainties are also provided with each result.
  }
  \label{fig:stxs-stage-1}
\end{figure*}

For the STXS stage-1.2 measurements, the \ggH process is split in 7 (8) POIs for 
the (CB-) \NNana; the \qqH and \VH processes are split in 4 POIs each, resulting 
in a total of up to 16 POIs, which have an exact correspondence to the gray boxes 
shown in Figs.~\ref{fig:stxs_ggh}--\ref{fig:stxs_vh}. We note that, for the \CBana, 
the POIs for the $\ggH\,\Njet=0$ bins have been combined into one, due to a lack 
of sensitivity. For the CB- and \VHanapl, the STXS stage-1.2 estimates are obtained 
from the same set of input distributions as the inclusive and STXS stage-0 
estimates. For the \NNana, two different sets of input distributions are used for 
the inclusive and STXS stage-0 estimates, and for the STXS stage-1.2 estimates, 
as stated above and discussed in more detail in Section~\ref{sec:nn-application}. 
In Fig.~\ref{fig:stxs-stage-1}, the results for the STXS stage-1.2 measurement 
are shown. Tabulated values of the inclusive, STXS stage-0, and -1.2 results are 
given in Table~\ref{tab:tabulated-results}. On average the constraints achieved 
by the \NNana are 30 (40)\% stronger in the \ggH (\qqH) bins when compared to the 
\CBana. In particular in the \qqH STXS bins an increased sensitivity by up to 
200\% is observed. This can be understood in the following way: 

\begin{enumerate}
\item for the more differential analysis, statistical uncertainties again dominate 
over systematic uncertainties, hence the NN training setup is optimally suited 
to serve the measurement target; 
\item by the kinematic restrictions the signal can be more clearly distinguished 
from backgrounds and competing signals; 
\item at stage-1.2, the signal in general obtains higher emphasis in the training, 
which is performed without prevalence. This is implied by the fact that we have 
more signal classes relative to the background classes in stage-1.2, compared to 
stage-0. 
\end{enumerate}

\begin{table*}[htbp]
  \centering
  \topcaption{
    Tabulated values of the STXS stage-0 and -1.2 signal strengths for the 
    combination of the (CB) CB-, resp.\ (NN) \NNana with the \VHana. The upper 
    four lines refer to the inclusive and STXS stage-0 measurements. The values 
    in braces correspond to the expected 68\% confidence intervals for an assumed 
    SM signal. The products of cross sections and branching fraction to \PGt 
    leptons as expected from the SM with the uncertainties as discussed in 
    Section~\ref{sec:signal-uncertainties} are also given.
  }
  \renewcommand{\arraystretch}{1.3}
  \label{tab:tabulated-results}

    \cmsTable{\begin{tabular}{lcr@{--}lr@{--}lr@{$\pm$}lr@{$\pm$}lr@{$\pm$}l}
      \hline
      & & \multicolumn{2}{c}{} & \multicolumn{2}{c}{} &
      \multicolumn{2}{c}{SM (fb)} & 
      \multicolumn{2}{c}{$\mu_{s}$ (CB)} & 
      \multicolumn{2}{c}{$\mu_{s}$ (NN)} \\
      \hline 
      Inclusive & & \multicolumn{2}{c}{} & \multicolumn{2}{c}{} 
      & $3422.28$ & $0.05$ 
      & $0.93$ & $^{0.12}_{0.12}\,(^{0.13}_{0.13})$ 
      & $0.82$ & $^{0.11}_{0.10}\,(^{0.12}_{0.11})$ \\ [\cmsTabSkip]
      \ggH & & \multicolumn{2}{c}{} & \multicolumn{2}{c}{} 
      & $3051.34$ & $0.05$ 
      & $0.97$ & $^{0.20}_{0.18}\,(^{0.24}_{0.22})$ 
      & $0.67$ & $^{0.20}_{0.18}\,(^{0.27}_{0.23})$ \\ [\cmsTabSkip]
      \qqH & & \multicolumn{2}{c}{} & \multicolumn{2}{c}{} 
      & $328.68$&$0.03$ 
      & $0.68$ & $^{0.24}_{0.23}\,(^{0.24}_{0.23})$ 
      & $0.81$ & $^{0.17}_{0.16}\,(^{0.17}_{0.17})$ \\ [\cmsTabSkip]
      \VH & & \multicolumn{2}{c}{} & \multicolumn{2}{c}{} 
      & $44.19$&$0.03$ 
      & $1.80$ & $^{0.46}_{0.42}\,(^{0.41}_{0.37})$ 
      & $1.79$ & $^{0.47}_{0.42}\,(^{0.41}_{0.37})$ \\ [\cmsTabSkip]
      \hline
      & \Njet & \multicolumn{2}{c}{\ptH (\GeVns)} & \multicolumn{2}{c}{} &
      \multicolumn{2}{c}{} & 
      \multicolumn{2}{c}{} & 
      \multicolumn{2}{c}{}  \\
      \hline
      & \multirow{2}{*}{$=0$} & 0 & 10 & \multicolumn{2}{c}{} 
      & $423.58$ & $0.13$ 
      & \multicolumn{2}{c}{\multirow{2}{*}{$-0.18\pm^{0.46}_{0.46}\,(^{0.45}_{0.44})$}} 
      & $-1.65$&$^{0.82}_{0.84}\,(^{0.84}_{0.83})$  \\ [\cmsTabSkip]
      & & 10 & 200 & \multicolumn{2}{c}{} 
      & $1329.36$ & $0.07$ & \multicolumn{2}{c}{} 
      & $-0.43$ & $^{0.48}_{0.48}\,(^{0.49}_{0.48})$ \\ [\cmsTabSkip]  
      & & 0 & 60 & \multicolumn{2}{c}{} 
      & $451.09$ &$0.14$ 
      & $-0.87$ & $^{1.21}_{1.21}\,(^{1.06}_{0.99})$ 
      & $0.73$ & $^{0.69}_{0.66}\,(^{0.68}_{0.64})$ \\ [\cmsTabSkip] 
      \multirow{2}{*}{\ggH} & $=1$ & 60 & 120 & \multicolumn{2}{c}{} 
      & $287.68$ & $0.14$ 
      & $3.37$ & $^{1.23}_{1.13}\,(^{0.90}_{0.83})$ 
      & $2.10$ & $^{0.65}_{0.57}\,(^{0.54}_{0.50})$ \\ [\cmsTabSkip] 
      & & 120 & 200 & \multicolumn{2}{c}{} 
      & $50.04$ & $0.19$ 
      & $1.94$ & $^{1.21}_{1.24}\,(^{1.04}_{0.90})$ 
      & $1.61$ & $^{0.78}_{0.65}\,(^{0.68}_{0.60})$ \\ [\cmsTabSkip] 
      & $\geq2$ & 0 & 200 & \multicolumn{2}{c}{} 
      & $306.26$ & $0.23$ 
      & $0.05$ & $^{0.88}_{1.53}\,(^{0.83}_{0.71})$ 
      & $1.49$ & $^{0.67}_{0.56}\,(^{0.66}_{0.56})$ \\ [\cmsTabSkip] 
      & & 200 & 300 & \multicolumn{2}{c}{} 
      & $27.51$ & $0.42$ 
      & $0.70$ & $^{0.89}_{1.29}\,(^{0.91}_{0.77})$ 
      & $1.55$ & $^{0.83}_{0.68}\,(^{0.76}_{0.65})$ \\ [\cmsTabSkip] 
      & & 300 & $\infty$ & \multicolumn{2}{c}{} 
      & $7.19$ & $0.47$ 
      & $1.65$ & $^{1.28}_{1.46}\,(^{1.20}_{0.96})$ 
      & $2.31$ & $^{1.46}_{0.98}\,(^{1.10}_{0.86})$ \\ [\cmsTabSkip] 
      \hline
      & \Njet & \multicolumn{2}{c}{\ptH (\GeVns)} & \multicolumn{2}{c}{\mjj (\GeVns)} 
      &\multicolumn{2}{c}{}&\multicolumn{2}{c}{}&\multicolumn{2}{c}{}  \\
      \hline
      & & 0 & 200 & 350 &700 
      & $34.43$ & $0.04$ 
      & $-0.29$ & $^{1.77}_{1.44}\,(^{1.31}_{1.32})$ 
      & $1.47$ & $^{1.05}_{1.04}\,(^{1.03}_{1.03})$  \\ [\cmsTabSkip] 
      \multirow{2}{*}{\qqH} & $\geq2$ & 0 & 200 & 700 & $\infty$ 
      & $47.48$ & $0.04$ 
      & $0.68$ & $^{0.39}_{0.38}\,(^{0.39}_{0.38})$ 
      & $1.24$ & $^{0.32}_{0.31}\,(^{0.31}_{0.30})$ \\ [\cmsTabSkip] 
      & & 200 & $\infty$ & 350 & $\infty$ 
      & $9.90$ & $0.03$ 
      & $0.69$ & $^{0.58}_{0.45}\,(^{0.45}_{0.43})$ 
      & $0.16$ & $^{0.35}_{0.34}\,(^{0.37}_{0.35})$ \\ [\cmsTabSkip] 
      & \multicolumn{5}{c}{$\Njet<2$ or $\mjj\,[0,350]\GeV$} 
      & $209.46$ & $0.03$ 
      & $1.94$ & $^{4.55}_{2.93}\,(^{2.15}_{2.16})$ 
      & $-0.99$ & $^{1.21}_{1.19}\,(^{1.23}_{1.18})$ \\ [\cmsTabSkip] 
      \hline
      & & \multicolumn{2}{c}{\ptV (\GeVns)} & \multicolumn{2}{c}{} 
      &\multicolumn{2}{c}{}&\multicolumn{2}{c}{}&\multicolumn{2}{c}{}  \\
      \hline
      \multirow{2}{*}{\WH} & \multirow{3}{*}{} & 0 & 150 & \multicolumn{2}{c}{}  
      & $20.57$ & $0.03$ 
      & $0.77$ & $^{0.95}_{0.91}\,(^{0.90}_{0.85})$ 
      & $0.79$ & $^{0.94}_{0.91}\,(^{0.90}_{0.85})$  \\ [\cmsTabSkip] 
      & & 150 & $\infty$ & \multicolumn{2}{c}{}  
      & $3.30$ & $0.05$ 
      & $2.65$ & $^{1.38}_{1.26}\,(^{1.26}_{1.15})$ 
      & $2.65$ & $^{1.37}_{1.25}\,(^{1.26}_{1.15})$ \\ [\cmsTabSkip] 
      \multirow{2}{*}{\ZH} & \multirow{3}{*}{} & 0 & 150 & \multicolumn{2}{c}{}  
      & $11.99$ & $0.06$ 
      & $1.97$ & $^{0.90}_{0.81}\,(^{0.79}_{0.71})$ 
      & $2.00$ & $^{0.91}_{0.81}\,(^{0.79}_{0.71})$ \\ [\cmsTabSkip] 
      & & 150 & $\infty$ & \multicolumn{2}{c}{}   
      & $2.55$ & $0.10$ 
      & $2.23$ & $^{1.01}_{0.82}\,(^{0.78}_{0.61})$ 
      & $2.18$ & $^{1.00}_{0.82}\,(^{0.78}_{0.61})$ \\
      \hline
    \end{tabular}}
\end{table*}

An important point for the assessment of the STXS stage-1.2 results of the CB- and 
\NNanapl, and their constraining power relates to the separation of individual 
STXS bins. The \CBana primarily targets the separation of signal from background, 
while the distinction of STXS bins is left to the selection based event 
categorization. In contrast to this strategy the multiclassification ansatz of the 
\NNana targets the best possible separation of individual STXS bins and backgrounds 
in the 14D input space to the NNs. In this way, the \NNana generally achieves 
smaller migration effects, increased separation of individual STXS bins, and 
reduced vulnerability to variations of individual features within systematic 
uncertainties, compared to the \CBana.     

Both, the CB- and \NNanapl observe the same trends across STXS stage-1.2 bins. In 
particular, both analyses observe no signal in the 0-jet STXS bins, while their 
sensitivity to observe a signal larger than zero for a signal as expected from 
the SM is at the 2--3 s.d.\ level. In all other STXS bins, a signal compatible with 
the expectation of the SM is observed.

\begin{figure*}[htbp]
\centering
  \includegraphics[width=0.60\textwidth]{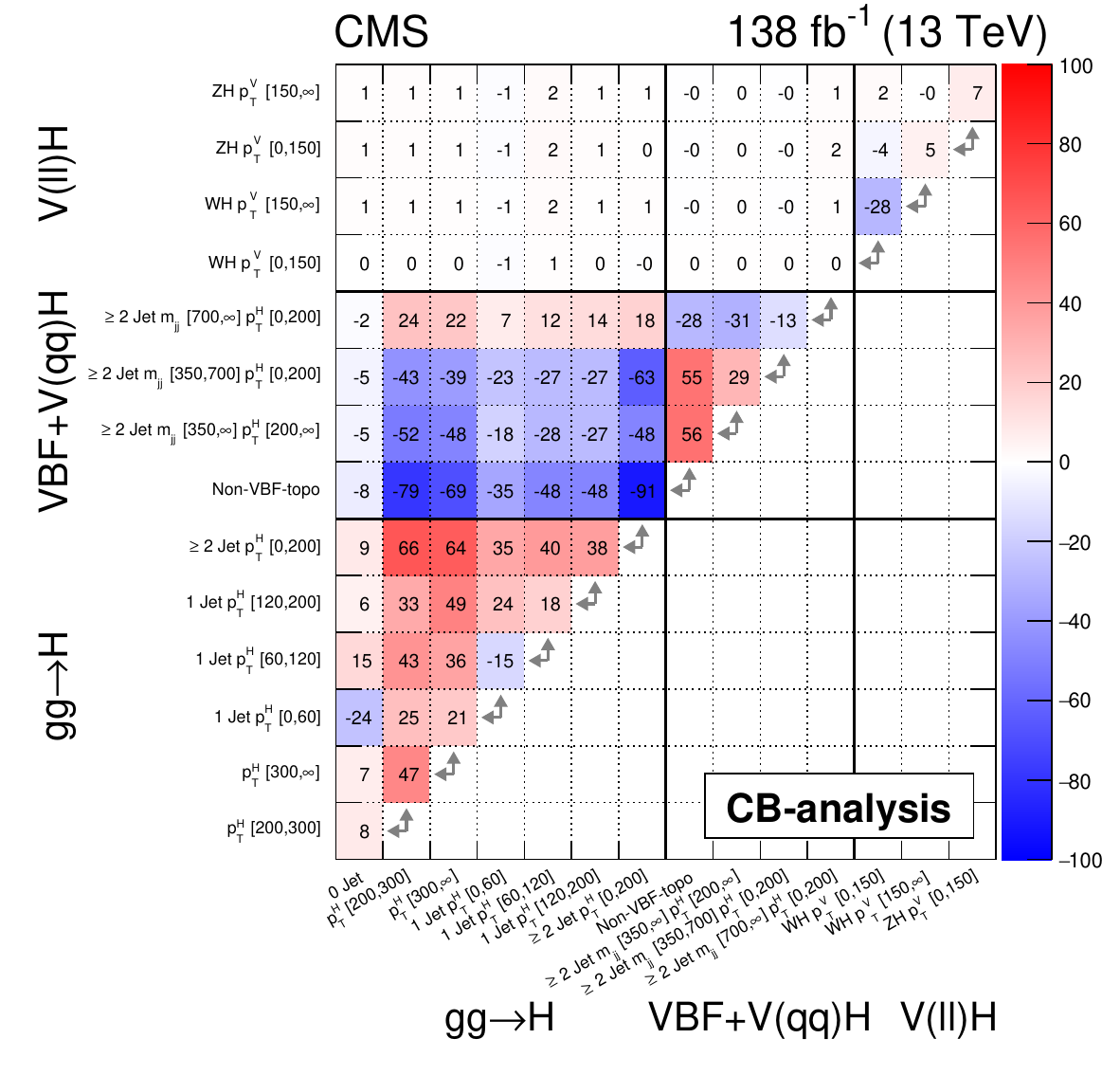}
  \includegraphics[width=0.60\textwidth]{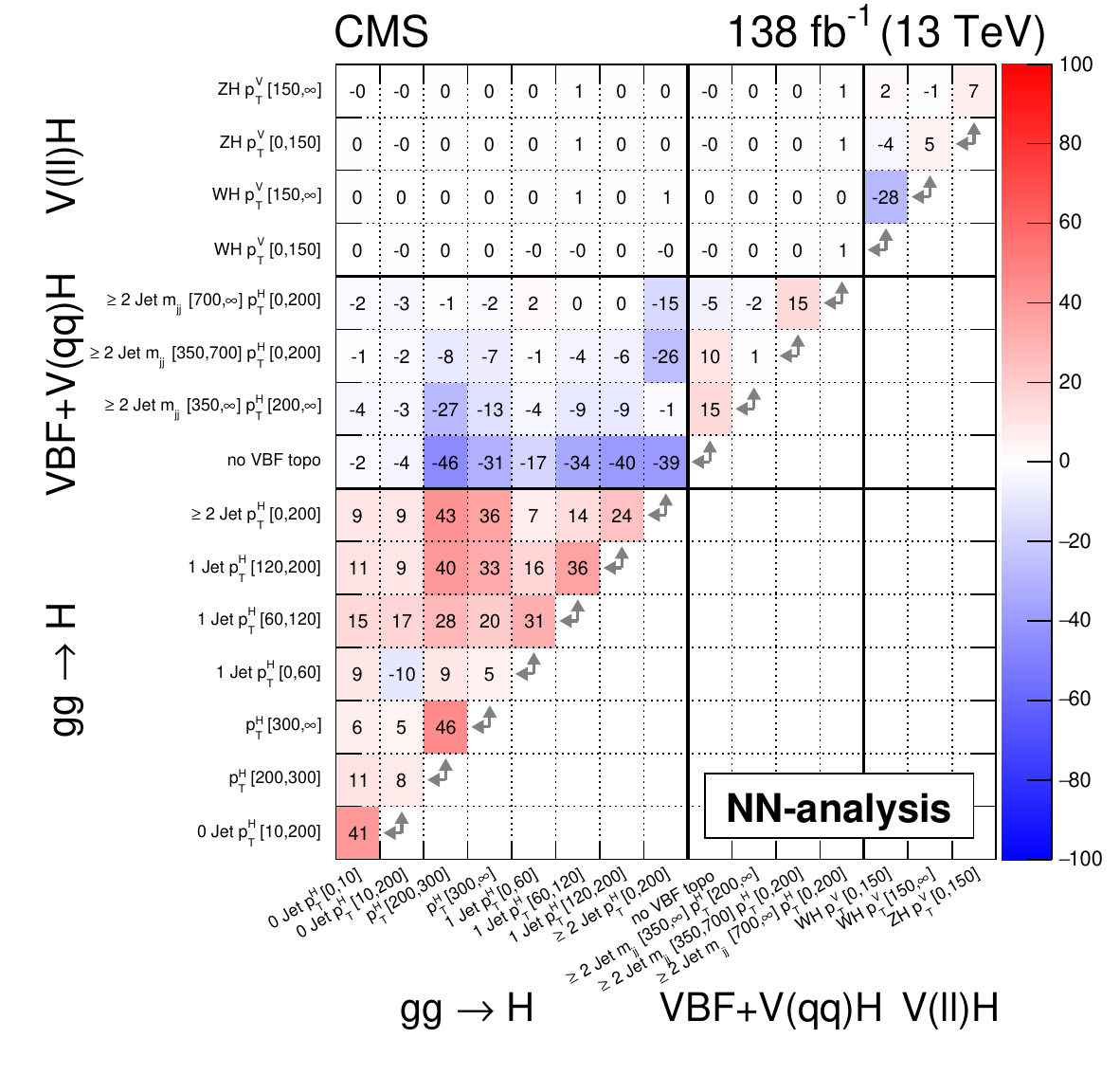}
  \caption{
    Correlation matrices of the POIs of the measured STXS stage-1.2 signal strengths 
    for the combination of the (upper) CB-, resp.\ (lower) \NNana with the \VHana. 
  }
  \label{fig:stxs-correlation-matrices}
\end{figure*}

The linear correlation matrices of the STXS stage-1.2 measurements are shown in 
Fig.~\ref{fig:stxs-correlation-matrices}. For the \NNana the observed correlations 
are generally below 0.4 with trends of a more global anticorrelation of \ggH 
$\Njet=1$ and $\Njet\geq2$ STXS bins with the \qqH STXS bins, reflecting the 
difficulty of experimentally distinguishing processes in these topological 
regions, especially for $\ptH>200\GeV$. As expected, no correlation of the \ggH 
and \qqH with the \VH STXS bins is observed. For the \CBana, global and local 
trends are again similar, which confirms that both analyses capture the same 
features in data. Single correlation coefficients are generally significantly 
higher for the \CBana, demonstrating the \NNana's ability to separate the 
individual STXS bins.

\begin{figure*}[htbp]
  \centering
    \includegraphics[width=0.49\textwidth]{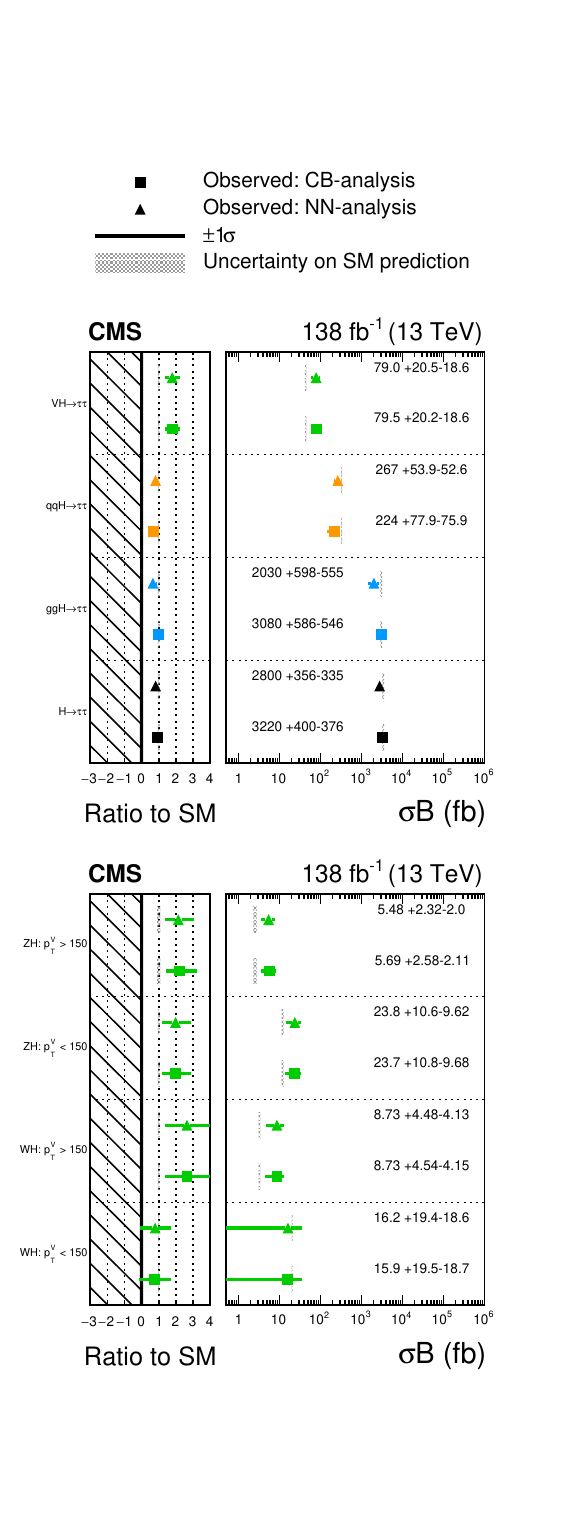}
    \includegraphics[width=0.49\textwidth]{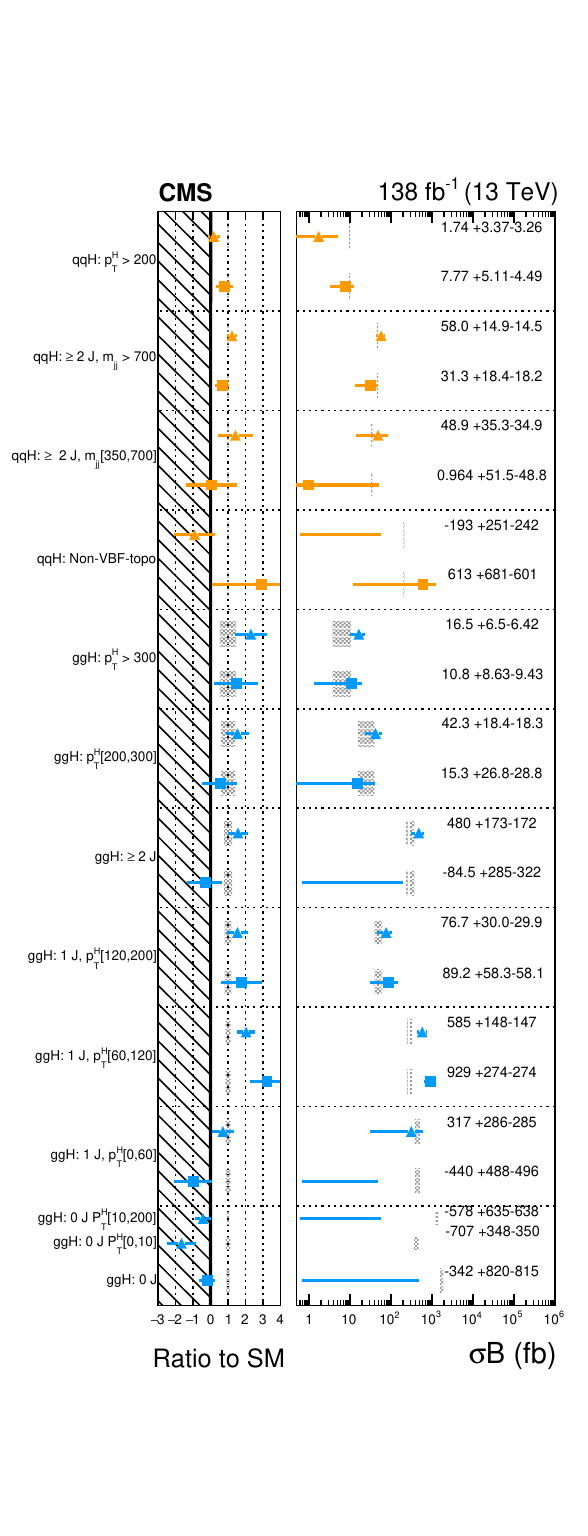}
    \caption{
      Cross section measurements in the (upper left) stage-0 bins, and in the 
      stage-1.2 bins related to the (lower left) \VH, (upper right) \qqH, and 
      (lower right) \ggH processes. The combination of the CB- and \VHanapl is 
      labeled by CB, the combination of the NN- and \VHanapl is labeled by NN. 
      Central values and combined statistical and systematic uncertainties are 
      given for each measurement. 
    }
    \label{fig:stxs-xsec}
\end{figure*}

The same results, as shown in Figs.~\ref{fig:stxs-stage-0} 
and~\ref{fig:stxs-stage-1}, may be obtained as measurements of the product of 
cross sections and branching fraction to \PGt leptons maximizing the same 
likelihood function, where those theoretical uncertainties related to pure 
normalization changes of the signal templates for individual processes or within 
individual STXS stage-1.2 bins are dropped from the uncertainty model. A summary 
of these measurements is shown in Fig.~\ref{fig:stxs-xsec}. The upper left panel 
of the figure refers to the inclusive and STXS stage-0 results, the lower left 
panel to the STXS stage-1.2 measurements in the \VH, and the right panel to the 
STXS stage-1.2 measurements in the \ggH and \qqH bins. The measured cross sections 
span several orders of magnitude and apart from that follow the same trends, as 
discussed in the previous paragraphs. 

\begin{figure}[htbp]
  \centering
  \includegraphics[width=\cmsFigWidthA]{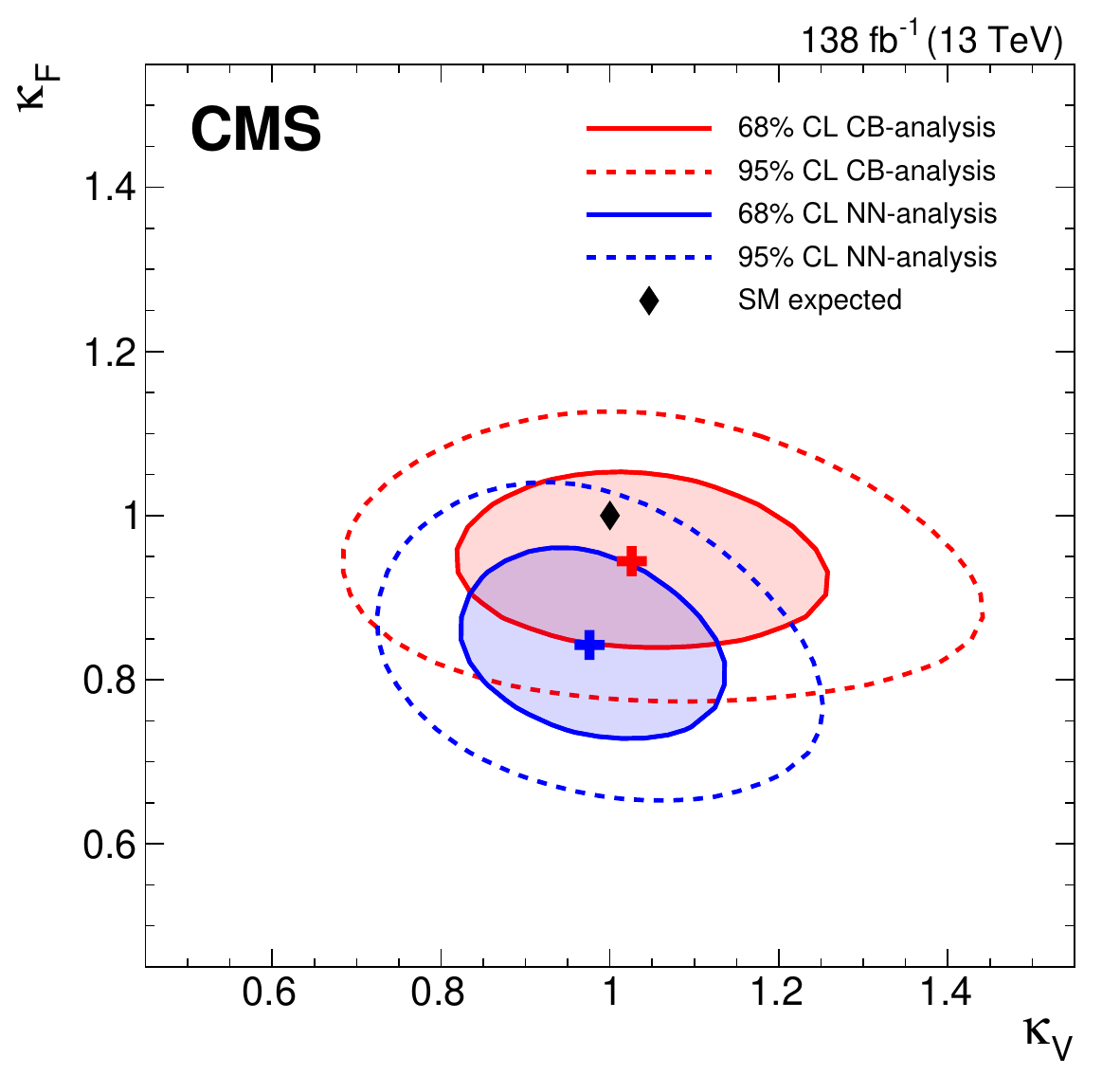}
  \caption{
    Most probable values, 68, and 95\% confidence interval contours obtained from 
    a negative log-likelihood scan, for a model treating the \PH couplings to 
    vector bosons (\kappaV) and fermions (\kappaF) as POIs. The combination 
    of the CB- (NN-) with \VHanapl is shown in light red (dark blue). For the 
    likelihood evaluation all nuisance parameters are profiled in each point 
    in the plane. The \HWW decay is treated as signal. 
  }
  \label{fig:kVkF}
\end{figure}

Figure~\ref{fig:kVkF} shows the most probable value, as well as the 68 and 95\% 
confidence interval contours in the plane of multiplicative modifiers of the 
\PH coupling strengths to vector bosons and fermions, \kappaV--\kappaF. This 
figure has been extensively referred to in the past~\cite{Heinemeyer:2016spz}. 
For this figure, \HWW decays have been treated as signal, giving increased 
sensitivity to the \emu final state in the CB- and \NNanapl, and the \VHana. 
Also here the measurement, as obtained from the \NNana, is compatible with the 
expectation from the SM, within the 95\% confidence interval. The measured value 
of \kappaV turns out to be close to 1, while the value of \kappaF is roughly 15\% 
lower than the SM expectation. These findings coincide with the previously 
discussed findings for \muggh and \muqqh, shown in Fig.~\ref{fig:stxs-stage-0}. 

Typically less than 10\% of the events in the most signal-sensitive bins of the 
CB-analysis are shared with the NN-analysis. The fraction of shared events with 
respect to the events used by the NN-analysis typically amounts to not more than 
20\%. Given this small overlap we conclude that both analyses are consistent with 
each other in their trends and general results. The measurement of the \CBana is 
less precise but closer to the SM expectation. These findings again coincide with 
the $p$-values under the assumption of the SM hypothesis, previously reported 
for the corresponding inclusive measurements. A complete set of the reported 
measurements is available in the HepData database~\cite{hepdata}.

\section{Summary}
\label{sec:summary}

Measurements of Higgs boson production, where the Higgs boson decays into a pair 
of \PGt leptons, have been presented. The analyzed data are selected from 
proton-proton collisions at a center-of-mass energy of 13\TeV collected by the 
CMS experiment at the LHC from 2016--2018, corresponding to an integrated 
luminosity of 138\fbinv. Results are presented in the form of signal strengths 
relative to the standard model predictions and products of cross sections and 
branching fraction to \PGt leptons, in up to 16 different kinematic regions, 
following the simplified template cross section scheme of the LHC Higgs Working 
Group. For the simultaneous measurements of a neural network based analysis and 
an analysis targeting vector boson associated Higgs boson production signal 
strengths are found to be $0.82\pm0.11$ for inclusive Higgs boson production, 
$0.67\pm0.19$ ($0.81\pm0.17$) for the production mainly via gluon fusion (vector 
boson fusion), and $1.79\pm0.45$ for vector boson associated Higgs boson production. 
The latter result significantly improves an earlier measurement performed by the 
CMS Collaboration on a smaller data set.  

\begin{acknowledgments}
We congratulate our colleagues in the CERN accelerator departments for the excellent performance of the LHC and thank the technical and administrative staffs at CERN and at other CMS institutes for their contributions to the success of the CMS effort. In addition, we gratefully acknowledge the computing centers and personnel of the Worldwide LHC Computing Grid and other centers for delivering so effectively the computing infrastructure essential to our analyses. Finally, we acknowledge the enduring support for the construction and operation of the LHC, the CMS detector, and the supporting computing infrastructure provided by the following funding agencies: BMBWF and FWF (Austria); FNRS and FWO (Belgium); CNPq, CAPES, FAPERJ, FAPERGS, and FAPESP (Brazil); MES and BNSF (Bulgaria); CERN; CAS, MoST, and NSFC (China); MINCIENCIAS (Colombia); MSES and CSF (Croatia); RIF (Cyprus); SENESCYT (Ecuador); MoER, ERC PUT and ERDF (Estonia); Academy of Finland, MEC, and HIP (Finland); CEA and CNRS/IN2P3 (France); BMBF, DFG, and HGF (Germany); GSRI (Greece); NKFIH (Hungary); DAE and DST (India); IPM (Iran); SFI (Ireland); INFN (Italy); MSIP and NRF (Republic of Korea); MES (Latvia); LAS (Lithuania); MOE and UM (Malaysia); BUAP, CINVESTAV, CONACYT, LNS, SEP, and UASLP-FAI (Mexico); MOS (Montenegro); MBIE (New Zealand); PAEC (Pakistan); MES and NSC (Poland); FCT (Portugal); MESTD (Serbia); MCIN/AEI and PCTI (Spain); MOSTR (Sri Lanka); Swiss Funding Agencies (Switzerland); MST (Taipei); MHESI and NSTDA (Thailand); TUBITAK and TENMAK (Turkey); NASU (Ukraine); STFC (United Kingdom); DOE and NSF (USA).

\hyphenation{Rachada-pisek} Individuals have received support from the Marie-Curie program and the European Research Council and Horizon 2020 Grant, contract Nos.\ 675440, 724704, 752730, 758316, 765710, 824093, 884104, and COST Action CA16108 (European Union); the Leventis Foundation; the Alfred P.\ Sloan Foundation; the Alexander von Humboldt Foundation; the Belgian Federal Science Policy Office; the Fonds pour la Formation \`a la Recherche dans l'Industrie et dans l'Agriculture (FRIA-Belgium); the Agentschap voor Innovatie door Wetenschap en Technologie (IWT-Belgium); the F.R.S.-FNRS and FWO (Belgium) under the ``Excellence of Science -- EOS" -- be.h project n.\ 30820817; the Beijing Municipal Science \& Technology Commission, No. Z191100007219010; the Ministry of Education, Youth and Sports (MEYS) of the Czech Republic; the Hellenic Foundation for Research and Innovation (HFRI), Project Number 2288 (Greece); the Deutsche Forschungsgemeinschaft (DFG), under Germany's Excellence Strategy -- EXC 2121 ``Quantum Universe" -- 390833306, and under project number 400140256 - GRK2497; the Hungarian Academy of Sciences, the New National Excellence Program - \'UNKP, the NKFIH research grants K 124845, K 124850, K 128713, K 128786, K 129058, K 131991, K 133046, K 138136, K 143460, K 143477, 2020-2.2.1-ED-2021-00181, and TKP2021-NKTA-64 (Hungary); the Council of Science and Industrial Research, India; the Latvian Council of Science; the Ministry of Education and Science, project no. 2022/WK/14, and the National Science Center, contracts Opus 2021/41/B/ST2/01369 and 2021/43/B/ST2/01552 (Poland); the Funda\c{c}\~ao para a Ci\^encia e a Tecnologia, grant CEECIND/01334/2018 (Portugal); MCIN/AEI/10.13039/501100011033, ERDF ``a way of making Europe", and the Programa Estatal de Fomento de la Investigaci{\'o}n Cient{\'i}fica y T{\'e}cnica de Excelencia Mar\'{\i}a de Maeztu, grant MDM-2017-0765 and Programa Severo Ochoa del Principado de Asturias (Spain); the Chulalongkorn Academic into Its 2nd Century Project Advancement Project, and the National Science, Research and Innovation Fund via the Program Management Unit for Human Resources \& Institutional Development, Research and Innovation, grant B05F650021 (Thailand); the Kavli Foundation; the Nvidia Corporation; the SuperMicro Corporation; the Welch Foundation, contract C-1845; and the Weston Havens Foundation (USA).
\end{acknowledgments}

\ifthenelse{\boolean{cms@external}}{\clearpage}{}

\bibliography{auto_generated}
\cleardoublepage \appendix\section{The CMS Collaboration \label{app:collab}}\begin{sloppypar}\hyphenpenalty=5000\widowpenalty=500\clubpenalty=5000\input{HIG-19-010-public-authorlist.tex}\end{sloppypar}
\end{document}

%% file: HIG-19-010-public-authorlist.tex
\cmsinstitute{Yerevan Physics Institute, Yerevan, Armenia}
{\tolerance=6000
A.~Tumasyan\cmsAuthorMark{1}\cmsorcid{0009-0000-0684-6742}
\par}
\cmsinstitute{Institut f\"{u}r Hochenergiephysik, Vienna, Austria}
{\tolerance=6000
W.~Adam\cmsorcid{0000-0001-9099-4341}, J.W.~Andrejkovic, T.~Bergauer\cmsorcid{0000-0002-5786-0293}, S.~Chatterjee\cmsorcid{0000-0003-2660-0349}, K.~Damanakis\cmsorcid{0000-0001-5389-2872}, M.~Dragicevic\cmsorcid{0000-0003-1967-6783}, A.~Escalante~Del~Valle\cmsorcid{0000-0002-9702-6359}, R.~Fr\"{u}hwirth\cmsAuthorMark{2}\cmsorcid{0000-0002-0054-3369}, M.~Jeitler\cmsAuthorMark{2}\cmsorcid{0000-0002-5141-9560}, N.~Krammer\cmsorcid{0000-0002-0548-0985}, L.~Lechner\cmsorcid{0000-0002-3065-1141}, D.~Liko\cmsorcid{0000-0002-3380-473X}, I.~Mikulec\cmsorcid{0000-0003-0385-2746}, P.~Paulitsch, F.M.~Pitters, J.~Schieck\cmsAuthorMark{2}\cmsorcid{0000-0002-1058-8093}, R.~Sch\"{o}fbeck\cmsorcid{0000-0002-2332-8784}, D.~Schwarz\cmsorcid{0000-0002-3821-7331}, S.~Templ\cmsorcid{0000-0003-3137-5692}, W.~Waltenberger\cmsorcid{0000-0002-6215-7228}, C.-E.~Wulz\cmsAuthorMark{2}\cmsorcid{0000-0001-9226-5812}
% ...
\par}
\vskip\cmsinstskip
\dag:~Deceased\\
$^{1}$Also at Yerevan State University, Yerevan, Armenia\\
$^{2}$Also at TU Wien, Vienna, Austria\\
% ...


%% file: HIG-19-010_temp.bbl
\providecommand{\href}[2]{#2}\begingroup\raggedright\begin{thebibliography}{100}%
\makeatletter
\providecommand{\hrefCMSnoop }[0]{\@secondoftwo}%
\makeatother
\providecommand{\doi}{\texttt{doi:}\begingroup \urlstyle{tt}\Url}

\bibitem{Glashow:1961tr}
\hrefCMSnoop {}{S.~L. Glashow, ``Partial-symmetries of weak interactions'',}
  \textit{ Nucl. Phys.} \textbf{ 22} (1961) 579,
\href{http://dx.doi.org/10.1016/0029-5582(61)90469-2}{\doi{10.1016/0029-5582(61)90469-2}}.

\bibitem{Weinberg:1967tq}
\hrefCMSnoop {}{S.~Weinberg, ``A model of leptons'',} \textit{ Phys. Rev.
  Lett.} \textbf{ 19} (1967) 1264,
  \href{http://dx.doi.org/10.1103/PhysRevLett.19.1264}{\doi{10.1103/PhysRevLett.19.1264}}.

\bibitem{Salam:1968rm}
\hrefCMSnoop {}{A.~Salam, ``Weak and electromagnetic interactions'',} \textit{
  Conf. Proc. C} \textbf{ 680519} (1968) 367,
  \href{http://dx.doi.org/10.1142/9789812795915_0034}{\doi{10.1142/9789812795915_0034}}.

\bibitem{Englert:1964et}
\hrefCMSnoop {}{F.~Englert and R.~Brout, ``Broken symmetry and the mass of
  gauge vector mesons'',} \textit{ Phys. Rev. Lett.} \textbf{ 13} (1964) 321,
  \href{http://dx.doi.org/10.1103/PhysRevLett.13.321}{\doi{10.1103/PhysRevLett.13.321}}.

\bibitem{Higgs:1964ia}
\hrefCMSnoop {}{P.~W. Higgs, ``Broken symmetries, massless particles and gauge
  fields'',} \textit{ Phys. Lett.} \textbf{ 12} (1964) 132,
  \href{http://dx.doi.org/10.1016/0031-9163(64)91136-9}{\doi{10.1016/0031-9163(64)91136-9}}.

\bibitem{Higgs:1964pj}
\hrefCMSnoop {}{P.~W. Higgs, ``Broken symmetries and the masses of gauge
  bosons'',} \textit{ Phys. Rev. Lett.} \textbf{ 13} (1964) 508,
  \href{http://dx.doi.org/10.1103/PhysRevLett.13.508}{\doi{10.1103/PhysRevLett.13.508}}.

\bibitem{Guralnik:1964eu}
\hrefCMSnoop {}{G.~S. Guralnik, C.~R. Hagen, and T.~W.~B. Kibble, ``Global
  conservation laws and massless particles'',} \textit{ Phys. Rev. Lett.}
  \textbf{ 13} (1964) 585,
  \href{http://dx.doi.org/10.1103/PhysRevLett.13.585}{\doi{10.1103/PhysRevLett.13.585}}.

\bibitem{Higgs:1966ev}
\hrefCMSnoop {}{P.~W. Higgs, ``Spontaneous symmetry breakdown without massless
  bosons'',} \textit{ Phys. Rev.} \textbf{ 145} (1966) 1156,
  \href{http://dx.doi.org/10.1103/PhysRev.145.1156}{\doi{10.1103/PhysRev.145.1156}}.

\bibitem{Kibble:1967sv}
\hrefCMSnoop {}{T.~W.~B. Kibble, ``Symmetry breaking in non-abelian gauge
  theories'',} \textit{ Phys. Rev.} \textbf{ 155} (1967) 1554,
  \href{http://dx.doi.org/10.1103/PhysRev.155.1554}{\doi{10.1103/PhysRev.155.1554}}.

\bibitem{Aad:2012tfa}
\hrefCMSnoop {}{{ATLAS Collaboration}, ``Observation of a new particle in the
  search for the standard model {Higgs} boson with the {ATLAS} detector at the
  {LHC}'',} \textit{ Phys. Lett. B} \textbf{ 716} (2012) 1,
  \href{http://dx.doi.org/10.1016/j.physletb.2012.08.020}{\doi{10.1016/j.physletb.2012.08.020}},
  \href{http://www.arXiv.org/abs/1207.7214}{\texttt{arXiv:1207.7214}}.

\bibitem{Chatrchyan:2012ufa}
\hrefCMSnoop {}{{CMS Collaboration}, ``Observation of a new boson at a mass of
  125 {GeV} with the {CMS} experiment at the {LHC}'',} \textit{ Phys. Lett. B}
  \textbf{ 716} (2012) 30,
  \href{http://dx.doi.org/10.1016/j.physletb.2012.08.021}{\doi{10.1016/j.physletb.2012.08.021}},
  \href{http://www.arXiv.org/abs/1207.7235}{\texttt{arXiv:1207.7235}}.

\bibitem{Chatrchyan:2013lba}
\hrefCMSnoop {}{{CMS Collaboration}, ``Observation of a new boson with mass
  near 125 {GeV in $\Pp\Pp$} collisions at {$\sqrt{s}$ = 7 and 8 TeV}'',}
  \textit{ JHEP} \textbf{ 06} (2013) 081,
  \href{http://dx.doi.org/10.1007/JHEP06(2013)081}{\doi{10.1007/JHEP06(2013)081}},
  \href{http://www.arXiv.org/abs/1303.4571}{\texttt{arXiv:1303.4571}}.

\bibitem{CMS:2020xrn}
\hrefCMSnoop {}{{CMS Collaboration}, ``A measurement of the {Higgs} boson mass
  in the diphoton decay channel'',} \textit{ Phys. Lett. B} \textbf{ 805}
  (2020) 135425,
  \href{http://dx.doi.org/10.1016/j.physletb.2020.135425}{\doi{10.1016/j.physletb.2020.135425}},
  \href{http://www.arXiv.org/abs/2002.06398}{\texttt{arXiv:2002.06398}}.

\bibitem{CMS:2013fjq}
\hrefCMSnoop {}{{CMS Collaboration}, ``Measurement of the properties of a
  {Higgs} boson in the four-lepton final state'',} \textit{ Phys. Rev. D}
  \textbf{ 89} (2014) 092007,
  \href{http://dx.doi.org/10.1103/PhysRevD.89.092007}{\doi{10.1103/PhysRevD.89.092007}},
  \href{http://www.arXiv.org/abs/1312.5353}{\texttt{arXiv:1312.5353}}.

\bibitem{Khachatryan:2016vau}
\hrefCMSnoop {}{{ATLAS and CMS Collaborations}, ``Measurements of the {Higgs}
  boson production and decay rates and constraints on its couplings from a
  combined {ATLAS} and {CMS} analysis of the {LHC} pp collision data at
  {$\sqrt{s}=7$ and 8 TeV}'',} \textit{ JHEP} \textbf{ 08} (2016) 045,
  \href{http://dx.doi.org/10.1007/JHEP08(2016)045}{\doi{10.1007/JHEP08(2016)045}},
  \href{http://www.arXiv.org/abs/1606.02266}{\texttt{arXiv:1606.02266}}.

\bibitem{Sirunyan:2018koj}
\hrefCMSnoop {}{{CMS Collaboration}, ``Combined measurements of {Higgs} boson
  couplings in {proton\textendash{}proton} collisions at
  {$\sqrt{s}=13\,\text{Te}\text{V}$}'',} \textit{ Eur. Phys. J. C} \textbf{ 79}
  (2019) 421,
  \href{http://dx.doi.org/10.1140/epjc/s10052-019-6909-y}{\doi{10.1140/epjc/s10052-019-6909-y}},
  \href{http://www.arXiv.org/abs/1809.10733}{\texttt{arXiv:1809.10733}}.

\bibitem{Aad:2019mbh}
\hrefCMSnoop {}{{ATLAS Collaboration}, ``Combined measurements of {Higgs} boson
  production and decay using up to {$80$ fb$^{-1}$} of proton-proton collision
  data at {$\sqrt{s}=$ 13 TeV} collected with the {ATLAS experiment}'',}
  \textit{ Phys. Rev. D} \textbf{ 101} (2020) 012002,
  \href{http://dx.doi.org/10.1103/PhysRevD.101.012002}{\doi{10.1103/PhysRevD.101.012002}},
  \href{http://www.arXiv.org/abs/1909.02845}{\texttt{arXiv:1909.02845}}.

\bibitem{ATLAS:2020rej}
\hrefCMSnoop {}{{ATLAS Collaboration}, ``{Higgs} boson production cross-section
  measurements and their {EFT} interpretation in the {$4\ell $} decay channel
  at {$\sqrt{s}=$13 TeV} with the {ATLAS detector}'',} \textit{ Eur. Phys. J.
  C} \textbf{ 80} (2020) 957,
  \href{http://dx.doi.org/10.1140/epjc/s10052-020-8227-9}{\doi{10.1140/epjc/s10052-020-8227-9}},
  \href{http://www.arXiv.org/abs/2004.03447}{\texttt{arXiv:2004.03447}}.
  [Erratum: \DOI{10.1140/epjc/s10052-020-08644-x}, Erratum:
  \DOI{10.1140/epjc/s10052-021-09116-6}].

\bibitem{CMS:2019ekd}
\hrefCMSnoop {}{{CMS Collaboration}, ``Measurements of the {Higgs} boson width
  and anomalous {$\PH\PV\PV$} couplings from on-shell and off-shell production
  in the four-lepton final state'',} \textit{ Phys. Rev. D} \textbf{ 99} (2019)
  112003,
  \href{http://dx.doi.org/10.1103/PhysRevD.99.112003}{\doi{10.1103/PhysRevD.99.112003}},
  \href{http://www.arXiv.org/abs/1901.00174}{\texttt{arXiv:1901.00174}}.

\bibitem{CMS:2020cga}
\hrefCMSnoop {}{{CMS Collaboration}, ``Measurements of {$\mathrm{t\bar{t}H}$}
  production and the {CP} structure of the {Yukawa} interaction between the
  {Higgs} boson and top quark in the diphoton decay channel'',} \textit{ Phys.
  Rev. Lett.} \textbf{ 125} (2020) 061801,
  \href{http://dx.doi.org/10.1103/PhysRevLett.125.061801}{\doi{10.1103/PhysRevLett.125.061801}},
  \href{http://www.arXiv.org/abs/2003.10866}{\texttt{arXiv:2003.10866}}.

\bibitem{ATLAS:2020ior}
\hrefCMSnoop {}{{ATLAS Collaboration}, ``{CP} properties of {Higgs} boson
  interactions with top quarks in the {$\mathrm{t\bar{t}H}$} and
  {$\mathrm{tH}$} processes using {$\mathrm{H} \rightarrow \gamma\gamma$} with
  the {ATLAS} detector'',} \textit{ Phys. Rev. Lett.} \textbf{ 125} (2020)
  061802,
  \href{http://dx.doi.org/10.1103/PhysRevLett.125.061802}{\doi{10.1103/PhysRevLett.125.061802}},
  \href{http://www.arXiv.org/abs/2004.04545}{\texttt{arXiv:2004.04545}}.

\bibitem{CMS:2021nnc}
\hrefCMSnoop {}{{CMS Collaboration}, ``Constraints on anomalous {Higgs} boson
  couplings to vector bosons and fermions in its production and decay using the
  four-lepton final state'',} \textit{ Phys. Rev. D} \textbf{ 104} (2021)
  052004,
  \href{http://dx.doi.org/10.1103/PhysRevD.104.052004}{\doi{10.1103/PhysRevD.104.052004}},
  \href{http://www.arXiv.org/abs/2104.12152}{\texttt{arXiv:2104.12152}}.

\bibitem{ATLAS:2021pkb}
\hrefCMSnoop {}{{ATLAS Collaboration}, ``Constraints on {Higgs} boson
  properties using {$WW^{*}(\rightarrow e\nu\mu\nu) jj$} production in {36.1
  fb$^{-1}$} of {$\sqrt{s}$=13 TeV} pp collisions with the {ATLAS detector}'',}
  2021. \href{http://www.arXiv.org/abs/2109.13808}{\texttt{arXiv:2109.13808}}.
  Submitted to \textit{Eur. Phys. J. C}.

\bibitem{ATLAS:2020evk}
\hrefCMSnoop {}{{ATLAS Collaboration}, ``Test of {CP} invariance in
  vector-boson fusion production of the {Higgs} boson in the {$H\to\Pgt\Pgt$}
  channel in proton-proton collisions at {$\sqrt{s}=13\TeV$} with the {ATLAS
  detector}'',} \textit{ Phys. Lett. B} \textbf{ 805} (2020) 135426,
  \href{http://dx.doi.org/10.1016/j.physletb.2020.135426}{\doi{10.1016/j.physletb.2020.135426}},
  \href{http://www.arXiv.org/abs/2002.05315}{\texttt{arXiv:2002.05315}}.

\bibitem{CMS:2021sdq}
\hrefCMSnoop {}{{CMS Collaboration}, ``Analysis of the {$CP$} structure of the
  {Yukawa} coupling between the {Higgs} boson and {$\tau$} leptons in
  proton-proton collisions at {$\sqrt{s}$ = 13 TeV}'',} \textit{ JHEP} \textbf{
  06} (2022) 012,
  \href{http://dx.doi.org/10.1007/JHEP06(2022)012}{\doi{10.1007/JHEP06(2022)012}},
  \href{http://www.arXiv.org/abs/2110.04836}{\texttt{arXiv:2110.04836}}.

\bibitem{Golfand:1971iw}
\hrefCMSnoop {}{{\relax Yu}.~A. Golfand and E.~P. Likhtman, ``Extension of the
  algebra of {Poincar\'{e}} group generators and violation of p invariance'',}
  \textit{ JETP Lett.} \textbf{ 13} (1971)
323.

\bibitem{Wess:1974tw}
\hrefCMSnoop {}{J.~Wess and B.~Zumino, ``Supergauge transformations in
  four-dimensions'',} \textit{ Nucl. Phys. B} \textbf{ 70} (1974) 39,
\href{http://dx.doi.org/10.1016/0550-3213(74)90355-1}{\doi{10.1016/0550-3213(74)90355-1}}.

\bibitem{ATLAS:2017eiz}
\hrefCMSnoop {}{{ATLAS Collaboration}, ``Search for additional heavy neutral
  {H}iggs and gauge bosons in the ditau final state produced in {36 fb$^{~1}$}
  of pp collisions at {$\sqrt{s}=13$ TeV} with the {ATLAS} detector'',}
  \textit{ JHEP} \textbf{ 01} (2018) 055,
  \href{http://dx.doi.org/10.1007/JHEP01(2018)055}{\doi{10.1007/JHEP01(2018)055}},
  \href{http://www.arXiv.org/abs/1709.07242}{\texttt{arXiv:1709.07242}}.

\bibitem{Sirunyan:2018zut}
\hrefCMSnoop {}{{CMS Collaboration}, ``Search for additional neutral {MSSM
  Higgs} bosons in the {$\tau\tau$} final state in proton-proton collisions at
  {$\sqrt{s}=$ 13 TeV}'',} \textit{ JHEP} \textbf{ 09} (2018) 007,
  \href{http://dx.doi.org/10.1007/JHEP09(2018)007}{\doi{10.1007/JHEP09(2018)007}},
  \href{http://www.arXiv.org/abs/1803.06553}{\texttt{arXiv:1803.06553}}.

\bibitem{Chatrchyan:2014nva}
\hrefCMSnoop {}{{CMS Collaboration}, ``Evidence for the {125 GeV Higgs} boson
  decaying to a pair of {$\tau$} leptons'',} \textit{ JHEP} \textbf{ 05} (2014)
  104,
  \href{http://dx.doi.org/10.1007/JHEP05(2014)104}{\doi{10.1007/JHEP05(2014)104}},
  \href{http://www.arXiv.org/abs/1401.5041}{\texttt{arXiv:1401.5041}}.

\bibitem{Chatrchyan:2014vua}
\hrefCMSnoop {}{{CMS Collaboration}, ``Evidence for the direct decay of the
  {125 GeV Higgs} boson to fermions'',} \textit{ Nature Phys.} \textbf{ 10}
  (2014) 557,
  \href{http://dx.doi.org/10.1038/nphys3005}{\doi{10.1038/nphys3005}},
  \href{http://www.arXiv.org/abs/1401.6527}{\texttt{arXiv:1401.6527}}.

\bibitem{Aad:2015vsa}
\hrefCMSnoop {}{{ATLAS Collaboration}, ``Evidence for the {Higgs-boson Yukawa}
  coupling to tau leptons with the {ATLAS} detector'',} \textit{ JHEP} \textbf{
  04} (2015) 117,
  \href{http://dx.doi.org/10.1007/JHEP04(2015)117}{\doi{10.1007/JHEP04(2015)117}},
  \href{http://www.arXiv.org/abs/1501.04943}{\texttt{arXiv:1501.04943}}.

\bibitem{Sirunyan:2017khh}
\hrefCMSnoop {}{{CMS Collaboration}, ``Observation of the {Higgs} boson decay
  to a pair of $\tau$ leptons with the {CMS} detector'',} \textit{ Phys. Lett.
  B} \textbf{ 779} (2018) 283,
  \href{http://dx.doi.org/10.1016/j.physletb.2018.02.004}{\doi{10.1016/j.physletb.2018.02.004}},
  \href{http://www.arXiv.org/abs/1708.00373}{\texttt{arXiv:1708.00373}}.

\bibitem{deFlorian:2016spz}
\hrefCMSnoop {}{{LHC Higgs Cross Section Working Group}, ``Handbook of {LHC}
  {H}iggs cross sections: 4. {D}eciphering the nature of the {H}iggs sector'',}
  CERN Report CERN-2017-002-M, 2016.
\newblock
  \href{http://dx.doi.org/10.23731/CYRM-2017-002}{\doi{10.23731/CYRM-2017-002}},
  \href{http://www.arXiv.org/abs/1610.07922}{\texttt{arXiv:1610.07922}}.

\bibitem{Berger:2019wnu}
\hrefCMSnoop {}{N.~Berger {et~al.}, ``Simplified template cross sections -
  stage 1.1'',} {LHC Higgs Cross Section Working Group Report}
  LHCHXSWG-2019-003, DESY-19-070, 2019.
\newblock
  \href{http://www.arXiv.org/abs/1906.02754}{\texttt{arXiv:1906.02754}}.

\bibitem{Aaboud:2018pen}
\hrefCMSnoop {}{{ATLAS Collaboration}, ``Cross-section measurements of the
  {Higgs} boson decaying into a pair of {$\tau$}-leptons in proton-proton
  collisions at {$\sqrt{s}=13$ TeV} with the {ATLAS detector}'',} \textit{
  Phys. Rev. D} \textbf{ 99} (2019) 072001,
  \href{http://dx.doi.org/10.1103/PhysRevD.99.072001}{\doi{10.1103/PhysRevD.99.072001}},
  \href{http://www.arXiv.org/abs/1811.08856}{\texttt{arXiv:1811.08856}}.

\bibitem{ATLAS:2022yrq}
\hrefCMSnoop {}{{ATLAS Collaboration}, ``Measurements of {Higgs} boson
  production cross-sections in the {$H\to\tau^{+}\tau^{-}$} decay channel in
  {$\Pp\Pp$} collisions at {$\sqrt{s}=13\,\text{TeV}$} with the {ATLAS
  detector}'',} 2022.
  \href{http://www.arXiv.org/abs/2201.08269}{\texttt{arXiv:2201.08269}}.
  Submitted to \textit{JHEP}.

\bibitem{CMS:2021gxc}
\hrefCMSnoop {}{{CMS Collaboration}, ``Measurement of the inclusive and
  differential {Higgs} boson production cross sections in the decay mode to a
  pair of {$\tau$} leptons in pp collisions at {$\sqrt{s} = $ 13 TeV}'',}
  \textit{ Phys. Rev. Lett.} \textbf{ 128} (2022) 081805,
  \href{http://dx.doi.org/10.1103/PhysRevLett.128.081805}{\doi{10.1103/PhysRevLett.128.081805}},
  \href{http://www.arXiv.org/abs/2107.11486}{\texttt{arXiv:2107.11486}}.

\bibitem{Sirunyan:2020zal}
\hrefCMSnoop {}{{CMS Collaboration}, ``Performance of the {CMS} {Level-1}
  trigger in proton-proton collisions at {$\sqrt{s} = 13$\,TeV}'',} \textit{
  JINST} \textbf{ 15} (2020) P10017,
  \href{http://dx.doi.org/10.1088/1748-0221/15/10/P10017}{\doi{10.1088/1748-0221/15/10/P10017}},
  \href{http://www.arXiv.org/abs/2006.10165}{\texttt{arXiv:2006.10165}}.

\bibitem{Khachatryan:2016bia}
\hrefCMSnoop {}{{CMS Collaboration}, ``The {CMS} trigger system'',} \textit{
  JINST} \textbf{ 12} (2017) P01020,
  \href{http://dx.doi.org/10.1088/1748-0221/12/01/P01020}{\doi{10.1088/1748-0221/12/01/P01020}},
\href{http://www.arXiv.org/abs/1609.02366}{\texttt{arXiv:1609.02366}}.

\bibitem{Chatrchyan:2008zzk}
\hrefCMSnoop {}{{CMS Collaboration}, ``The {CMS} experiment at the {CERN}
  {LHC}'',} \textit{ JINST} \textbf{ 3} (2008) S08004,
  \href{http://dx.doi.org/10.1088/1748-0221/3/08/S08004}{\doi{10.1088/1748-0221/3/08/S08004}}.

\bibitem{Sirunyan:2017ulk}
\hrefCMSnoop {}{{CMS Collaboration}, ``Particle-flow reconstruction and global
  event description with the {CMS} detector'',} \textit{ JINST} \textbf{ 12}
  (2017) P10003,
  \href{http://dx.doi.org/10.1088/1748-0221/12/10/P10003}{\doi{10.1088/1748-0221/12/10/P10003}},
  \href{http://www.arXiv.org/abs/1706.04965}{\texttt{arXiv:1706.04965}}.

\bibitem{CMS-TDR-15-02}
\href {http://cds.cern.ch/record/2020886}{{CMS Collaboration}, ``Technical
  proposal for the {Phase-II} upgrade of the {Compact Muon Solenoid}'',} CMS
  Technical Proposal CERN-LHCC-2015-010, CMS-TDR-15-02, 2015.

\bibitem{Khachatryan:2015hwa}
\hrefCMSnoop {}{{CMS Collaboration}, ``Performance of electron reconstruction
  and selection with the {CMS} detector in proton-proton collisions at
  {$\sqrt{s} = 8$\TeV}'',} \textit{ JINST} \textbf{ 10} (2015) P06005,
  \href{http://dx.doi.org/10.1088/1748-0221/10/06/P06005}{\doi{10.1088/1748-0221/10/06/P06005}},
\href{http://www.arXiv.org/abs/1502.02701}{\texttt{arXiv:1502.02701}}.

\bibitem{CMS:2020uim}
\hrefCMSnoop {}{{CMS Collaboration}, ``Electron and photon reconstruction and
  identification with the {CMS} experiment at the {CERN LHC}'',} \textit{
  JINST} \textbf{ 16} (2021) P05014,
  \href{http://dx.doi.org/10.1088/1748-0221/16/05/P05014}{\doi{10.1088/1748-0221/16/05/P05014}},
  \href{http://www.arXiv.org/abs/2012.06888}{\texttt{arXiv:2012.06888}}.

\bibitem{CMS:2012nsv}
\hrefCMSnoop {}{{CMS Collaboration}, ``Performance of {CMS} muon reconstruction
  in {$\Pp\Pp$} collision events at {$\sqrt{s}=7$ TeV}'',} \textit{ JINST}
  \textbf{ 7} (2012) P10002,
  \href{http://dx.doi.org/10.1088/1748-0221/7/10/P10002}{\doi{10.1088/1748-0221/7/10/P10002}},
  \href{http://www.arXiv.org/abs/1206.4071}{\texttt{arXiv:1206.4071}}.

\bibitem{CMS:2018rym}
\hrefCMSnoop {}{{CMS Collaboration}, ``Performance of the {CMS} muon detector
  and muon reconstruction with proton-proton collisions at {$\sqrt{s}=$ 13
  TeV}'',} \textit{ JINST} \textbf{ 13} (2018) P06015,
  \href{http://dx.doi.org/10.1088/1748-0221/13/06/P06015}{\doi{10.1088/1748-0221/13/06/P06015}},
  \href{http://www.arXiv.org/abs/1804.04528}{\texttt{arXiv:1804.04528}}.

\bibitem{Cacciari:2008gp}
\hrefCMSnoop {}{M.~Cacciari, G.~P. Salam, and G.~Soyez, ``The anti-\kt jet
  clustering algorithm'',} \textit{ JHEP} \textbf{ 04} (2008) 063,
  \href{http://dx.doi.org/10.1088/1126-6708/2008/04/063}{\doi{10.1088/1126-6708/2008/04/063}},
  \href{http://www.arXiv.org/abs/0802.1189}{\texttt{arXiv:0802.1189}}.

\bibitem{Cacciari:2011ma}
\hrefCMSnoop {}{M.~Cacciari, G.~P. Salam, and G.~Soyez, ``{FastJet} user
  manual'',} \textit{ Eur. Phys. J. C} \textbf{ 72} (2012) 1896,
  \href{http://dx.doi.org/10.1140/epjc/s10052-012-1896-2}{\doi{10.1140/epjc/s10052-012-1896-2}},
\href{http://www.arXiv.org/abs/1111.6097}{\texttt{arXiv:1111.6097}}.

\bibitem{Sirunyan:2017ezt}
\hrefCMSnoop {}{{CMS Collaboration}, ``Identification of heavy-flavour jets
  with the {CMS} detector in pp collisions at {13 TeV}'',} \textit{ JINST}
  \textbf{ 13} (2018) P05011,
  \href{http://dx.doi.org/10.1088/1748-0221/13/05/P05011}{\doi{10.1088/1748-0221/13/05/P05011}},
  \href{http://www.arXiv.org/abs/1712.07158}{\texttt{arXiv:1712.07158}}.

\bibitem{Bols:2020bkb}
E.~Bols\hrefCMSnoop {}{ {et~al.}, ``Jet flavour classification using
  {DeepJet}'',} \textit{ JINST} \textbf{ 15} (2020) P12012,
  \href{http://dx.doi.org/10.1088/1748-0221/15/12/P12012}{\doi{10.1088/1748-0221/15/12/P12012}},
  \href{http://www.arXiv.org/abs/2008.10519}{\texttt{arXiv:2008.10519}}.

\bibitem{CMS-DP-2018-058}
\href {http://cds.cern.ch/record/2646773}{{CMS Collaboration}, ``Performance of
  the {DeepJet} b tagging algorithm using 41.9/fb of data from proton-proton
  collisions at 13 {TeV} with {Phase 1 CMS detector}'',} CMS Detector
  Performance Note CMS-DP-2018-058, CERN, 2018.

\bibitem{Sirunyan:2018pgf}
\hrefCMSnoop {}{{CMS Collaboration}, ``Performance of reconstruction and
  identification of {$\tau$} leptons decaying to hadrons and {$\nu_\tau$} in pp
  collisions at {$\sqrt{s}=$ 13 TeV}'',} \textit{ JINST} \textbf{ 13} (2018)
  P10005,
  \href{http://dx.doi.org/10.1088/1748-0221/13/10/P10005}{\doi{10.1088/1748-0221/13/10/P10005}},
  \href{http://www.arXiv.org/abs/1809.02816}{\texttt{arXiv:1809.02816}}.

\bibitem{CMS:2022prd}
\hrefCMSnoop {}{{CMS Collaboration}, ``{Identification of hadronic tau lepton
  decays using a deep neural network}'',} \textit{ JINST} \textbf{ 17} (2022)
  P07023,
  \href{http://dx.doi.org/10.1088/1748-0221/17/07/P07023}{\doi{10.1088/1748-0221/17/07/P07023}},
  \href{http://www.arXiv.org/abs/2201.08458}{\texttt{arXiv:2201.08458}}.

\bibitem{Khachatryan:2014gga}
\hrefCMSnoop {}{{CMS Collaboration}, ``Performance of the {CMS} missing
  transverse momentum reconstruction in pp data at {$\sqrt{s}$ = 8 TeV}'',}
  \textit{ JINST} \textbf{ 10} (2015) P02006,
  \href{http://dx.doi.org/10.1088/1748-0221/10/02/P02006}{\doi{10.1088/1748-0221/10/02/P02006}},
  \href{http://www.arXiv.org/abs/1411.0511}{\texttt{arXiv:1411.0511}}.

\bibitem{Bertolini:2014bba}
\hrefCMSnoop {}{D.~Bertolini, P.~Harris, M.~Low, and N.~Tran, ``Pileup per
  particle identification'',} \textit{ JHEP} \textbf{ 10} (2014) 059,
  \href{http://dx.doi.org/10.1007/JHEP10(2014)059}{\doi{10.1007/JHEP10(2014)059}},
\href{http://www.arXiv.org/abs/1407.6013}{\texttt{arXiv:1407.6013}}.

\bibitem{Sirunyan:2019kia}
\hrefCMSnoop {}{{CMS Collaboration}, ``Performance of missing transverse
  momentum reconstruction in proton-proton collisions at {$\sqrt{s} = 13$\,TeV}
  using the {CMS} detector'',} \textit{ JINST} \textbf{ 14} (2019) P07004,
  \href{http://dx.doi.org/10.1088/1748-0221/14/07/P07004}{\doi{10.1088/1748-0221/14/07/P07004}},
\href{http://www.arXiv.org/abs/1903.06078}{\texttt{arXiv:1903.06078}}.

\bibitem{Bianchini:2014vza}
\hrefCMSnoop {}{L.~Bianchini, J.~Conway, E.~K. Friis, and C.~Veelken,
  ``Reconstruction of the {Higgs} mass in {$H\to\tau\tau$} events by dynamical
  likelihood techniques'',} \textit{ J. Phys. Conf. Ser.} \textbf{ 513} (2014)
  022035,
\href{http://dx.doi.org/10.1088/1742-6596/513/2/022035}{\doi{10.1088/1742-6596/513/2/022035}}.

\bibitem{Bullock:1992yt}
\hrefCMSnoop {}{B.~K. Bullock, K.~Hagiwara, and A.~D. Martin, ``Tau
  polarization and its correlations as a probe of new physics'',} \textit{
  Nucl. Phys. B} \textbf{ 395} (1993) 499,
  \href{http://dx.doi.org/10.1016/0550-3213(93)90045-Q}{\doi{10.1016/0550-3213(93)90045-Q}}.

\bibitem{PDG2020}
\hrefCMSnoop {}{{Particle Data Group}, P.~A. Zyla {et~al.}, ``Review of
  particle physics'',} \textit{ Prog. Theor. Exp. Phys.} \textbf{ 2020} (2020)
  083C01,
  \href{http://dx.doi.org/10.1093/ptep/ptaa104}{\doi{10.1093/ptep/ptaa104}}.

\bibitem{CMS-DP-2016-026}
\href {http://cds.cern.ch/record/2161024}{{CMS Collaboration}, ``Electron
  performance using first data collected by {CMS} in 2016'',} CMS Detector
  Performance Note CMS-DP-2016-026, 2016.

\bibitem{CMS-DP-2016-067}
\href {https://cds.cern.ch/record/2229697}{{CMS Collaboration}, ``Performance
  of muon reconstruction including alignment position errors for 2016 collision
  data'',} CMS Detector Performance Note CMS-DP-2016-067, 2016.

\bibitem{CMS-DP-2019-012}
\href {https://cds.cern.ch/record/2678958}{{CMS Collaboration}, ``Tau lepton
  {Run} 2 trigger performance'',} CMS Detector Performance Note
  CMS-DP-2019-012, 2019.

\bibitem{Sirunyan:2019drn}
\hrefCMSnoop {}{{CMS Collaboration}, ``An embedding technique to determine
  {$\tau\tau$} backgrounds in proton-proton collision data'',} \textit{ JINST}
  \textbf{ 14} (2019) P06032,
  \href{http://dx.doi.org/10.1088/1748-0221/14/06/P06032}{\doi{10.1088/1748-0221/14/06/P06032}},
  \href{http://www.arXiv.org/abs/1903.01216}{\texttt{arXiv:1903.01216}}.

\bibitem{Sirunyan:2018qio}
\hrefCMSnoop {}{{CMS Collaboration}, ``Measurement of the
  {$\mathrm{Z}\gamma^{*}\to\tau\tau$} cross section in pp collisions at
  {$\sqrt{s}=$ 13 TeV} and validation of {$\tau$} lepton analysis
  techniques'',} \textit{ Eur. Phys. J. C} \textbf{ 78} (2018) 708,
  \href{http://dx.doi.org/10.1140/epjc/s10052-018-6146-9}{\doi{10.1140/epjc/s10052-018-6146-9}},
  \href{http://www.arXiv.org/abs/1801.03535}{\texttt{arXiv:1801.03535}}.

\bibitem{Alwall:2011uj}
J.~Alwall\hrefCMSnoop {}{ {et~al.}, ``{MadGraph 5}: Going beyond'',} \textit{
  JHEP} \textbf{ 06} (2011) 128,
  \href{http://dx.doi.org/10.1007/JHEP06(2011)128}{\doi{10.1007/JHEP06(2011)128}},
\href{http://www.arXiv.org/abs/1106.0522}{\texttt{arXiv:1106.0522}}.

\bibitem{MadGraph}
J.~Alwall\hrefCMSnoop {}{ {et~al.}, ``The automated computation of tree-level
  and next-to-leading order differential cross sections, and their matching to
  parton shower simulations'',} \textit{ JHEP} \textbf{ 07} (2014) 079,
  \href{http://dx.doi.org/10.1007/JHEP07(2014)079}{\doi{10.1007/JHEP07(2014)079}},
  \href{http://www.arXiv.org/abs/1405.0301}{\texttt{arXiv:1405.0301}}.

\bibitem{Alwall:2007fs}
\hrefCMSnoop {}{J.~Alwall {et~al.}, ``Comparative study of various algorithms
  for the merging of parton showers and matrix elements in hadronic
  collisions'',} \textit{ Eur. Phys. J. C} \textbf{ 53} (2008) 473,
  \href{http://dx.doi.org/10.1140/epjc/s10052-007-0490-5}{\doi{10.1140/epjc/s10052-007-0490-5}},
  \href{http://www.arXiv.org/abs/0706.2569}{\texttt{arXiv:0706.2569}}.

\bibitem{Nason:2004rx}
\hrefCMSnoop {}{P.~Nason, ``A new method for combining {NLO QCD} with shower
  {Monte} {Carlo} algorithms'',} \textit{ JHEP} \textbf{ 11} (2004) 040,
  \href{http://dx.doi.org/10.1088/1126-6708/2004/11/040}{\doi{10.1088/1126-6708/2004/11/040}},
\href{http://www.arXiv.org/abs/hep-ph/0409146}{\texttt{arXiv:hep-ph/0409146}}.

\bibitem{Frixione:2007vw}
\hrefCMSnoop {}{S.~Frixione, P.~Nason, and C.~Oleari, ``Matching {NLO QCD}
  computations with parton shower simulations: the {POWHEG} method'',} \textit{
  JHEP} \textbf{ 11} (2007) 070,
  \href{http://dx.doi.org/10.1088/1126-6708/2007/11/070}{\doi{10.1088/1126-6708/2007/11/070}},
\href{http://www.arXiv.org/abs/0709.2092}{\texttt{arXiv:0709.2092}}.

\bibitem{Alioli:2008tz}
\hrefCMSnoop {}{S.~Alioli, P.~Nason, C.~Oleari, and E.~Re, ``{NLO Higgs} boson
  production via gluon fusion matched with shower in {POWHEG}'',} \textit{
  JHEP} \textbf{ 04} (2009) 002,
  \href{http://dx.doi.org/10.1088/1126-6708/2009/04/002}{\doi{10.1088/1126-6708/2009/04/002}},
\href{http://www.arXiv.org/abs/0812.0578}{\texttt{arXiv:0812.0578}}.

\bibitem{Alioli:2010xd}
\hrefCMSnoop {}{S.~Alioli, P.~Nason, C.~Oleari, and E.~Re, ``A general
  framework for implementing {NLO} calculations in shower {Monte Carlo}
  programs: the {POWHEG BOX}'',} \textit{ JHEP} \textbf{ 06} (2010) 043,
  \href{http://dx.doi.org/10.1007/JHEP06(2010)043}{\doi{10.1007/JHEP06(2010)043}},
\href{http://www.arXiv.org/abs/1002.2581}{\texttt{arXiv:1002.2581}}.

\bibitem{Alioli:2010xa}
S.~Alioli\hrefCMSnoop {}{ {et~al.}, ``Jet pair production in {POWHEG}'',}
  \textit{ JHEP} \textbf{ 04} (2011) 081,
  \href{http://dx.doi.org/10.1007/JHEP04(2011)081}{\doi{10.1007/JHEP04(2011)081}},
\href{http://www.arXiv.org/abs/1012.3380}{\texttt{arXiv:1012.3380}}.

\bibitem{Bagnaschi:2011tu}
\hrefCMSnoop {}{E.~Bagnaschi, G.~Degrassi, P.~Slavich, and A.~Vicini, ``Higgs
  production via gluon fusion in the {POWHEG} approach in the {SM} and in the
  {MSSM}'',} \textit{ JHEP} \textbf{ 02} (2012) 088,
  \href{http://dx.doi.org/10.1007/JHEP02(2012)088}{\doi{10.1007/JHEP02(2012)088}},
\href{http://www.arXiv.org/abs/1111.2854}{\texttt{arXiv:1111.2854}}.

\bibitem{Nason:2009ai}
\hrefCMSnoop {}{P.~Nason and C.~Oleari, ``{NLO Higgs} boson production via
  vector-boson fusion matched with shower in {POWHEG}'',} \textit{ JHEP}
  \textbf{ 02} (2010) 037,
  \href{http://dx.doi.org/10.1007/JHEP02(2010)037}{\doi{10.1007/JHEP02(2010)037}},
  \href{http://www.arXiv.org/abs/0911.5299}{\texttt{arXiv:0911.5299}}.

\bibitem{Luisoni:2013cuh}
\hrefCMSnoop {}{G.~Luisoni, P.~Nason, C.~Oleari, and F.~Tramontano,
  ``{$\PH\PWpm$/$\PH\PZ$} + 0 and 1 jet at {NLO} with the {POWHEG BOX}
  interfaced to {GoSam} and their merging within {MiNLO}'',} \textit{ JHEP}
  \textbf{ 10} (2013) 083,
  \href{http://dx.doi.org/10.1007/JHEP10(2013)083}{\doi{10.1007/JHEP10(2013)083}},
  \href{http://www.arXiv.org/abs/1306.2542}{\texttt{arXiv:1306.2542}}.

\bibitem{Hartanto:2015uka}
\hrefCMSnoop {}{H.~B. Hartanto, B.~Jager, L.~Reina, and D.~Wackeroth, ``{Higgs}
  boson production in association with top quarks in the {POWHEG BOX}'',}
  \textit{ Phys. Rev. D} \textbf{ 91} (2015) 094003,
  \href{http://dx.doi.org/10.1103/PhysRevD.91.094003}{\doi{10.1103/PhysRevD.91.094003}},
  \href{http://www.arXiv.org/abs/1501.04498}{\texttt{arXiv:1501.04498}}.

\bibitem{Hamilton:2013fea}
\hrefCMSnoop {}{K.~Hamilton, P.~Nason, E.~Re, and G.~Zanderighi, ``{NNLOPS}
  simulation of {Higgs} boson production'',} \textit{ JHEP} \textbf{ 10} (2013)
  222,
  \href{http://dx.doi.org/10.1007/JHEP10(2013)222}{\doi{10.1007/JHEP10(2013)222}},
  \href{http://www.arXiv.org/abs/1309.0017}{\texttt{arXiv:1309.0017}}.

\bibitem{Hamilton:2015nsa}
\hrefCMSnoop {}{K.~Hamilton, P.~Nason, and G.~Zanderighi, ``Finite quark-mass
  effects in the {NNLOPS POWHEG+MiNLO Higgs} generator'',} \textit{ JHEP}
  \textbf{ 05} (2015) 140,
  \href{http://dx.doi.org/10.1007/JHEP05(2015)140}{\doi{10.1007/JHEP05(2015)140}},
  \href{http://www.arXiv.org/abs/1501.04637}{\texttt{arXiv:1501.04637}}.

\bibitem{Heinemeyer:2016spz}
\hrefCMSnoop {}{{LHC Higgs Cross Section Working Group}, ``Handbook of {LHC
  Higgs} cross sections: 3. {Higgs} properties: Report of the {LHC Higgs Cross
  Section Working Group}'',} CERN Report CERN-2013-004, 2013.
\newblock
  \href{http://dx.doi.org/10.5170/CERN-2013-004}{\doi{10.5170/CERN-2013-004}},
  \href{http://www.arXiv.org/abs/1307.1347}{\texttt{arXiv:1307.1347}}.

\bibitem{Ball:2014uwa}
\hrefCMSnoop {}{{NNPDF} Collaboration, ``Parton distributions for the {LHC} run
  {II}'',} \textit{ JHEP} \textbf{ 04} (2015) 040,
  \href{http://dx.doi.org/10.1007/JHEP04(2015)040}{\doi{10.1007/JHEP04(2015)040}},
\href{http://www.arXiv.org/abs/1410.8849}{\texttt{arXiv:1410.8849}}.

\bibitem{Ball:2017nwa}
\hrefCMSnoop {}{{NNPDF} Collaboration, ``Parton distributions from
  high-precision collider data'',} \textit{ Eur. Phys. J. C} \textbf{ 77}
  (2017) 663,
  \href{http://dx.doi.org/10.1140/epjc/s10052-017-5199-5}{\doi{10.1140/epjc/s10052-017-5199-5}},
  \href{http://www.arXiv.org/abs/1706.00428}{\texttt{arXiv:1706.00428}}.

\bibitem{Sjostrand:2014zea}
T.~Sj{\"o}strand\hrefCMSnoop {}{ {et~al.}, ``An introduction to {PYTHIA
  8.2}'',} \textit{ Comput. Phys. Commun.} \textbf{ 191} (2015) 159,
  \href{http://dx.doi.org/10.1016/j.cpc.2015.01.024}{\doi{10.1016/j.cpc.2015.01.024}},
  \href{http://www.arXiv.org/abs/1410.3012}{\texttt{arXiv:1410.3012}}.

\bibitem{Khachatryan:2015pea}
\hrefCMSnoop {}{{CMS Collaboration}, ``Event generator tunes obtained from
  underlying event and multiparton scattering measurements'',} \textit{ Eur.
  Phys. J. C} \textbf{ 76} (2016) 155,
  \href{http://dx.doi.org/10.1140/epjc/s10052-016-3988-x}{\doi{10.1140/epjc/s10052-016-3988-x}},
  \href{http://www.arXiv.org/abs/1512.00815}{\texttt{arXiv:1512.00815}}.

\bibitem{Sirunyan:2019dfx}
\hrefCMSnoop {}{{CMS Collaboration}, ``{Extraction and validation of a new set
  of CMS PYTHIA8 tunes from underlying-event measurements}'',} \textit{ Eur.
  Phys. J. C} \textbf{ 80} (2020) 4,
  \href{http://dx.doi.org/10.1140/epjc/s10052-019-7499-4}{\doi{10.1140/epjc/s10052-019-7499-4}},
  \href{http://www.arXiv.org/abs/1903.12179}{\texttt{arXiv:1903.12179}}.

\bibitem{Melnikov:2006kv}
\hrefCMSnoop {}{K.~Melnikov and F.~Petriello, ``Electroweak gauge boson
  production at hadron colliders through
  {$\mathcal{O}(\alpha_\text{s}^{2})$}'',} \textit{ Phys. Rev. D} \textbf{ 74}
  (2006) 114017,
  \href{http://dx.doi.org/10.1103/PhysRevD.74.114017}{\doi{10.1103/PhysRevD.74.114017}},
\href{http://www.arXiv.org/abs/hep-ph/0609070}{\texttt{arXiv:hep-ph/0609070}}.

\bibitem{Czakon:2011xx}
\hrefCMSnoop {}{M.~Czakon and A.~Mitov, ``{Top++: A} program for the
  calculation of the top-pair cross-section at hadron colliders'',} \textit{
  Comput. Phys. Commun.} \textbf{ 185} (2014) 2930,
  \href{http://dx.doi.org/10.1016/j.cpc.2014.06.021}{\doi{10.1016/j.cpc.2014.06.021}},
\href{http://www.arXiv.org/abs/1112.5675}{\texttt{arXiv:1112.5675}}.

\bibitem{Kidonakis:2013zqa}
\hrefCMSnoop {}{N.~Kidonakis, ``Top quark production'',} in \textit{ {Helmholtz
  International Summer School on Physics of Heavy Quarks and Hadrons}}, p.~139.
\newblock 2014.
\newblock \href{http://www.arXiv.org/abs/1311.0283}{\texttt{arXiv:1311.0283}}.
\newblock
  \href{http://dx.doi.org/10.3204/DESY-PROC-2013-03/Kidonakis}{\doi{10.3204/DESY-PROC-2013-03/Kidonakis}}.

\bibitem{Campbell:2011bn}
\hrefCMSnoop {}{J.~M. Campbell, R.~K. Ellis, and C.~Williams, ``Vector boson
  pair production at the {LHC}'',} \textit{ JHEP} \textbf{ 07} (2011) 018,
  \href{http://dx.doi.org/10.1007/JHEP07(2011)018}{\doi{10.1007/JHEP07(2011)018}},
\href{http://www.arXiv.org/abs/1105.0020}{\texttt{arXiv:1105.0020}}.

\bibitem{Gehrmann:2014fva}
T.~Gehrmann\hrefCMSnoop {}{ {et~al.}, ``{$\PWp\PWm$} production at hadron
  colliders in next to next to leading order {QCD}'',} \textit{ Phys. Rev.
  Lett.} \textbf{ 113} (2014) 212001,
  \href{http://dx.doi.org/10.1103/PhysRevLett.113.212001}{\doi{10.1103/PhysRevLett.113.212001}},
\href{http://www.arXiv.org/abs/1408.5243}{\texttt{arXiv:1408.5243}}.

\bibitem{Agostinelli:2002hh}
\hrefCMSnoop {}{S.~Agostinelli {et~al.}, ``{\GEANTfour}---a simulation
  toolkit'',} \textit{ Nucl. Instrum. Meth. A} \textbf{ 506} (2003) 250,
\href{http://dx.doi.org/10.1016/S0168-9002(03)01368-8}{\doi{10.1016/S0168-9002(03)01368-8}}.

\bibitem{GoodBengCour16}
I.~J. Goodfellow, Y.~Bengio, and A.~Courville, ``Deep Learning''.
\newblock MIT Press, Cambridge, MA, USA, 2016.
\newblock \url{http://www.deeplearningbook.org}.

\bibitem{Gritsan:2016hjl}
\hrefCMSnoop {}{A.~V. Gritsan, R.~R{\"o}ntsch, M.~Schulze, and M.~Xiao,
  ``Constraining anomalous {Higgs} boson couplings to the heavy flavor fermions
  using matrix element techniques'',} \textit{ Phys. Rev. D} \textbf{ 94}
  (2016) 055023,
  \href{http://dx.doi.org/10.1103/PhysRevD.94.055023}{\doi{10.1103/PhysRevD.94.055023}},
  \href{http://www.arXiv.org/abs/1606.03107}{\texttt{arXiv:1606.03107}}.

\bibitem{Wunsch:2018oxb}
\hrefCMSnoop {}{S.~Wunsch, R.~Friese, R.~Wolf, and G.~Quast, ``Identifying the
  relevant dependencies of the neural network response on characteristics of
  the input space'',} \textit{ Comput. Softw. Big Sci.} \textbf{ 2} (2018) 5,
  \href{http://dx.doi.org/10.1007/s41781-018-0012-1}{\doi{10.1007/s41781-018-0012-1}},
  \href{http://www.arXiv.org/abs/1803.08782}{\texttt{arXiv:1803.08782}}.

\bibitem{glorot2010}
\href {https://proceedings.mlr.press/v9/glorot10a.html}{X.~Glorot and
  Y.~Bengio, ``Understanding the difficulty of training deep feedforward neural
  networks'',} in \textit{ Proceedings of the thirteenth international
  conference on artificial intelligence and statistics}, p.~249.
\newblock 2010.

\bibitem{Shimizu2018}
R.~{Shimizu}\hrefCMSnoop {}{ {et~al.}, ``Balanced mini-batch training for
  imbalanced image data classification with neural network'',} in \textit{ 2018
  {F}irst International Conference on Artificial Intelligence for Industries
  {(AI4I)}}, p.~27.
\newblock 2018.
\newblock
  \href{http://dx.doi.org/10.1109/AI4I.2018.8665709}{\doi{10.1109/AI4I.2018.8665709}}.

\bibitem{Kingma:2014vow}
\hrefCMSnoop {}{D.~P. Kingma and J.~Ba, ``Adam: {A} method for stochastic
  optimization'',} 2014.
\href{http://www.arXiv.org/abs/1412.6980}{\texttt{arXiv:1412.6980}}.

\bibitem{Tikhonov:1963}
\hrefCMSnoop {}{A.~N. Tikhonov, ``Solution of incorrectly formulated problems
  and the regularization method'',} \textit{ Soviet Math. Dokl.} \textbf{ 4}
  (1963) 1035.

\bibitem{Baker:1983tu}
\hrefCMSnoop {}{S.~Baker and R.~D. Cousins, ``Clarification of the use of
  chi-square and likelihood functions in fits to histograms'',} \textit{ Nucl.
  Instrum. Meth.} \textbf{ 221} (1984) 437,
\href{http://dx.doi.org/10.1016/0167-5087(84)90016-4}{\doi{10.1016/0167-5087(84)90016-4}}.

\bibitem{Abulencia:2005kq}
\hrefCMSnoop {}{{CDF} Collaboration, ``Search for neutral {MSSM Higgs} bosons
  decaying to tau pairs in {$\Pp\Pap$} collisions at {$\sqrt{s} = 1.96$
  TeV}'',} \textit{ Phys. Rev. Lett.} \textbf{ 96} (2006) 011802,
  \href{http://dx.doi.org/10.1103/PhysRevLett.96.011802}{\doi{10.1103/PhysRevLett.96.011802}},
  \href{http://www.arXiv.org/abs/hep-ex/0508051}{\texttt{arXiv:hep-ex/0508051}}.

\bibitem{Sirunyan:vh}
\hrefCMSnoop {}{{CMS Collaboration}, ``Search for the associated production of
  the {H}iggs boson and a vector boson in proton-proton collisions at
  {$\sqrt{s}=$ 13 TeV} via {H}iggs boson decays to {$\tau$} leptons'',}
  \textit{ JHEP} \textbf{ 06} (2019) 093,
  \href{http://dx.doi.org/10.1007/JHEP06(2019)093}{\doi{10.1007/JHEP06(2019)093}},
  \href{http://www.arXiv.org/abs/1809.03590}{\texttt{arXiv:1809.03590}}.

\bibitem{Barlow:1993dm}
\hrefCMSnoop {}{R.~J. Barlow and C.~Beeston, ``Fitting using finite {Monte
  Carlo} samples'',} \textit{ Comput. Phys. Commun.} \textbf{ 77} (1993) 219,
  \href{http://dx.doi.org/10.1016/0010-4655(93)90005-W}{\doi{10.1016/0010-4655(93)90005-W}}.

\bibitem{Khachatryan:2010xn}
\hrefCMSnoop {}{{CMS Collaboration}, ``Measurements of inclusive {$\PW$} and
  {$\PZ$} cross sections in {$\Pp\Pp$} collisions at {$\sqrt{s}=7\TeV$}'',}
  \textit{ JHEP} \textbf{ 01} (2011) 080,
  \href{http://dx.doi.org/10.1007/JHEP01(2011)080}{\doi{10.1007/JHEP01(2011)080}},
\href{http://www.arXiv.org/abs/1012.2466}{\texttt{arXiv:1012.2466}}.

\bibitem{Khachatryan:2016kdb}
\hrefCMSnoop {}{{CMS Collaboration}, ``Jet energy scale and resolution in the
  {CMS} experiment in {$\Pp\Pp$} collisions at 8 {TeV}'',} \textit{ JINST}
  \textbf{ 12} (2017) P02014,
  \href{http://dx.doi.org/10.1088/1748-0221/12/02/P02014}{\doi{10.1088/1748-0221/12/02/P02014}},
\href{http://www.arXiv.org/abs/1607.03663}{\texttt{arXiv:1607.03663}}.

\bibitem{Khachatryan:2015oqa}
\hrefCMSnoop {}{{CMS Collaboration}, ``Measurement of the differential cross
  section for top quark pair production in pp collisions at {$\sqrt{s} =
  8\TeV$}'',} \textit{ Eur. Phys. J. C} \textbf{ 75} (2015) 542,
  \href{http://dx.doi.org/10.1140/epjc/s10052-015-3709-x}{\doi{10.1140/epjc/s10052-015-3709-x}},
\href{http://www.arXiv.org/abs/1505.04480}{\texttt{arXiv:1505.04480}}.

\bibitem{CMS:2021xjt}
\hrefCMSnoop {}{{CMS Collaboration}, ``Precision luminosity measurement in
  proton-proton collisions at {$\sqrt{s} =$ 13 TeV} in 2015 and 2016 at
  {CMS}'',} \textit{ Eur. Phys. J. C} \textbf{ 81} (2021) 800,
  \href{http://dx.doi.org/10.1140/epjc/s10052-021-09538-2}{\doi{10.1140/epjc/s10052-021-09538-2}},
  \href{http://www.arXiv.org/abs/2104.01927}{\texttt{arXiv:2104.01927}}.

\bibitem{CMS-PAS-LUM-17-004}
\href {https://cds.cern.ch/record/2621960}{{CMS Collaboration}, ``{{CMS}
  luminosity measurement for the 2017 data-taking period at $\sqrt{s} =
  13{\TeV}$}'',} {CMS Physics Analysis Summary} {CMS-PAS-LUM-17-004}, 2018.

\bibitem{CMS-PAS-LUM-18-002}
\href {https://cds.cern.ch/record/2676164}{{CMS Collaboration}, ``{{CMS}
  luminosity measurement for the 2018 data-taking period at $\sqrt{s} =
  13~{\mathrm{TeV}}$}'',} {CMS Physics Analysis Summary} {CMS-PAS-LUM-18-002},
  2019.

\bibitem{Butterworth:2015oua}
\hrefCMSnoop {}{J.~Butterworth {et~al.}, ``{PDF4LHC} recommendations for {LHC}
  {R}un {II}'',} \textit{ J. Phys. G} \textbf{ 43} (2016) 023001,
  \href{http://dx.doi.org/10.1088/0954-3899/43/2/023001}{\doi{10.1088/0954-3899/43/2/023001}},
  \href{http://www.arXiv.org/abs/1510.03865}{\texttt{arXiv:1510.03865}}.

\bibitem{hepdata}
\hrefCMSnoop {}{}{HEPD}ata record for this analysis, 2022.
\newblock
  \href{http://dx.doi.org/10.17182/hepdata.127974}{\doi{10.17182/hepdata.127974}}.

\end{thebibliography}\endgroup
